\documentclass[11pt]{article}
\usepackage{times}
\usepackage{amsmath,amssymb,amsthm,amsfonts,latexsym,bbm,xspace,graphicx,float,mathtools,dsfont}
\usepackage[backref,colorlinks,citecolor=blue,bookmarks=true]{hyperref}
\usepackage[dvips, paper=letterpaper, top=1in, bottom=.75in, left=1in, right=1in, nohead, includefoot, footskip=.25in]{geometry}

\usepackage{color}
\usepackage{comment}
\usepackage{algorithm}
\usepackage{algorithmic}
\newcommand{\ind}{\mathds{1}}
\usepackage{tweaklist}

\usepackage{fullpage}
\usepackage{appendix}
\usepackage[nomessages]{fp}
\usepackage{multirow}

\usepackage{enumitem}
\usepackage{booktabs}

\usepackage{flushend}

\newtheorem{theorem}{Theorem}
\newtheorem{corollary}{Corollary}
\newtheorem{lemma}{Lemma}

\newtheorem{claim}{Claim}

\newtheorem{definition}{Definition}

\newtheorem{example}{Example}

\newtheorem{informaltheorem}{Informal Theorem}
\newcommand{\single}{\textsc{Single}}
\newcommand{\nf}{\textsc{Non-Favorite}}
\newcommand{\core}{\textsc{Core}}
\newcommand{\tail}{\textsc{Tail}}
\newcommand{\rev}{\textsc{Rev}}

\newcommand{\copies}{\textsc{OPT}^{\textsc{Copies-UD}}}

\newcommand{\srev}{\textsc{SRev}}
\newcommand{\brev}{\textsc{BRev}}
\newcommand{\aperev}{\textsc{APostEnRev}}
\newcommand{\prev}{\textsc{PostRev}}
\newcommand{\Med}{\textsc{Median}}

\newcommand{\sold}{\textsc{SOLD}}
\newenvironment{prevproof}[2]{\noindent {\em {Proof of {#1}~\ref{#2}:}}}{$\Box$\vskip \belowdisplayskip}

\newcommand{\Bs}{\bold{s_i}}
\newcommand{\Br}{\bold{r_i}}
\newcommand{\FF}{\mathcal{F}}
\newcommand{\SI}{\bf{\sigma}}
\newcommand{\R}{\ensuremath{\mathbb{R}}} 

\newcommand{\notshow}[1]{{}}

\DeclareMathOperator{\E}{E}

\def \E  {{\mathbb{E}}}

\def \polytope{{P(D)}}
\def \L{{\mathcal{L}}}

\DeclareMathOperator{\argmax}{argmax}
\DeclareMathOperator{\argmin}{argmin}

\def\tp{{\tilde{\varphi}}}

\definecolor{MyGray}{rgb}{0.8,0.8,0.8}

\begin{document}
\title{Simple Mechanisms for Subadditive Buyers via Duality}
\author{Yang Cai
\\ Department of Computer Science\\
Yale University\\
yang.cai@yale.edu \and Mingfei Zhao
\\ School of Computer Science\\ McGill University\\ mingfei.zhao@mail.mcgill.ca}
\addtocounter{page}{-1}

\maketitle

\begin{abstract}
We provide simple and approximately revenue-optimal mechanisms in the multi-item multi-bidder settings. We unify and improve all previous results, as well as generalize the results to broader cases. In particular, we prove that the better of the following two simple, deterministic and Dominant Strategy Incentive Compatible mechanisms, a sequential posted price mechanism or an anonymous sequential posted price mechanism with entry fee, achieves a constant fraction of the optimal revenue among all randomized, Bayesian Incentive Compatible mechanisms, when buyers' valuations are XOS over independent items. If the buyers' valuations are subadditive over independent items, the approximation factor degrades to $O(\log m)$, where $m$ is the number of items. We obtain our results by first extending the Cai-Devanur-Weinberg duality framework
to derive an effective benchmark of the optimal revenue for subadditive bidders, and then analyzing this upper bound with new techniques.


\end{abstract}
\thispagestyle{empty}

\newpage
\section{Introduction}
In Mechanism Design, we aim to design a mechanism/system such that a group of strategic participants, who are only interested in optimizing their own utilities, are incentivized to choose actions that also help achieve the designer's objective. Clearly, the quality of the solution with respect to the designer's objective is crucial. However, perhaps one should also pay equal attention to another criterion of a mechanism, that is, its simplicity. When facing a complicated mechanism, participants may be confused by the rules and thus unable to optimize their actions and react in unpredictable ways instead. This may lead to undesirable outcomes and poor performance of the mechanism. An ideal mechanism would be optimal and simple. However, such cases of simple mechanisms being optimal only exist in single-item auctions, with the seminal examples of auctions by Vickrey~\cite{Vickrey61} and Myerson~\cite{Myerson81}, while none has been discovered in broader settings. Indeed, we now know that even in fairly simple settings the optimal mechanisms suffer many undesirable properties including randomization, non-monotonicity, and others~\cite{RochetC98, Tha04, Pavlov11a, HartN13, HartR12, BriestCKW10, DaskalakisDT13, DaskalakisDT14}. 
To move forward, one has to compromise -- either settle with optimal but somewhat complex mechanisms or turn to simple but approximately optimal solutions.

Recently, there has been extensive research effort focusing on the latter approach, that is, studying the performance of simple mechanisms through the lens of approximation. In particular, a central problem on this front is how to design simple and approximately revenue-optimal mechanisms in multi-item settings. For instance, when bidders have unit-demand valuations, we know sequential posted price mechanisms approximates the optimal revenue due to a line of work initiated by Chawla et al.~\cite{ChawlaHK07, ChawlaHMS10, ChawlaMS15, CaiDW16}. When buyers have additive valuations, we know that either selling the items separately or running a VCG mechanism with per bidder entry fee approximates the optimal revenue due to a series of work initiated by Hart and Nisan~\cite{HartN12, CaiH13, LiY13, BabaioffILW14, Yao15, CaiDW16}. Recently, Chawla and Miller~\cite{ChawlaM16} generalized the two lines of work described above to matroid rank functions\footnote{{Here is a hierarchy of the valuation functions. additive \& unit-demand $\subseteq$ matroid rank $\subseteq$ constrained additive \&  submodular
 $\subseteq$ XOS $\subseteq$ subadditive. A function is constrained additive if it is additive up to some downward closed feasibility constraints. The class of submodular functions is neither a superset nor a subset of the class of constrained additive functions.} See Definition~\ref{def:valuation classes} for the formal definition. }. They show that a simple mechanism, the sequential two-part tariff mechanism, suffices to extract a constant fraction of the optimal revenue. For subadditive valuations beyond matroid rank functions, we only know how to handle a single buyer~\cite{RubinsteinW15}\footnote{All results mentioned above assume that the buyers' valuation distributions are over independent items. For additive and unit-demand valuations, this means a bidder's values for the items are independent. The definition is generalized to subadditive valuations by Rubinstein and Weinberg~\cite{RubinsteinW15}. See Definition~\ref{def:subadditive independent}.}. It is a major open problem to extend this result to multiple subadditive buyers.

In this paper, we unify and strengthen all the results mentioned above via an extension of the duality framework proposed by Cai et al.~\cite{CaiDW16}. Moreover, we show that even when there are multiple buyers with XOS valuation functions, there exists a simple, deterministic and Dominant Strategy Incentive Compatible (DSIC) mechanism that achieves a constant fraction of the optimal Bayesian Incentive Compatible (BIC) revenue\footnote{A mechanism is Bayesian Incentive Compatible (BIC) if it is in every bidder's interest to tell the truth, assuming that all other bidders' reported their values. A mechanism is Dominant Strategy Incentive Compatible (DSIC) if it is in every bidder's interest to tell the truth no matter what reports the other bidders make.}. For subadditive valuations, our approximation ratio degrades to $O(\log m)$.

\begin{informaltheorem}
	There exists a simple, deterministic and DSIC mechanism that achieves a constant fraction of the optimal BIC revenue in multi-item settings, when the buyers' valuation distributions are XOS over independent items. When the buyers' valuation distributions are subadditive over independent items, our mechanism achieves at least $\Omega(\frac{1}{\log m})$ of the optimal BIC revenue, where $m$ is the number of items.
\end{informaltheorem}
The original paper by Cai et al.~\cite{CaiDW16} provided a unified treatment  for additive and unit-demand valuations. However, it is inadequate to provide an analyzable benchmark for even a single subadditive bidder. In this paper, we show how to extend their duality framework to accommodate general subadditive valuations. Using this extended framework, we substantially improve the approximation ratios for many of the settings discussed above, and in the meantime generalize the results to broader cases. See Table~\ref{table:comp} for the comparison between the best ratios reported in the literature and the new ratios obtained in this work. 

\begin{table*}

\centering
\begin{tabular}{|c|l|p{2.1cm}|c|p{2cm}|c|c|}
\hline
	& &\centering Additive or Unit-demand& \multirow{2}{*}{\rotatebox[origin=c]{0}{\parbox[c]{2.2cm}{\centering Matroid-Rank}}}& \centering Constrained Additive&\multirow{2}{*}{\rotatebox[origin=c]{0}{\parbox[c]{1cm}{\centering XOS}}}&\multirow{2}{*}{\rotatebox[origin=c]{0}{\parbox[c]{2cm}{\centering Subadditive}}} \\
\hline
\multirow{2}{*}{\rotatebox[origin=c]{0}{\parbox[c]{1.3cm}{\centering Single Buyer}}} & Previous&      6~\cite{BabaioffILW14} or  4~\cite{ChawlaMS15}& 31.1* &\centering 31.1~\cite{ChawlaM16} & 338* & 338~\cite{RubinsteinW15} \\\cline{2-7}
		&This Paper & \centering - &{11*} &\centering{11}&{40*}&  {40} \\\cline{2-7}
\hline\hline
\multirow{2}{*}{\rotatebox[origin=c]{0}{\parbox[c]{1.3cm}{\centering Multiple Buyer}}} & Previous&     8~\cite{CaiDW16}  or 24~\cite{CaiDW16}& 133~\cite{ChawlaM16}&\centering ? & ? &? \\ \cline{2-7}
		& This Paper& \centering - & 70*  &\centering 70 & 268 &$O(\log m)$  \\
\hline
\end{tabular}\\
* The result is implied by another result for a more general setting.
\caption{Comparison of approximation ratios between previous and current work.}
		  \label{table:comp}
\end{table*}

\notshow{
\begin{table}
	\begin{tabular}{|l|c|c|c|c|c|}
	\hline
		                  & Additive or Unit-demand& Matroid-Rank& Constrained Additive&XOS&Subadditive \\
		\hline
		Single Buyer&      6~\cite{BabaioffILW14} or  4~\cite{ChawlaMS15}& 33.1~\cite{ChawlaM16} &$\rightarrow$ &$\rightarrow$ & 338~\cite{RubinsteinW15} \\\hline
		Multiple Buyers&     8~\cite{CaiDW16}  or 24~\cite{CaiDW16}& 133~\cite{ChawlaM16}& ? & ? &? \\
		\hline
		\end{tabular}
		  \caption{The best approximation ratios known prior to this work.}
		  \label{table:old}
\end{table}

\begin{table}\centering
	\begin{tabular}{|l|c|c|c|c|}
	\hline
		                   &Matroid-Rank& Constrained Additive&XOS&Subadditive \\
		\hline
		Single Buyer&  $\rightarrow$ &14& $\rightarrow$& 48 \\\hline
		Multiple Buyers& $\rightarrow$  & 70 & 268 &$O(\log m)$ \\
		\hline
		\end{tabular}
		  \caption{New approximation ratios obtained in this work.}
		  \label{table:new}
\end{table}

}
Our mechanism is either a \emph{rationed sequential posted price mechanism} (\textbf{RSPM}) or an \emph{anonymous sequential posted price with entry fee mechanism} (\textbf{ASPE}). In an RSPM, there is a price $p_{ij}$ for buyer $i$ if she wants to buy item $j$, and she is allowed to purchase at most one item. We visit the buyers in some arbitrary order and the buyer takes her favorite item among the available items given the item prices for her. Here we allow personalized prices, that is, $p_{ij}$ could be different from $p_{kj}$ if $i\neq k$. In an ASPE, every buyer faces the same collection of item prices $\{p_j\}_{j\in[m]}$. Again, we visit the buyers in some arbitrary order. For each buyer, we show her the available items and the associated price for each item. Then we ask her to pay the entry fee to enter the mechanism, which may depend on what items are still available and the identity of the buyer. If the buyer accepts the entry fee, she can proceed to purchase any item at the given prices; if she rejects the entry fee, then she will leave the mechanism without receiving anything. Given the entry fee and item prices, the decision making for the buyer is straightforward, as she only accepts the entry fee when the surplus for winning her favorite bundle is larger than the entry fee. Therefore, both RSPM and ASPE are DSIC and ex-post Individually Rational (ex-post IR).

\subsection{Our Contributions}
To obtain the new generalizations, we provide important extensions to the duality framework in~\cite{CaiDW16}, as well as novel analytic techniques and new simple mechanisms.

\vspace{.05in}
\noindent \textbf{1. Accommodating subadditive valuations:} the original duality framework in~\cite{CaiDW16} already unified the additive case and unit-demand case by providing an approximately tight upper bound for the optimal revenue using a single dual solution. A trivial upper bound for the revenue is the social welfare, which may be arbitrarily bad in the worst case. The duality based upper bound in~\cite{CaiDW16} improves this trivial upper bound, the social welfare, by substituting the value of each buyer's favorite item with the corresponding Myerson's virtual value. However, the substitution is viable only when the following condition holds -- the buyer's marginal gain for adding an item solely depends on her value for that item (assuming it's feasible to add that item\footnote{WLOG, we can reduce any constrained additive valuation to an additive valuation with a  feasibility constraint (see Definition~\ref{def:valuation classes})}), but not the set of items she has already received. This applies to valuations that are additive, unit-demand and more generally constrained additive, but breaks under more general valuation functions, e.g., submodular, XOS or subadditive valuations. As a consequence, the original dual solution from~\cite{CaiDW16} fails to provide a nice upper bound for more general valuations. To overcome this difficulty, we take a different approach. Instead of directly studying the dual of the original problem, we first relax the valuations and argue that the optimal revenue of the relaxed valuation is comparable to the original one. Then, since we choose the relaxation in a particular way, by applying a dual solution similar to the one in~\cite{CaiDW16} to the relaxed valuation, we recover an upper bound of the optimal revenue for the relaxed valuation resembling the appealing format of the one in~\cite{CaiDW16}. Combining these two steps, we obtain an upper bound for subadditive valuations that is easy to analyze. Indeed, we use our new upper bound to improve the approximation ratio for a single subadditive buyer from $338$~\cite{RubinsteinW15}  to $40$. See Section~\ref{sec:valuation relaxation} for more details.

\vspace{.05in}
\noindent\textbf{2. An adaptive dual:}  our second major change to the framework is that we choose the dual in an adaptive manner. In~\cite{CaiDW16}, a dual solution $\lambda$ is chosen up front inducing a virtual value function $\Phi(\cdot)$, then the corresponding optimal virtual welfare is used as a benchmark for the optimal revenue. Finally, it is shown that the revenue of some simple mechanism is within a constant factor of the optimal virtual welfare. Unfortunately, when the valuations are beyond additive and unit-demand, the optimal virtual welfare for this particular choice of virtual value function becomes extremely complex and hard to analyze. Indeed, it is already challenging to bound when the buyers' valuations are $k$-demand. In this paper, we take a more flexible approach. For any particular allocation rule $\sigma$, we tailor a special dual $\lambda^{(\sigma)}$ based on $\sigma$ in a fashion that is inspired by Chawla and Miller's ex-ante relaxation~\cite{ChawlaM16}. Therefore, the induced virtual valuation $\Phi^{(\sigma)}$ also depends on $\sigma$. By duality, we can show that the optimal revenue obtainable by $\sigma$ is still upper bounded by the virtual welfare with respect to $\Phi^{(\sigma)}$ under allocation rule $\sigma$. Since the virtual valuation is designed specifically for allocation $\sigma$, the induced virtual welfare is much easier to analyze. Indeed, we manage to prove that for any allocation $\sigma$ the induced virtual welfare is within a constant factor of the revenue of some simple mechanism, when bidders have XOS valuations. See Section~\ref{sec:virtual for relaxed} and~\ref{sec:choice of beta} for more details.

\vspace{.05in}

\noindent\textbf{3. A novel analysis and new mechanism:} with the two contributions above, we manage to derive an upper bound of the optimal revenue similar to the one in \cite{CaiDW16} but for subadditive bidders. The third major contribution of this paper is a novel approach to analyzing this upper bound. The analysis in~\cite{CaiDW16} essentially breaks the upper bound into three different terms-- \single, \tail~ and \core, and bound them separately. All three terms are relatively simple to bound for additive and unit-demand buyers, but for more general settings the $\core$ becomes much more challenging to handle. Indeed, the analysis in~\cite{CaiDW16} was  insufficient to tackle the $\core$ even when the buyers have $k$-demand valuations\footnote{The class of $k$-demand valuations is a generalization of unit-demand valuations, where the buyer's value is additive up to $k$ items.}-- a very special case of matroid rank valuations, which itself is a special case of XOS or subadditive valuations. Rubinstein and Weinberg~\cite{RubinsteinW15} showed how to approximate the $\core$ for a single subadditive bidder using grand bundling, but their approach does not apply to multiple bidders. Yao~\cite{Yao15} showed how to approximate the $\core$ for multiple additive bidders using a VCG with per bidder entry fee mechanism, but again it is unclear how his approach can be extended to multiple k-demand bidders. A recent paper by Chawla and Miller~\cite{ChawlaM16} finally broke the barrier of analyzing the $\core$ for multiple $k$-demand buyers. They showed how to bound the $\core$ for matroid rank valuations  using a sequential posted price mechanism by applying the \emph{online contention resolution scheme (OCRS)} developed by Feldman et al.~\cite{FeldmanSZ16}. The connection with OCRS is an elegant observation, and one might hope the same technique applies to more general valuations. Unfortunately, OCRS is only known to exist for special cases of downward closed constraints, and as we show in Section~\ref{sec:core comparison}, the approach by Chawla and Miller cannot yield any constant factor approximation for general constrained additive valuations.

We take an entirely different approach to bound the $\core$. Here we provide some intuition behind our mechanism and analysis. The $\core$ is essentially the optimal social welfare induced by some truncated valuation $v'$, and our goal is to design a mechanism that extracts a constant fraction of the welfare as revenue. Let $M$ be any sequential posted price mechanism. A key observation is that when bidder $i$'s valuation is subadditive over independent items, her utility in $M$, which is the largest surplus she can achieve from the unsold items, is also subadditive over independent items. If we can argue that her utility function is $a$-Lipschitz (Definition~\ref{def:Lipschitz}) with some small $a$,  Talagrand's concentration inequality~\cite{Talagrand1995concentration,Schechtman2003concentration} allows us to set an entry fee for the bidder so that we can extract a constant fraction of her utility just through the entry fee. 
If we modify $M$ by introducing an entry fee for every bidder, according to Talagrand's concentration inequality, the new mechanism $M'$ should intuitively have revenue that is a constant fraction of the social welfare obtained by $M$~\footnote{$M$'s welfare is simply its revenue plus the sum of utilities of the bidders, and $M'$ can extract some extra revenue from the entry fee, which is a constant fraction of the total utility from the bidders.}. Therefore, if there exists a sequential posted price mechanism $M$ that achieves a constant fraction of the optimal social welfare under the truncated valuation $v'$, the modified mechanism $M'$ can obtain a constant fraction of $\core$ as revenue. Surprisingly, when the bidders have XOS valuations, Feldman et al.~\cite{FeldmanGL15} showed that there exists an anonymous sequential posted price mechanism that always obtains at least half of the optimal social welfare. Hence, an anonymous sequential posted price with per bidder entry fee mechanism should approximate the $\core$ well, and this is exactly the intuition behind our ASPE mechanism.

 To turn the intuition into a theorem, there are two technical difficulties that we need to address: (i) the Lipschitz constants of the bidders' utility functions turn out to be too large (ii) we deliberately neglected the difference in bidders' behavior under $M$ and $M'$ in hope to keep our discussion in the previous paragraph intuitive. However, due to the entry fee, bidders may end up purchasing completely different items under $M$ and $M'$, so it is not straightforward to see how one can relate the revenue of $M'$ to the welfare obtained by $M$. 
  See Section~\ref{sec:core comparison} for a more detailed discussion on how we overcome these two difficulties.

\subsection{Related Work}
In recent years, we have witnessed several breakthroughs in designing (approximately) optimal mechanisms in multi-dimensional settings. The black-box reduction by Cai et al.~\cite{CaiDW12a,CaiDW12b,CaiDW13a,CaiDW13b} shows that we can reduce any Bayesian mechanism design problem to a similar algorithm design problem via convex optimization. Through their reduction, it is proved that all optimal mechanisms can be characterized as a distribution of virtual welfare maximizers, where the virtual valuations are computed by an LP. Although this characterization provides important insights about the structure of the optimal mechanism, the optimal allocation rule is unavoidably randomized and might still be complex as the virtual valuations are only a solution of an LP.

Another line of work considers the ``Simple vs. Optimal'' auction design problem. For instance, a sequence of results~\cite{ChawlaHK07,ChawlaHMS10,ChawlaMS10,ChawlaMS15} show that sequential posted price mechanism can achieve $\frac{1}{33.75}$ of the optimal revenue, whenever the buyers have unit-demand valuations over independent items. Another series of results~\cite{HartN12,CaiH13,LiY13,BabaioffILW14,Yao15} show that the better of selling the items separately and running the VCG mechanism with per bidder entry fee achieves $\frac{1}{69}$ of the optimal revenue, whenever the buyers' valuations are additive over independent items. Cai et al.~\cite{CaiDW16} unified the two lines of results and improved the approximation ratios to $\frac{1}{8}$ for the additive case and $\frac{1}{24}$ for the unit-demand case using their duality framework.

Some recent works have shown that simple mechanisms can approximate the optimal revenue even when buyers have more sophisticated valuations. For instance, Chawla and Miller~\cite{ChawlaM16} showed that the sequential two-part tariff mechanism can approximate the optimal revenue when buyers have matroid rank valuation functions over independent items. Their mechanism requires every buyer to pay an entry fee up front, and then run a sequential posted price mechanism on buyers who have accepted the entry fee. Our ASPE is similar to their mechanism, but with two major differences: (i) since buyers are asked to pay the entry fee before the seller visits them, the buyers have to make their decisions based on the expected utility (assuming every other buyer behaves truthfully) they can receive. Hence, the mechanism is only guaranteed to be BIC and interim IR. While in our mechanism, the buyers can see what items are still available before paying the entry fee, therefore the decision making is straightforward and the ASPE is DSIC and ex-post IR; (ii) the item prices in the ASPE are anonymous, while in the sequential two-part tariff mechanism, personalized prices are allowed. For valuations beyond matroid rank functions, Rubinstein and Weinberg~\cite{RubinsteinW15} showed that for a single buyer whose valuation is subadditive over independent items, either grand bundling or selling the items separately achieves at least $\frac{1}{338}$ of the optimal revenue.

The Cai-Devanur-Weinberg duality framework~\cite{CaiDW16} has been applied to other intriguing Mechanism Design problems. For example, Eden et al. showed that the better of selling separately and bundling together gets an $O(d)$-approximation for a single bidder with ``complementarity-$d$ valuations over independent items''~\cite{EdenFFTW16a}. The same authors also proved a Bulow-Klemperer result for regular i.i.d. and constrained additive bidders~\cite{EdenFFTW16b}. Liu and Psomas provided a Bulow-Klemperer result for {dynamic auctions}~\cite{LiuP16}. Finally, Brustle et al.~\cite{BrustleCWZ17} extended the duality framework to two-sided markets and used it to design simple mechanisms for approximating the Gains from Trade.

Strong duality frameworks have recently been developed for one additive buyer~\cite{DaskalakisDT13,DaskalakisDT15,Giannakopoulos14a,GiannakopoulosK14,GiannakopoulosK15}. These frameworks show that the dual problem of revenue maximization can be viewed as an optimal transport/bipartite matching problem. Hartline and Haghpanah provided an alternative duality framework in~\cite{HartlineH15}. They showed that if certain paths exist, these paths provide a witness of the optimality of a certain Myerson-type mechanism, but these paths are not guaranteed to exist in general. Similar to the Cai-Devanur-Weinberg framework, Carroll~\cite{Carroll15} independently made use of a partial Lagrangian over incentive constraints. These duality frameworks have been successfully provide conditions under which a certain type of mechanism is optimal when there is a single unit-demand or additive bidder. However, none of these frameworks succeeds in yielding any approximately optimal results in multi-buyer settings.

\section{Preliminaries}\label{sec:prelim}

We focus on revenue maximization in the combinatorial auction with $n$ independent bidders and $m$ heterogenous items. Each bidder has a valuation that is \textbf{subadditive over independent items} (see Definition~\ref{def:subadditive independent}). We denote bidder $i$'s type $t_i$ as $\langle t_{ij}\rangle_{j=1}^m$, where $t_{ij}$ is bidder $i$'s private information about item $j$. For each $i$, $j$, we assume $t_{ij}$ is drawn independently from the distribution $D_{ij}$. Let $D_i=\times_{j=1}^m D_{ij}$ be the distribution of bidder $i$'s type and $D=\times_{i=1}^n D_i$ be the distribution of the type profile. We use $T_{ij}$ (or $T_i, T$) and $f_{ij}$ (or $f_i, f$) to denote the support and density function of $D_{ij}$ (or $D_i, D$). For notational convenience, we let  $t_{-i}$ to be the types of all bidders except $i$ and $t_{<i}$ (or $t_{\leq i})$ to be the types of the first $i-1$ (or $i$) bidders. Similarly, we define $D_{-i}$, $T_{-i}$
  and $f_{-i}$ for the corresponding  distributions, support sets and density functions. When bidder $i$'s type is $t_i$, her valuation for a set of items $S$ is denoted by $v_i(t_i,S)$. 

\begin{definition}~\cite{RubinsteinW15}\label{def:subadditive independent}
For every bidder $i$, whose type is drawn from a product distribution $F_i=\prod_j F_{ij}$, her distribution $\mathcal{V}_i$ of valuation function $v_i(t_i,\cdot)$ is \textbf{subadditive over independent items} if:

\begin{itemize}[leftmargin=0.7cm]
\item \textbf{- $v_i(\cdot,\cdot)$ has no externalities}, i.e., for each $t_i\in T_i$ and $S\subseteq [m]$, $v_i(t_i,S)$ only depends on $\langle t_{ij}\rangle_{j\in S}$, formally, for any $t_i'\in T_i$ such that $t_{ij}'=t_{ij}$ for all $j\in S$, $v_i(t_i',S)=v_i(t_i,S)$.

\item \textbf{- $v_i(\cdot,\cdot)$ is monotone}, i.e., for all $t_i\in T_i$ and $U\subseteq V\subseteq [m]$, $v_i(t_i,U)\leq v_i(t_i,V)$.

\item \textbf{- $v_i(\cdot,\cdot)$ is subadditive}, i.e., for all $t_i\in T_i$ and $U, V\subseteq [m]$, $v_i(t_i,U\cup V)\leq v_i(t_i,U)+ v_i(t_i,V)$.
    
\end{itemize}

 We use $V_i(t_{ij})$ to denote $v_i(t_i,\{j\})$, as it only depends on $t_{ij}$. When $v_i(t_i,\cdot)$ is XOS (or constrained additive) for all $i$ and $t_i\in T_i$, we say $\mathcal{V}_i$ is XOS (or constrained additive) over independent items.
\end{definition}

We first formally define various valuation classes.
\begin{definition}\label{def:valuation classes}
We define several classes of valuations formally. Let $t$ be the type and $v(t,S)$ be the value for bundle $S\in[m]$.
	
\begin{itemize}[leftmargin=0.7cm]
	\item  \textbf{Constrained Additive:}
$v(t,S) =\max_{R\subseteq S, R\in \mathcal{I}}\sum_{j\in R} v(t, \{j\})$, where $\mathcal{I}\subseteq 2^{[m]}$ is a downward closed set system over the items specifying the feasible bundles. In particular, when $\mathcal{I}=2^{[m]}$, the valuation is an \textbf{additive function}; when $\mathcal{I}=\{\{j\}\ |\ j\in[m] \}$, the valuation is a \textbf{unit-demand function}; when $\mathcal{I}$ is a matroid, the valuation is a \textbf{matroid-rank function}. An equivalent way to represent any constrained additive valuations is to view the function as additive but the bidder is only allowed to receive bundles that are feasible, i.e., bundles in $\mathcal{I}$. To ease notations, we interpret $t$ as an $m$-dimensional vector $(t_1, t_2,\cdots, t_m)$ such that $t_j =  v(t, \{j\})$.

\item \textbf{XOS/Fractionally Subadditive:}
 $v(t,S) = \max_{i\in [{K}]} v^{(i)}(t, S)$, where ${K}$ is some finite number and $v^{(i)}(t,\cdot)$ is an additive function for any $i\in[{K}]$.

\item \textbf{Subadditive:}
$v(t,S_1\cup S_2)\leq v(t,S_1)+v(t,S_2)$ for any $S_1, S_2\subseteq [m]$.
\end{itemize}

\end{definition}

{The following are a few examples of various valuation distributions which are over independent items (Definition~\ref{def:subadditive independent}):

\begin{example}\label{eg:valuation}\cite{RubinsteinW15}
$t=\{t_j\}_{j\in[m]}$ where $t$ is drawn from $\prod_j D_j$,
\begin{itemize}[leftmargin=0.7cm]
\item Additive: $t_j$ is the value of item $j$. $v(t,S)=\sum_{j\in S}t_j$.
\item Unit-demand: $t_j$ is the value of item $j$. $v(t,S)=\max_{j\in S}t_j$.
\item Constrained Additive: $t_j$ is the value of item $j$.
 $v(t,S)=$\\
 \noindent$\max_{R\subseteq S, R\in \mathcal{I}}\sum_{j\in R} t_j$.
\item XOS/Fractionally Subadditive: $t_j=\{t_{j}^{(k)}\}_{k\in[K]}$ encodes all the possible values associated with item $j$, and $v(t,S)=$\\
    \noindent$\max_{k\in[K]}\sum_{j\in S}t_{j}^{(k)}$.
\end{itemize}
\end{example}

}

Given  $D$ and $v=\{v_i(\cdot,\cdot)\}_{i\in [n]}$, we use \textbf{$\textsc{Rev}(M,v,D)$} to denote the expected revenue of a BIC mechanism $M$. Throughout the paper, we use the following notations for the simple mechanisms we consider.

\vspace{.05in}
\noindent\textbf{Single-Bidder Mechanisms:}
\vspace{.03in}

\noindent \textbf{- $\srev(v,D)$} denotes the optimal expected revenue achievable by any posted price mechanism that only allows the buyer to purchase at most one item, and we use $\srev$ for short if there is no confusion\footnote{The mechanism is slightly different from selling separately, as we only allow the buyer to purchase at most one item.}.

\noindent \textbf{- $\brev(v,D)$} denotes the optimal expected revenue achievable by selling a grand bundle and we use $\brev$ for short if there is no confusion.

\vspace{.05in}
\noindent\textbf{Multi-Bidder Mechanisms:}
\vspace{.03in}

\noindent\textbf{- $\prev(v,D)$} denotes the optimal expected revenue achievable by selling the items via an RSPM to the bidders, and we use $\prev$ for short when there is no confusion.

\noindent\textbf{- $\aperev(v,D)$} denotes the optimal expected revenue achievable by selling the items via an ASPE to the bidders, and we use $\aperev$ for short when there is no confusion.

\vspace{.05in}
\noindent\textbf{Single-Dimensional Copies Setting:} In the analysis for unit-demand bidders in~\cite{ChawlaHMS10, CaiDW16}, the optimal revenue is upper bounded by the optimal revenue in the single-dimensional copies setting defined in~\cite{ChawlaHMS10}. We use the same technique. We construct $nm$ agents, where agent $(i,j)$ has value $V_i(t_{ij})$ of being served with $t_{ij}\sim D_{ij}$, and we are only allow to use matchings, that is, for each $i$ at most one agent $(i,k)$ is served and for each $j$ at most one agent $(k,j)$ is served\footnote{This is exactly the copies setting used in~\cite{ChawlaHMS10}, if every bidder $i$ is unit-demand and has value $V_i(t_{ij})$ with type $t_i$. Notice that this unit-demand multi-dimensional setting is equivalent as adding an extra constraint, each buyer can purchase at most one item, to the original setting with subadditive bidders.}. 
Notice that this is a single-dimensional setting, as each agent's type is specified by a single number. Let $\copies$ be the optimal BIC revenue in this copies setting.

\vspace{.05in}
\noindent\textbf{Continuous vs. Discrete Distributions:} We explicitly assume that the input distributions are discrete. Nevertheless, it is known that every $D$ can be discretized into $D^{+}$ such that the optimal revenue for $D$ and $D^{+}$ are within $(1\pm\epsilon)$ of each other~\cite{CaiDW16}. So our results also apply to  continuous distributions.  
\subsection{Our Mechanisms}
In this section, we introduce a class of mechanisms called \emph{Sequential Posted Price with Entry Fee}. For each bidder $i$, the mechanism first determines a posted price $\xi_{ij}$ for each item $j$ and an entry fee function ${\delta_i}(\cdot):2^{[m]}\to {\mathbb{R}_{\geq 0}}$ for each bidder $i$ that maps the set of available items to a real value entry fee. The seller visits the bidders sequentially in some arbitrary order. For simplicity, we assume the bidders are visited in the lexicographical order. When bidder $i$ is visited, let $S_i(t_{<i})$ be the set of items that are still available. Clearly, this set only depends on the types of bidders who are visited before $i$. The mechanism shows the set $S_i(t_{<i})$ to bidder $i$ and asks her for an entry fee ${\delta_i}(S_i(t_{<i}))$. If she accepts the entry fee, she can enter the mechanism and take her favorite bundle $S_i^{*}$ by paying $\sum_{j\in S_i^{*}} \xi_{ij}$.

 If there exist multiple bundles with the same maximum surplus, the bidder can break ties arbitrarily. Sometimes, there is a feasibility constraint $\mathcal{F}$ on what items a buyer can purchase. In particular, if we say the mechanism is rationed, then $\mathcal{F}=\{\emptyset\} \cup \{\{j\}\ |\ j\in[m] \}$, i.e., a buyer can purchase at most one item. Formally, the favorite bundle $S_i^{*}$ is defined as follows:
$S_i^{*}={\argmax}_{S\subseteq S_i(t_{<i})\land S\in \mathcal{F}} v_i(t_i,S)-\sum_{j\in S}\xi_{ij}$.

\begin{algorithm}[ht]
\begin{algorithmic}[1]
\REQUIRE $\xi_{ij}$ is the price for bidder $i$ to purchase item $j$ and $\delta_i(\cdot)$ is bidder $i$'s entry fee function.
\STATE $S\gets [m]$
\FOR{$i \in [n]$}
	\STATE Show bidder $i$ {the} set of available items $S$, and define entry fee as ${\delta_i}(S)$.
    \IF{Bidder $i$ pays the entry fee ${\delta_i}(S)$}
        \STATE $i$ receives her favorite bundle $S_i^{*}$, paying $\sum_{j\in S_i^{*}}\xi_{ij}$.
        \STATE $S\gets S\backslash S_i^{*}$.
    \ELSE
        \STATE $i$ gets nothing and pays $0$.
    \ENDIF
\ENDFOR
\end{algorithmic}
\caption{{\sf Sequential Posted Price with Entry Fee Mechanism}}
\label{alg:seq-mech}
\end{algorithm}

See Algorithm \ref{alg:seq-mech} for the formal specification of the above mechanism. Notice that before the bidder decides whether to pay the entry fee, she is aware of the set $S_i(t_{<i})$ which contains all available items. Thus, she can compute her favorite bundle $S_i^{*}$ and the corresponding utility if she chooses to enter the mechanism. She can then compare that utility with the entry fee and accept the entry fee if the former is greater than the latter. The mechanism described above is therefore deterministic and DSIC. Throughout this paper, we  focus on the following two special cases of this class of mechanisms:

\vspace{.05in}
 \noindent\textbf{-Rationed Sequential Posted Price Mechanism (RSPM):}  Every buyer can purchase at most one item and the mechanism always charges $0$ entry fee, i.e., $\mathcal{F}=\{\emptyset\} \cup \{\{j\}\ |\ j\in[m] \}$ and ${\delta_i}(S)=0$ for all $i$ and $S$.

 \vspace{.05in}
\noindent\textbf{-Anonymous Sequential Posted Price with Entry Fee Mechanism (ASPE):}  The mechanism uses anonymous posted prices, i.e., ${\xi}_{ij} = \xi_{kj}$ for any item $j$ and bidders $i\neq k$, but may charge positive and personalized entry fee. Also, any buyer can purchase any bundle available once she has paid the entry fee, i.e., $\mathcal{F} = 2^{[m]}$.



\section{Paper Organization}\label{sec:roadmap}
In this section, we provide the roadmap to our paper. In Section~\ref{sec:duality}, we review the Duality framework of~\cite{CaiDW16}.

In Section~\ref{sec:flow}, we derive an upper bound of the optimal revenue for subadditive bidders by combining the duality framework with our new techniques, i.e. valuation relaxation and adaptive dual variables. Our main result in this section,  Theorem~\ref{thm:revenue upperbound for subadditive}, shows that the revenue can be upper bounded by two terms -- $\nf$ and $\single$ defined in Lemma~\ref{lem:upper bound the revenue of the relaxed mechanism}.

In Section~\ref{sec:single}, we use the single bidder case to familiarize the readers with some basic ideas and techniques used to bound $\single$ and $\nf$. The main result of this section, Theorem~\ref{thm:single}, shows that the optimal revenue for a single subadditive bidder is upper bounded by $24\srev$ and $16\brev$.

Section~\ref{sec:multi} contains the main result of this paper. We show how to upper bound the optimal revenue for XOS (or subadditive) bidders with a constant number of (or $O(\log m)$) $\prev$ (the optimal revenue obtainable by an RSPM) and $\aperev$ ((the optimal revenue obtainable by an ASPE). In particular, $\single$ can be upper bounded by the optimal revenue  $\copies$ in the copies setting which is again upper bounded by $6\prev$. We further decompose $\nf$ into two terms $\tail$ and $\core$, and show how to bound $\tail$ in Section~\ref{subsection:tail} and how to bound $\core$ in Section~\ref{subsection:core}.


\section{Duality}\label{sec:duality}

 The focus of \cite{CaiDW16} was on additive and unit-demand valuations and their respective dual was derived from an LP that is only meaningful for constrained additive valuations. In order to tackle general valuations, we need to apply the duality framework to an LP that is meaningful for general valuations. Instead of using the ``implicit forms'' LP from~\cite{CaiDW13b, CaiDW16}, we choose a slightly different and more intuitive LP formulation (see Figure~\ref{fig:LPRevenue}). For all bidders $i$ and types $t_i \in T_i$, we use $p_i(t_i)$ as the interim price paid by bidder $i$ and $\sigma_{iS}(t_i)$ as the interim probability of receiving the exact bundle $S$. To ease the notation, we use a special type $\varnothing$ to represent the choice of not participating in the mechanism. More specifically, ${\sigma}_{iS}(\varnothing)=0$ for any $S$ and $p_{i}(\varnothing)=0$. Now a Bayesian IR (BIR) constraint is simply another BIC constraint: for any type $t_{i}$, bidder $i$ will not want to lie to type $\varnothing$.  We let $T_{i}^{+}=T_{i}\cup \{\varnothing\}$.

Following the recipe provided by~\cite{CaiDW16}, we take the partial Lagrangian dual of the LP in Figure~\ref{fig:LPRevenue} by lagrangifying the BIC constraints. Let $\lambda_i(t_i,t_i')$ be the Lagrange multiplier associated with the BIC constraint that if bidder $i$'s true type is $t_i$ she will not prefer to lie to type $t_i'$
 (see Figure~\ref{fig:Lagrangian} and Definition~\ref{def:Lagrangian}). As shown in~\cite{CaiDW16}, the dual solution has finite value if and only if the dual variables $\lambda_i$ form a valid flow for every bidder $i$. The reason is that the payments $p_i(t_i)$ are unconstrained variables, therefore the corresponding coefficients must be $0$ in order for the dual solution to have finite value. It turns out when all these coefficients are $0$, the dual variables $\lambda$ can be interpreted as a flow described in Lemma~\ref{lem:useful dual}. We refer the readers to~\cite{CaiDW16} for a complete proof. From now on, we only consider $\lambda$ that corresponds to a flow.

\begin{figure}[ht]
\colorbox{MyGray}{
\begin{minipage}{\textwidth} {
\noindent\textbf{Variables:}
\begin{itemize}[leftmargin=0.7cm]
\item $p_i(t_i)$, for all bidders $i$ and types $t_i \in T_i$, denoting the expected price paid by bidder $i$ when reporting type $t_i$ over the randomness of the mechanism and the other bidders' types.
\item $\sigma_{iS}(t_i)$, for all bidders $i$, all bundles of items $S\subseteq[m]$, and types $t_i \in T_i$, denoting the probability that bidder $i$ receives \textbf{exactly} the bundle $S$ when reporting type $t_i$ over the randomness of the mechanism and the other bidders' types.
\end{itemize}
\textbf{Constraints:}
\begin{itemize}[leftmargin=0.7cm]
\item $\sum_{S\subseteq[m]} {\sigma}_{iS}(t_i) \cdot v_i(t_i,S) - p_i(t_i) \geq\sum_{S\subseteq[m]}{\sigma}_{iS}(t'_i) \cdot v_i(t_i, S) - p_i(t'_i) $, for all bidders $i$, and types $t_i \in T_i, t'_i \in T_i^+$, guaranteeing that the reduced form mechanism $({\SI},{p})$ is BIC and Bayesian IR.
\item ${\SI} \in \polytope$, guaranteeing ${\sigma}$ is feasible.
\end{itemize}
\textbf{Objective:}
\begin{itemize}[leftmargin=0.7cm]
\item $\displaystyle\max \sum_{i=1}^{n} \sum_{t_i \in T_i} f_{i}(t_{i})\cdot p_i(t_i)$, the expected revenue.\\
\end{itemize}}
\end{minipage}}
\caption{A Linear Program (LP) for Revenue Optimization.}
\label{fig:LPRevenue}
\end{figure}

\begin{definition}\label{def:Lagrangian}
Let $\L(\lambda, \sigma, p)$ be the partial Lagrangian defined as follows:
\begin{align*}
& \L(\lambda, \sigma, p)\\\stepcounter{equation}\tag{\theequation} \label{eq:primal lagrangian}
=&\sum_{i=1}^{n} \left(\sum_{t_i \in T_i} f_{i}(t_{i})\cdot p_i(t_i)+\sum_{t_{i}\in T_{i},t_{i}'\in T_i^{+}} \lambda_{i}(t_{i},t_{i}')\cdot \left(\sum_{S\subseteq[m]} v_i(t_{i},S)\cdot\left(\sigma_{iS}(t_{i})-\sigma_{iS}({t_{i}'})\right)-\left((p_{i}(t_{i})-p_{i}(t_{i}')\right)\right)\right)\\
=& \sum_{i=1}^{n}\left(\sum_{t_{i}\in T_{i}} p_{i}(t_{i})\cdot\left(f_{i}(t_{i})+\sum_{t_{i}'\in T_{i}} \lambda_{i}(t_{i}',t_{i})-\sum_{t_{i}'\in T_{i}^{+}} \lambda_{i}(t_{i},t_{i}')\right)\right)\\
&+\sum_{i=1}^{n}\left(\sum_{t_{i}\in T_{i}}\sum_{S\subseteq[m]}\sigma_{iS}(t_{i})\cdot \left(v_i(t_{i},S)\cdot \sum_{t_{i}'\in T_{i}^{+}}\lambda_{i}(t_{i},t_{i}')-\sum_{t'_i\in T_{i}}\left(v_i(t'_{i},S)\cdot \lambda_{i}(t_{i}',t_{i})\right)\right)\right)~~ ({\sigma}_i(\varnothing)=\textbf{0},\ p_{i}(\varnothing)=0)\stepcounter{equation}\tag{\theequation} \label{eq:dual lagrangian}
\end{align*}
\end{definition}

\begin{figure}[ht]
\colorbox{MyGray}{
\begin{minipage}{\textwidth} {
\noindent\textbf{Variables:}
\begin{itemize}[leftmargin=0.7cm]
\item $\lambda_i(t_{i},t_{i}')$ for all $i,t_{i}\in T_{i},t_{i}' \in T_i^{+}$, the Lagrangian multipliers for Bayesian IC and IR constraints.
\end{itemize}
\textbf{Constraints:}
\begin{itemize}[leftmargin=0.7cm]
\item $\lambda_i(t_{i},t_{i}')\geq 0$ for all $i,t_{i}\in T_{i},t_{i}' \in T_i^{+}$, guaranteeing that the Lagrangian multipliers are non-negative.
\end{itemize}
\textbf{Objective:}
\begin{itemize}[leftmargin=0.7cm]
\item $\displaystyle\min_{\lambda}\max_{\sigma\in \polytope, p} \L(\lambda, \sigma, p)$.\\
\end{itemize}}
\end{minipage}}
\caption{Partial Lagrangian of the Revenue Maximization LP.}
\label{fig:Lagrangian}
\end{figure}

\notshow{ \begin{definition}[Useful Dual Variables~\cite{CaiDW16}]
A set of feasible duals $\lambda$ is \textbf{useful} if $\max_{\sigma\in\polytope, p} \L(\lambda, \sigma, p)< \infty$.
\end{definition}}

\begin{lemma}[Useful Dual Variables~\cite{CaiDW16}]\label{lem:useful dual}
A set of feasible duals $\lambda$ is \textbf{useful} if $\max_{\sigma\in\polytope, p} \L(\lambda, \sigma, p)< \infty$. $\lambda$ is useful iff for each bidder $i$, $\lambda_{i}$ forms a valid flow, i.e., iff the following satisfies flow conservation at all nodes except the source and the sink:

 \noindent\textbf{\emph{1.}} Nodes: A super source $s$ and a super sink $\varnothing$, along with a node $t_{i}$ for every type $t_{i}\in T_{i}$.

\noindent \textbf{\emph{2.}} An edge from $s$ to $t_{i}$ with flow $f_i(t_{i})$, for all $t_{i}\in T_{i}$.

\noindent \textbf{\emph{3.}} An edge from $t_i$ to $t_i'$ with flow $\lambda_i(t_i,t_i')$ for all $t_i\in T_i$, and $t_i'\in T_{i}^{+}$ (including the sink).

\end{lemma}

\begin{definition}[Virtual Value Function]\label{def:virtual value}
For each flow $\lambda$, we define a corresponding virtual value function $\Phi(\cdot)$, such that for every bidder $i$, every type $t_{i}\in T_{i}$ and every set $S\subseteq[m]$,
$$\Phi_{i}(t_{i}, S)=v_i(t_{i},S)-{1\over f_{i}(t_{i})}\sum_{t_{i}'\in T_{i}} \lambda_{i}(t_{i}',t_{i})\left(v_i(t_{i}',S)-v_i(t_{i},S)\right).$$
\notshow{
\vspace{.1in}\noindent$~~~~~~~~~~~~~~~~~~~~~~~~~~~~~~~~~\Phi_{i}(t_{i}, S)=v_i(t_{i},S)-$

$\hspace{1.3cm}{1\over f_{i}(t_{i})}\sum_{t_{i}'\in T_{i}} \lambda_{i}(t_{i}',t_{i})\left(v_i(t_{i}',S)-v_i(t_{i},S)\right).$	
}
\end{definition}
The proof of Theorem~\ref{thm:revenue less than virtual welfare} is essentially the same as in~\cite{CaiDW16}. We include it in Appendix~\ref{sec:proof_duality} for completeness.
\begin{theorem}[Virtual Welfare $\geq$ Revenue~\cite{CaiDW16}]\label{thm:revenue less than virtual welfare}
For any flow $\lambda$ and any BIC mechanism $M=(\sigma,p)$, the revenue of $M$ is $\leq$  the virtual welfare of {$\sigma$} w.r.t. the virtual valuation $\Phi(\cdot)$ corresponding to $\lambda$.
$$\sum_{i=1}^{n} \sum_{t_i \in T_i} f_{i}(t_{i})\cdot p_i(t_i)\leq \sum_{i=1}^{n} \sum_{t_{i}\in T_{i}} f_{i}(t_{i}) \sum_{S\subseteq[m]}\sigma_{iS}(t_{i})\cdot\Phi_{i}(t_{i},S)$$

\notshow{
\vspace{.05in} $\sum_{i=1}^{n} \sum_{t_i \in T_i} f_{i}(t_{i})\cdot p_i(t_i)\leq $

$\hspace{1cm}\sum_{i=1}^{n} \sum_{t_{i}\in T_{i}} f_{i}(t_{i}) \sum_{S\subseteq[m]}\sigma_{iS}(t_{i})\cdot\Phi_{i}(t_{i},S)$
}
Let $\lambda^{*}$ be the optimal dual variables and $M^{*}=(\sigma^{*},p^{*})$ be the revenue optimal BIC mechanism, then the expected virtual welfare with respect to $\Phi^{*}$ (induced by $\lambda^{*}$) under  $\sigma^{*}$ equals to the expected revenue of $M^{*}$.
\end{theorem}

\section{Canonical Flow and Properties of the Virtual Valuations}\label{sec:flow}

In this section, we present a canonical way of setting the dual variables/flow that induces our benchmarks. A recap of the flow for additive valuations and the appealing properties of the corresponding virtual valuation functions can be found in Appendix~\ref{sec:flow_additive}. We refer readers to that Section for more intuition about the flow.

Although any flow can provide a finite upper bound of the optimal revenue, we focus on a particular class of flows, in which every flow $\lambda^{(\beta)}$ is parametrized by a set of parameters $\beta=\{\beta_{ij}\}_{i\in[n],j\in[m]}\in\R^{nm}_{\geq 0}$. Based on $\beta$, we partition the type set $T_i$ of each buyer $i$ into $m+1$ regions: \textbf{(i)} $R_{0}^{(\beta_i)}$ contains all types $t_i$ such that $V_i(t_{ij})<\beta_{ij}$ for all $j\in[m]$. \textbf{(ii)} $R_j^{(\beta_i)}$ contains all types $t_i$ such that $V_i(t_{ij})-\beta_{ij}\geq 0$ and $j$ is the smallest index in $\argmax_k\{V_i(t_{ik})-\beta_{ik}\}$. Intuitively, if we view $\beta_{ij}$ as the price of item $j$ for bidder $i$, then $R^{(\beta_i)}_0$ contains all types in $T_i$ that cannot afford any item, and any $R^{(\beta_i)}_j$ with $j>0$ contains all types in $T_i$ whose ``favorite'' item is $j$. We first provide a {\bf Partial Specification of the flow $\lambda^{(\beta)}$:}

\noindent\textbf{1.} For every type $t_{i}$ in region $R^{(\beta_{i})}_{0}$,  the flow goes directly to $\varnothing$ (the super sink).

\noindent \textbf{2.}  For all $j>0$, any flow entering $R^{(\beta_{i})}_{j}$ is from  $s$ (the super source) and any flow leaving $R^{(\beta_{i})}_{j}$ is to $\varnothing$.

\noindent \textbf{3.} For all $t_{i}$ and $t_{i}'$ in $R^{(\beta_{i})}_{j}$ ($j>0$), {$\lambda^{(\beta)}_{i}(t_{i},t_{i}')>0$} only if $t_{i}$ and $t_{i}'$ only differ in the $j$-th coordinate.

\notshow{
\begin{figure}
  \centering{\includegraphics[width=0.5\linewidth]{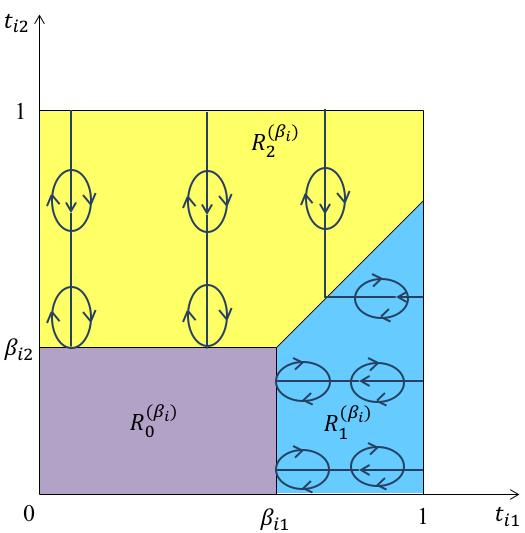}}
  \caption{An example of $\lambda^{(\beta)}_{i}$ for additive bidders with two items.}
  \label{fig:multiflow}
\end{figure}
}

For additive valuations and any type $t_i \in R_j^{(\beta_i)}$ , the contribution to the virtual value function $\Phi(t_i,S)$  from any type $t_i'\in R_j^{(\beta_i)}$ is either $0$ if $j\notin S$, or {$\lambda_i^{(\beta)}(t_i', t_i)(v_i(t_i',S)-v_i(t_i,S))=\lambda_i^{(\beta)}(t_i', t_i)(t_{ij}'-t_{ij})$} if $t_i$, $t_i'$ only differs on the $j$-th coordinate and $j\in S$. In either case, the contribution does not depend on $t_{ik}$ for any $k\neq j$. This is the key property that allows~\cite{CaiDW16} to choose a flow such that the value of the favorite item is replaced by the corresponding Myerson's ironed virtual value in the virtual value function $\Phi_i(t_i,\cdot)$. 
Unfortunately, this property no longer holds for subadditive valuations. When $j\in S$ and $\lambda_i^{(\beta)}(t_i',t_i)>0$, the contribution {$\lambda_i^{(\beta)}(t_i', t_i)(v_i(t_i',S)-v_i(t_i,S))$} heavily depends on $t_{ik}$ of all the other item $k\in S$. All we can conclude is that the contribution lies in the range {$[-\lambda_i^{(\beta)}(t_i', t_i)\cdot V_{i}(t_{ij}), \lambda_i^{(\beta)}(t_i', t_i)\cdot V_{i}(t_{ij}')]$}\footnote{$v_i(t,\cdot)$ is subadditive and monotone for every type $t\in T_i$, therefore $v_i(t_i,S)\in[v_i(t_i, S\backslash\{j\}),v_i(t_i, S\backslash\{j\})+V_{i}(t_{ij})]$ and $v_i(t'_i,S)\in[v_i(t'_i, S\backslash\{j\}),v_i(t'_i, S\backslash\{j\})+V_{i}(t'_{ij})]$.}, but this is not sufficient for us to convert the value of item $j$ into the corresponding Myerson's ironed virtual value.

\subsection{Valuation Relaxation}\label{sec:valuation relaxation}
This is the first major barrier for extending the duality framework to accommodate subadditive valuations. We overcome it by considering a relaxation of the valuation functions. More specifically, for any $\beta$, we construct another function $v_i^{(\beta_i)}(\cdot,\cdot): T_i\times 2^{[m]}\mapsto {\mathbb{R}_{\geq 0}}$ for every buyer $i$ such that: (i) for any $t_i$,  $v_i^{(\beta_i)}(t_i,\cdot)$ is subadditive and monotone, and for every bundle $S$ the new value $v_i^{(\beta_i)}(t_i,S)$ is no smaller than the original value $v_i(t_i,S)$; (ii) for any BIC mechanism $M$ with respect to the original valuations, there exists another mechanism $M^{(\beta)}$ that is BIC with respect to the new valuations and its revenue is comparable to the revenue of $M$; (iii) for the new valuations $v^{(\beta)}$, there exists a flow whose induced virtual value functions have properties similar to those in the additive case.
Property (ii) implies that the optimal revenue with respect to $v^{(\beta)}$ can serve as a proxy for the original optimal revenue. Moreover, due to Theorem~\ref{thm:revenue less than virtual welfare}, the optimal revenue for $v^{(\beta)}$ is upper bounded by the partial Lagrangian dual with respect to $v^{(\beta)}$, which has an appealing format similar to the additive case by property (iii). Thus, we obtain a benchmark for subadditive bidders that resembles the benchmark for additive bidders in~\cite{CaiDW16}.

\begin{definition}[Relaxed Valuation]\label{def:relaxed valuation}
	Given $\beta$, for any buyer $i$, define $v_i^{(\beta_i)}(t_i,S)=v_i(t_i,S\backslash\{j\})+V_i(t_{ij})$, if the ``favorite'' item is in $S$, i.e., $t_i\in R_j^{(\beta_i)} \text{ and } j\in S$. Otherwise, define $v_i^{(\beta_i)}(t_i,S)=v_i(t_i,S)$.

\end{definition}

In the next Lemma, we show that for any BIC mechanism $M$ for $v$, there exists a BIC mechanism $M^{(\beta)}$ for $v^{(\beta)}$ such that its revenue is comparable to the revenue of $M$ (property (ii)). Moreover, the ex-ante probability for any buyer $i$ to receive any item $j$ in $M^{(\beta)}$ is no greater than in $M$ (property (i)). We will see later that this is an important property for our analysis. The proof of Lemma~\ref{lem:relaxed valuation} is similar to the $\epsilon$-BIC to BIC reduction in~\cite{HartlineKM11, BeiH11,DaskalakisW12} and can be found in Appendix~\ref{sec:proof_relaxed_valuation}.

\begin{lemma}\label{lem:relaxed valuation}
	For any $\beta$ and any BIC mechanism $M$ for subadditive valuation  $\{v_i(t_i,\cdot)\}_{i\in[n]}$ with $t_i\sim D_i$ for all $i$, there exists a BIC mechanism $M^{(\beta)}$ for valuations $\{v_i^{(\beta_i)}(t_i,\cdot)\}_{i\in[n]}$ with $t_i\sim D_i$ for all $i$, such that

 \vspace{.1in}
  \noindent \emph{\textbf{(i)}} $\displaystyle\sum_{t_i\in T_i}f_i(t_i)\cdot\sum_{S: j\in S}\sigma^{(\beta)}_{iS}(t_i)\leq \sum_{t_i\in T_i}f_i(t_i)\cdot\sum_{S: j\in S}\sigma_{iS}(t_i)$, for all $i$ and $j$,

  \vspace{.1in}
 \noindent \emph{\textbf{(ii)}} $\displaystyle\rev(M, v, D)\leq2\cdot{\rev(M^{(\beta)},v^{(\beta)}, D)}\displaystyle+2\cdot\sum_i
\sum_{t_i\in T_i}\sum_{S\subseteq[m]} f_i(t_i)\cdot\sigma^{(\beta)}_{iS}(t_i)\cdot \left(v_i^{(\beta_i)}(t_i, S)-v_i(t_i, S)\right).$

\vspace{0.05in}
\noindent$\rev(M, v, D)$ (or $\rev(M^{(\beta)},v^{(\beta)}, D)$) is the revenue of the mechanism $M$ (or $M^{(\beta)}$) while the buyers' types are drawn from $D$ and buyer $i$'s valuation is $v_i(t_i,\cdot)$ (or $v_i^{(\beta_i)}(t_i,\cdot)$). $\sigma_{iS}(t_i)$ (or $\sigma^{(\beta)}_{iS}(t_i)$) is the probability of buyer $i$ receiving exactly bundle $S$ when her reported type is $t_i$ in mechanism $M$ (or $M^{(\beta)}$).
\end{lemma}
\notshow{
From now on, we fix $M^{(\beta)}$ to be the mechanism that is constructed by setting $\eta$ to be $1/2$ and $\epsilon$ be a extremely tiny positive constant $\epsilon_o$ in Lemma~\ref{lem:relaxed valuation}.
\begin{corollary}
	For any $\beta$, there exists a mechanism $M^{(\beta)}$ such that
	$$\rev(M, v, D)\leq 2\cdot{\rev(M^{(\beta)},v^{(\beta)}, D)}+2\cdot\sum_i
\sum_{t_i\in T_i}\sum_{S\subseteq[m]} f_i(t_i)\cdot\sigma^{(\beta)}_{iS}(t_i)\cdot \left(v_i^{(\beta_i)}(t_i, S)-v_i(t_i, S)\right)+\epsilon_0.$$
\end{corollary}}
\subsection{Virtual Valuation for the Relaxed Valuation}\label{sec:virtual for relaxed}
For any $\beta$, based on the same partition of the type sets as in the beginning of Section~\ref{sec:flow}, we construct a flow $\lambda^{(\beta)}$ that respects the partial specification, such that the corresponding virtual valuation function for $v^{(\beta)}$ has the same appealing properties as in the additive case.
For the relaxed valuation, as {$\lambda_i^{(\beta)}(t_i, t_i')$} is only positive for types $t_i$, $t_i'\in R_j^{(\beta_i)}$ that only differ in the $j$-th coordinate, the contribution from item $j$ to the virtual valuation solely depends on $t_{ij}$ and $t'_{ij}$ but not $t_{ik}$ for any other item $k\in S$
. Notice that this property does not hold for the original valuation, and it is the main reason why we choose the relaxed valuation as in Definition~\ref{def:relaxed valuation}. Moreover, we can choose $\lambda_i^{(\beta)}$ carefully so that the virtual valuation of $v^{(\beta)}$ has the following format:



\begin{lemma}\label{lem:subadditive flow properties}
	Let $F_{ij}$ be the distribution of $V_i(t_{ij})$ when $t_{ij}$ is drawn from $D_{ij}$. For any $\beta$, there exists a flow $\lambda^{(\beta)}_i$ such that the corresponding virtual value function $\Phi^{(\beta_i)}_{i}(t_{i}, \cdot)$ of valuation $v_i^{(\beta_i)}(t_i,\cdot)$ satisfies the following properties:

\vspace{.05in}	
\noindent 1. For any $t_{i}\in R^{(\beta_i)}_{0}$, $\Phi^{(\beta_i)}_{i}(t_{i},S) = v_i(t_i, S)$.

\vspace{.05in}
\noindent 2. For any $j>0$, $t_{i}\in R^{(\beta_i)}_{j}$,  $\Phi_{i}^{(\beta_i)}(t_{i},S)\leq  v_i (t_{i}, S)\cdot\ind[j\notin S]+\left(v_i (t_{i}, S\backslash\{j\})+\tp_{ij}(V_i(t_{ij}))\right)\cdot\ind[j\in S],$ where $\tp_{ij}(V_i(t_{ij}))$ is the Myerson's ironed virtual value for  $V_i(t_{ij})$ with respect to $F_{ij}$.
\end{lemma}

The proof of Lemma \ref{lem:subadditive flow properties} is postponed to Appendix~\ref{sec:proof_virtual_relaxation}.
Next, we use the virtual welfare of the allocation $\sigma^{(\beta)}$ to bound the revenue of $M^{(\beta)}$.

\begin{lemma}\label{lem:upper bound the revenue of the relaxed mechanism}
	For any $\beta$, \begin{align*} &\rev(M^{(\beta)},v^{(\beta)},D)\leq \sum_i\sum_{t_i\in T_i}f_i(t_i)\sum_{S\subseteq[m]}\sigma_{iS}^{(\beta)}(t_i)\cdot\Phi^{(\beta_i)}_i(t_i,S)\\
 \leq &	\sum_i\sum_{t_i\in T_i}f_i(t_i)\cdot \ind\left[t_i\in R_0^{(\beta_i)}\right]\cdot \sum_{S\subseteq[m]}\sigma_{iS}^{(\beta)}(t_i)\cdot v_i(t_i,S)\\
 &+ \sum_i\sum_{t_i\in T_i}f_i(t_i)\cdot \sum_{j\in[m]} \ind\left[t_i\in R_j^{(\beta_i)}\right]\cdot \left(\sum_{S:j\in S}\sigma_{iS}^{(\beta)}(t_i)\cdot v_i(t_i,S\backslash\{j\})+\sum_{S:j\notin S}\sigma_{iS}^{(\beta)}(t_i)\cdot v_i(t_i,S)\right)\\
 &+ \sum_i\sum_{t_i\in T_i}f_i(t_i)\cdot \sum_{j\in[m]} \ind\left[t_i\in R_j^{(\beta_i)}\right]\cdot\pi^{(\beta)}_{ij}(t_i)\cdot \tp_{ij}(t_{ij}),\end{align*}
 where $ \pi_{ij}^{(\beta)}(t_i)=\sum_{S:j\in S}\sigma_{iS}^{(\beta)}(t_i)$. {\bf \nf$(M, \beta)$} denotes the sum of the first two terms. {\bf \single$(M, \beta)$} denotes the last term.  \end{lemma}

\begin{proof}
The Lemma follows easily from the properties in Lemma~\ref{lem:subadditive flow properties} and Theorem~\ref{thm:revenue less than virtual welfare}.
\end{proof}

We obtain Theorem~\ref{thm:revenue upperbound for subadditive} by combining Lemma~\ref{lem:relaxed valuation} and~\ref{lem:upper bound the revenue of the relaxed mechanism}. 
 \begin{theorem}\label{thm:revenue upperbound for subadditive}
For any mechanism $M$ and any $\beta$,
$$\rev{(M,v,D)}\leq 4\cdot\textsc{Non-Favorite}(M, \beta)+2\cdot\textsc{Single}(M,\beta).$$
\end{theorem}

\begin{prevproof}{Theorem}{thm:revenue upperbound for subadditive}
First, let's look at the value of $v_i^{(\beta_i)}(t_i, S)-v_i(t_i, S)$. When $t_i\in R_j^{(\beta_i)}$ for some $j>0$ and $j\in S$, $v_i^{(\beta_i)}(t_i, S)-v_i(t_i, S)= v_i(t_i, S\backslash\{j\})+V_i(t_{ij})-v_i(t_i,S)\leq v_i(t_i, S\backslash\{j\}),$ because $V_i(t_{ij})\leq v_i(t_i,S)$. For the other cases, $v_i^{(\beta_i)}(t_i, S)-v_i(t_i, S)=0$. Therefore,
\begin{align*}\label{eq:bounding delta}
	&\sum_i
\sum_{t_i\in T_i}\sum_{S\subseteq[m]} f_i(t_i)\cdot\sigma^{(\beta)}_{iS}(t_i)\cdot \left(v_i^{(\beta_i)}(t_i, S)-v_i(t_i, S)\right)\nonumber\\
\leq &\sum_i\sum_{t_i} f_i(t_i)\sum_{j} \ind[t_i\in R_j^{(\beta_i)}]\sum_{S: j\in S}\sigma^{(\beta)}_{iS}(t_i)\cdot v_i(t_i, S\backslash\{j\})\nonumber\\
\leq &\nf(M,\beta)~~~~~~~~~\text{(Definition of $\nf(M,\beta)$)}
\end{align*}

Our statement follows from combining Lemma~\ref{lem:relaxed valuation}, Lemma~\ref{lem:upper bound the revenue of the relaxed mechanism} with the inequality above.
\end{prevproof}

\subsection{Upper Bound for the Revenue of Subadditive Buyers}~\label{sec:choice of beta}
 In Section~\ref{sec:valuation relaxation}, we have argued that for any $\beta$, there exists a mechanism $M^{(\beta)}$ such that its revenue with respect to the relaxed valuation $v^{(\beta)}$ is comparable to the revenue of $M$ with respect to the original valuation. In Section~\ref{sec:virtual for relaxed}, we have shown for any $\beta$ how to choose a flow to obtain an upper bound for $\rev(M^{(\beta)},v^{(\beta)},D)$ and also an upper bound for $\rev(M,v,D)$. Now we specify our choice of $\beta$.

In~\cite{CaiDW16}, the authors fixed a particular $\beta$, and shown that under any allocation rule, the corresponding benchmark can be bounded by the sum of the revenue of a few simple mechanisms. However, for valuations beyond additive and unit-demand, the benchmark becomes much more challenging to analyze\footnote{Indeed, the difficulties already arise for valuations as simple as $k$-demand. A bidder's valuation is $k$-demand if her valuation is additive subject to a uniform matroid with rank $k$.}. We adopt an alternative and more flexible approach to obtain a new upper bound. Instead of fixing a single $\beta$ for all mechanisms, we customize a different $\beta$ for every different mechanism $M$. Next, we relax the valuation and design the flow based on the chosen $\beta$ as specified in Section~\ref{sec:valuation relaxation} and \ref{sec:virtual for relaxed}.
 Then we upper bound the revenue of $M$ with the benchmark in Theorem~\ref{thm:revenue upperbound for subadditive} and argue that for any mechanism $M$, the corresponding benchmark can be upper bounded by the sum of the revenue of a few simple mechanisms.  As we allow $\beta$, in other words the flow $\lambda^{(\beta)}$, to depend on the mechanism, our new approach may provide a better upper bound. As it turns out, our new upper bound is indeed easier to analyze.

 Lemma~\ref{lem:requirement for beta} specifies the two properties of our $\beta$ that play the most crucial roles in our analysis. We construct such a $\beta$ in the proof of Lemma~\ref{lem:requirement for beta}, however the construction is not necessarily unique and any $\beta$ satisfying these two properties suffices. Note that our construction heavily relies on property \textbf{(i)} of Lemma~\ref{lem:relaxed valuation}.
\begin{lemma}\label{lem:requirement for beta}
	For any constant $b\in (0,1)$ and any mechanism $M$, there exists a $\beta$ such that: for the mechanism $M^{(\beta)} $ constructed in Lemma~\ref{lem:relaxed valuation} according to $\beta$, any $i\in[n]$ and $j\in[m]$,

\noindent\emph{\textbf{(i)}} $\sum_{k\neq i} \Pr_{t_{kj}}\left[V_k(t_{kj})\geq \beta_{kj}\right]\leq b$;

\noindent\emph{\textbf{(ii)}} $\sum_{t_i\in T_i}f_i(t_i)\cdot \pi_{ij}^{(\beta)}(t_i)\leq \Pr_{t_{ij}}\left[V_i(t_{ij})\geq \beta_{ij}\right]/ b$, where $\pi_{ij}^{(\beta)}(t_i) = \sum_{S: j\in S} \sigma^{(\beta)}_{iS}(t_i)$.
\end{lemma}

Before proving Lemma~\ref{lem:requirement for beta}, we provide some intuition behind the two required properties.
Property \textbf{(i)} is used to guarantee that if item $j$'s price for bidder $i$ is higher than $\beta_{ij}$ for all $i$ and $j$ in an RSPM, for any item $j'$ and any bidder $i'$, $j'$ is still available with probability at least $(1-b)$ when $i'$ is visited. As for any bidder $k\neq i'$ to purchase item $j'$,  $V_k(t_{kj'})$ must be greater than her price for item $j'$. By the union bound, the probability that there exists such a bidder is upper bounded by the LHS of property (i), and therefore is at most $b$. With this guarantee, we can easily show that the RSPM achieves good revenue (Lemma~\ref{lem:neprev}). Property \textbf{(ii)} states that the ex-ante probability for bidder $i$ to receive an item $j$ in $M^{(\beta)}$ is not much bigger than the probability that bidder $i$'s value is larger than item $j$. This is crucial for proving our key Lemma~\ref{lem:hat Q}, in which we argue that two different valuations provide comparable welfare under the same allocation rule $\sigma^{(\beta)}$. With Lemma~\ref{lem:hat Q}, we can show that the ASPE obtains good revenue.

\begin{prevproof}{Lemma}{lem:requirement for beta}
	When there is only one buyer, we can simply set every $\beta_j$ to be $0$ and both conditions are satisfied.
	When there are multiple players, we let $$\beta_{ij}:=\inf\{{x\geq 0}: \Pr_{t_{ij}}\left[V_i(t_{ij})\geq x\right] \leq b\cdot\sum_{t_i\in T_i}f_i(t_i)\cdot\pi_{ij}(t_i)\},$$ where $ \pi_{ij}(t_i)=\sum_{S:j\in S}\sigma_{iS}(t_i)$. Clearly, when the distribution of $V_i(t_{ij})$ is continuous, then
\begin{equation}\label{equ:beta_second_condition}
\Pr_{t_{ij}}\left[V_i(t_{ij})\geq \beta_{ij}\right]=b\cdot\sum_{t_i\in T_i}f_i(t_i)\cdot\pi_{ij}(t_i),
\end{equation}
and therefore for any $j$, $$\sum_i\Pr_{t_{ij}}\left[V_i(t_{ij})\geq \beta_{ij}\right]=b\cdot\sum_i\sum_{t_i\in T_i}f_i(t_i)\cdot\pi_{ij}(t_i)\leq b.$$

So the first condition is satisfied. The second condition holds because by the first property in Lemma~\ref{lem:relaxed valuation}, $\sum_{t_i\in T_i}f_i(t_i)\cdot \pi_{ij}^{(\beta)}(t_i)\leq \sum_{t_i\in T_i}f_i(t_i)\cdot\pi_{ij}(t_i)$.

When the distribution for $V_i(t_{ij})$ is discrete, it is possible that Equation~\ref{equ:beta_second_condition} does not hold, but this is essentially a tie breaking issue and not hard to fix. Let $\epsilon>0$ be an extremely small constant that is smaller than $\left|V_i(t_{ij})-V_i(t'_{ij})\right|$ for any $t_{ij},  t'_{ij}\in T_{ij}$, any $i$ and any $j$. Let $\zeta_{ij}$ be a random variable uniformly distributed on $[0,\epsilon]$, and think of it as a random rebate that the seller gives to bidder $i$ when she purchases item $j$. Now we modify the definition of $\beta_{ij}$ as $\beta_{ij}:=\inf\{{x\geq 0}: \Pr_{t_{ij},\zeta_{ij}}[V_i(t_{ij})+\zeta_{ij}\geq x] \leq b\cdot\sum_{t_i\in T_i}f_i(t_i)\cdot\pi_{ij}(t_i)\}.$

\notshow{\begin{equation}
\epsilon_1=\epsilon\cdot \frac{\Pr_{t_{ij}}\left[V_i(t_{ij})\geq \beta_{ij}\right]-b\cdot\sum_{t_i\in T_i}f_i(t_i)\cdot\pi_{ij}(t_i)}{\Pr_{t_{ij}}\left[V_i(t_{ij})= \beta_{ij}\right]}
\end{equation}

By the definition of $\beta_{ij}$, $\epsilon_1\in [0,\epsilon)$. Let $\zeta_{ij}$ be a random variable uniformly distributed on $[\epsilon_1-\epsilon,\epsilon_1]$.  It is not hard to verify that $\Pr_{t_{ij},\zeta_{ij}}\left[V_i(t_{ij})\geq \beta_{ij}+\zeta_{ij}\right]=b\cdot\sum_{t_i\in T_i}f_i(t_i)\cdot\pi_{ij}(t_i)$. For those probabilities related to $\beta_{ij}$ shown in the proofs below, which is written for simplification, we refer to this definition. With regard to this definition, we make essentially small changes for the mechanism described below. Instead of using fixed price, add a small disturbance $\zeta_{ij}$ on item $j$'s price for bidder $i$. Since $\epsilon$ can be chosen as small as possible, the revenue will only be affected by a small constant. All the argument maintain to be true.}

Both of the two properties in Lemma~\ref{lem:requirement for beta} hold if we replace $V_i(t_{ij})$ with $V_i(t_{ij})+\zeta_{ij}$. The only change we need to make in the mechanism is to actually give the bidders $\zeta_{ij}$ as the corresponding rebate. Since we can choose $\epsilon$ to be arbitrarily small, the sum of the rebate is also arbitrarily small.  For the simplicity of the presentation, we will omit $\zeta_{ij}$ and $\epsilon$ in the rest of the paper. 
The random rebate indeed makes our mechanism randomized(according to the random variable $\zeta_{ij}\sim [0,\epsilon]$). However, the randomized mechanism is a uniform distribution of deterministic DSIC mechanisms (after determining all $\zeta_{ij}$), and the expected revenue of the randomized mechanism is simply the average revenue of all these deterministic mechanisms. Therefore, there must be one realization of the rebates such that the corresponding deterministic mechanism has revenue above the expectation, i.e., the expected revenue of the randomized one. Thus if the randomized mechanism is proved to achieve some approximation ratio, there must exist a deterministic one that achieves the same ratio. The deterministic mechanism will use a fixed value $z_{ij}\in [0,\epsilon]$ as the rebate.

Similarly, the same issue about discrete distributions arises when we define some other crucial parameters later, e.g., in the Definition of $c$, $c_i$ and $\tau_i$. We can resolve all of them together using the trick (adding a random rebate) described above, and we will not include a detailed proof for those cases.
\end{prevproof}

\section{Warm Up: Single Bidder}\label{sec:single}
To warm up, we first study the case where there is a single subadditive buyer and show how to improve the approximation ratio from $338$ to $40$. Since there is only one buyer, we will drop the subscript $i$ in the notations. As specified in Section~\ref{sec:choice of beta}, we use a $\beta$ that satisfies both properties in Lemma~\ref{lem:requirement for beta}. For a single buyer, we can simply set $\beta_{j}$ to be $0$ for all $j$. We use $\single(M), \nf(M)$ in the following proof to denote the corresponding terms in Theorem~\ref{thm:revenue upperbound for subadditive} for $\beta=\textbf{0}$. Notice $R_0^{(\textbf{0})}=\emptyset$. Theorem~\ref{thm:single} shows that the optimal revenue is within a constant factor of the better of selling separately and grand bundling.

\begin{theorem}\label{thm:single}
For a single buyer whose valuation distribution is subadditive over independent items,
\[\rev(M,v,D)\leq 24\cdot\srev+16\cdot\brev\]
for any BIC mechanism $M$.
\end{theorem}

Recall that the revenue for mechanism $M$ is upper bounded by $4\cdot \nf(M)+2\cdot\single(M)$ (Theorem~\ref{thm:revenue upperbound for subadditive}). We first upper bound $\single(M)$ by $\copies$. Since $\sigma^{(\beta)}_{S}(t)$ is a feasible allocation in the original setting, $ \ind[t\in R_j^{(\beta)}]\cdot\pi^{(\beta)}_{j}(t)$ with $\pi^{(\beta)}_j(t)=\sum_{S:j\in S}\sigma^{(\beta)}_{S}(t)$ is a feasible allocation in the copies setting, and therefore $\single(M)$ is the Myerson Virtual Welfare of a certain allocation in the copies setting, which is upper bounded by $\copies$. By~\cite{ChawlaHMS10}, $\copies$ is at most $2\cdot\srev$.
\begin{lemma}\label{lem:single-single}
For any BIC mechanism $M$, $\single (M)\leq \copies\leq 2\cdot\srev.$
\end{lemma}

{For $\nf(M)$, we first bound it by the social welfare from all non-favorite items. Then we decompose the latter into two terms $\core(M)$ and $\tail(M)$, and bound them separately.} For every $t\in T$, define $\mathcal{C}(t)=\{j:V(t_j)< c\}$, $\mathcal{T}(t)=[m]\backslash \mathcal{C}(t)$. Here the threshold $c$ is chosen as
\begin{equation}\label{equ:single-def of c}
c:=\inf\left\{x\geq 0:\  \sum_j \Pr_{t_j}\left[V(t_j)\geq x\right]\leq 2\right\}.
\end{equation}
Since $v(t,\cdot)$ is subadditive for all $t\in T$ , we have for every $S\subseteq [m]$, $v(t,S)\leq v\left(t,S\cap \mathcal{C}(t)\right)+\sum_{j\in S\cap \mathcal{T}(t)}V(t_j)$. {We decompose $\nf(M)$ based on the inequality above.} Proof of Lemma~\ref{lem:single decomposition} can be found in Appendix~\ref{sec:single_appx}.



\begin{lemma}\label{lem:single decomposition}
\begin{align*}	
\nf(M)\leq& \sum_{t\in T}f(t)\cdot\sum_j\ind[t\in R_j^{(\beta)}]\cdot v(t, [m]\backslash \{j\})\\
\leq&	\sum_{t\in T}f(t)\cdot v(t,\mathcal{C}(t))~~~~~~~~~\quad(\core(M))\\
+&\sum_j\sum_{t_{j}:V(t_{j})\geq c}f_{j}(t_{j})\cdot V(t_{j})\cdot \Pr_{t_{-j}}\left[\exists k\not=j, V(t_k)\geq V(t_j)\right]~~~~\quad(\tail(M))
\end{align*}
\end{lemma}

Using the definition of $c$ and $\srev$, we can upper bound $\tail(M)$ with a similar argument as in~\cite{CaiDW16}.
\begin{lemma}\label{lem:single-tail}
For any BIC mechanism $M$, $\tail(M)\leq 2\cdot\srev$.
\end{lemma}

\begin{proof}
Since $\tail(M)=\sum_j\sum_{t_j:V(t_j)\geq c}f_j(t_j)\cdot V(t_j)\cdot \Pr_{t_{-j}}\left[\exists k\not=j, V(t_k)\geq V(t_j)\right]$, for each type $t_j\in T_j$ consider the mechanism that posts the same price $V(t_j)$ for each item but only allows the buyer to purchase at most one. Notice if there exists $k\not= j$ such that $V(t_k)\geq V(t_j)$, the mechanism is guaranteed to sell one item obtaining revenue $V(t_j)$. Thus, the revenue obtained by this mechanism
is at least $V(t_j)\cdot \Pr_{t_{-j}}\left[\exists k\not=j, V(t_k)\geq V(t_j)\right]$. By definition, this is no more than $\srev$.

\begin{equation}\label{equ:single-tail}
\tail(M)\leq \sum_j\sum_{t_j:V(t_j)\geq c}f_j(t_j)\cdot \srev\notshow{\leq}{=} 2\cdot \srev 
\end{equation}

{
The last equality is because by the definition of $c$,
\noindent$\sum_j \Pr_{t_j}[V(t_j)\geq c]=2$.\footnote{This clearly holds if $V(t_j)$ is drawn from a continuous distribution. When $V(t_j)$ is drawn from a discrete distribution, see the proof of Lemma~\ref{lem:requirement for beta} for a simple fix.}
}
\end{proof}

The $\core(M)$ is upper bounded by $\E_{t}[v'(t,[m])]$ where $v'(t,S)$
$= v(t,S\cap \mathcal{C}(t))$. We argue that $v'(t,\cdot)$ is drawn from a distribution that is subadditive over independent items and $v'(\cdot,\cdot)$ is $c$-Lipschitz (see Definition~\ref{def:Lipschitz}). Using a concentration bound by Schechtman~\cite{Schechtman2003concentration}, we show $\E_{t}[v'(t,[m])]$ is upper bounded by the median of random variable $v'(t,[m])$ and $c$, which are upper bounded by $\brev$ and $\srev$ respectively.
\begin{lemma}\label{lem:single-core}
For any BIC mechanism $M$, $\core(M) \leq 3\cdot\srev+4\cdot\brev$.
\end{lemma}

Recall that
\begin{equation}
\core(M)=\sum_{t\in T}f(t)\cdot v(t,\mathcal{C}(t))
\end{equation}

We will bound $\core(M)$ with a concentration inequality from~\cite{Schechtman2003concentration}. It requires the following definition:

\begin{definition}\label{def:Lipschitz}
A function $v(\cdot,\cdot)$ is \textbf{$a$-Lipschitz} if for any type $t,t'\in T$, and set $X,Y\subseteq [m]$,
$$\left|v(t,X)-v(t',Y)\right|\leq a\cdot \left(\left|X\Delta Y\right|+\left|\{j\in X\cap Y:t_j\not=t_j'\}\right|\right),$$ where $X\Delta Y=\left(X\backslash Y\right)\cup \left(Y\backslash X\right)$ is the symmetric difference between $X$ and $Y$.
\end{definition}

Define a new valuation function for the bidder as $v'(t,S)=v(t,S\cap \mathcal{C}(t))$, for all $t\in T$ and $S\subseteq [m]$. Then $v'(\cdot,\cdot)$ is $c-$ Lipschitz, and when $t$ is drawn from the product distribution $D=\prod_j D_j$, $v'(t,\cdot)$ remains to be a valuation drawn from a distribution that is subadditive over independent items. See Appendix~\ref{sec:single_appx} for the proof of Lemma~\ref{lem:single subadditive} and Lemma~\ref{lem:single Lipschitz}.

\begin{lemma}\label{lem:single subadditive}
For all $t\in T$, $v'(t,\cdot)$ satisfies monotonicity, subadditivity and no externalities defined in Definition~\ref{def:subadditive independent}.
\end{lemma}

\begin{lemma}\label{lem:single Lipschitz}
$v'(\cdot,\cdot)$ is $c-$Lipschitz.
\end{lemma}

Next, we apply the following concentration inequality to derive Corollary~\ref{corollary:concentrate}, which is useful to analyze the $\core(M)$.

\begin{lemma}~\cite{Schechtman2003concentration}\label{lem:schechtman}
Let $g(t,\cdot)$ with $t\sim D=\prod_j D_j$ be a function drawn from a distribution that is  subadditive over independent items of ground set $I$. If $g(\cdot,\cdot)$ is $c$-Lipschitz, then for all $a>0, k\in \{1,2,...,|I|\}, q\in \mathbb{N}$,
$$\Pr_t[g(t,I)\geq (q+1)a+k\cdot c]\leq \Pr_t[g(t,I)\leq a]^{-q}q^{-k}.$$
\end{lemma}

\begin{corollary}\label{corollary:concentrate}
Let $g(t,\cdot)$ with $t\sim D=\prod_j D_j$ be a function drawn from a distribution that is  subadditive over independent items of ground set $I$. If $g(\cdot,\cdot)$ is $c$-Lipschitz, then if we let $a$ be the median of the value of the grand bundle $g(t,I)$, i.e. $a=\inf\left\{x\geq 0: \Pr_t[g(t,I)\leq x]\geq \frac{1}{2}\right\}$,
$$\mathds{E}_t[g(t,I)]\leq 2a+\frac{5c}{2}.$$
\end{corollary}

\begin{proof}
Let $Y$ be $g(t,I)$. If we apply Lemma~\ref{lem:schechtman} to the case where $a$ is the median and $q=2$, we have

\begin{align*}
\Pr_t[Y\geq 3a]\cdot\E_{t}[Y|Y\geq 3a]&= 3a\cdot \Pr_t[Y\geq 3a]+\int_{y=0}^{\infty}\Pr_t[Y\geq 3a+y]dy\\
&\leq 3a\cdot \Pr_t[Y\geq 3a]+c\cdot\sum_{k=0}^{|I|} \Pr_t[Y\geq 3a+k\cdot c] \quad(Y\leq |I|\cdot c)\\
&\leq 3a\cdot \Pr_t[Y\geq 3a]+c\cdot \sum_{k=0}^2 \Pr_t[Y > a]+ c\cdot\sum_{k=3}^{|I|} 4\cdot 2^{-k}\quad(\text{Lemma~\ref{lem:schechtman}})\\
&\leq 3a\cdot \Pr_t[Y\geq 3a]+\frac{5}{2}c\\
\end{align*}

With the inequality above, we can upper bound the expected value of $Y$.
\begin{align*}
\E_{t}[Y]&\leq a\cdot \Pr_t[Y\leq a]+3a\cdot \Pr_{t}[Y\in (a,3a)]+\Pr_t[Y\geq 3a]\cdot\E_{t}[Y|Y\geq 3a]\\
&\leq a\cdot \Pr_t[Y\leq a]+3a\cdot \Pr_{t}[Y\in (a,3a)]+3a\cdot \Pr_t[Y\geq 3a]+\frac{5}{2}c\\
&= a+2a\cdot \Pr_{t}[Y>a]+\frac{5}{2}c\\
&\leq 2a+\frac{5}{2}c
\end{align*}
\end{proof}

Now, we are ready to prove Lemma~\ref{lem:single-core}.

\begin{prevproof}{Lemma}{lem:single-core}
Let $\delta$ be the median of $v'(t,[m])$ when $t$ is sampled from distribution $D$. Now consider the mechanism that sells the grand bundle with price $\delta$. Notice that the bidder's valuation for the grand bundle is $v(t,[m])\geq v'(t,[m])$. Thus with probability at least $\frac{1}{2}$, 
  the bidder purchases the bundle. Thus, $\brev\geq \frac{1}{2}\delta$.

According to Corollary~\ref{corollary:concentrate},

\begin{equation}\label{equ:single-core-prev}
\core(M)=\mathds{E}_{t\sim D}[v'(t,[m])]\leq 2\delta+\frac{5c}{2}
\end{equation}

It remains to argue that the Lipchitz constant $c$ can be upper bounded using $\srev$. Notice that by AM-GM Inequality,
\begin{align*}
&\Pr_t\left[\exists j\in [m], V(t_j)\geq c\right]= 1-\prod_{j}\Pr_{t_j}[V(t_j)< c]\\
\geq& 1-(\frac{\sum_j \Pr_{t_j}[V(t_j)< c]}{m})^m
= 1-(1-\frac{2}{m})^m
\geq 1-e^{-2}
\end{align*}

Consider the mechanism that posts price $c$ for each item but only allow the buyer to purchase one item. Then with probability at least $1-e^{-2}$, the mechanism sells one item obtaining expected revenue $(1-e^{-2})\cdot c$. Thus $c\leq \frac{1}{1-e^{-2}}\cdot\srev$. Inequality~\eqref{equ:single-core-prev} becomes

\begin{equation}\label{equ:single-core}
\core(M)\leq 2\delta+\frac{5c}{2}<4\cdot\brev+3\cdot\srev
\end{equation}

\end{prevproof}

\begin{prevproof}{Theorem}{thm:single}
Since $\copies\leq 2 \srev$ (Lemma~\ref{lem:single-single}) and $\nf(M)\leq 5\srev+4\brev$ (Lemma~\ref{lem:single-tail} and~\ref{lem:single-core}), $\rev(M,v,D)\leq 24\cdot\srev+16\cdot\brev$ according to Theorem~\ref{thm:revenue upperbound for subadditive}.
\end{prevproof}

\section{Multiple Bidders}\label{sec:multi}

In this section, we prove our main result -- simple mechanisms can approximate the optimal BIC revenue even when there are multiple XOS/subadditive bidders. 
First, we need the definition of supporting prices.
\begin{definition}[Supporting Prices~\cite{DobzinskiNS05}]\label{def:supporting price}
For any $\alpha\geq 1$, a type $t$ and a subset $S\subseteq[m]$, prices $\{p_j\}_{j\in S}$
are $\alpha$-supporting prices for $v(t,S)$ if \textbf{(i)}	$v(t,S') \geq \sum_{j\in S'} p_j$ for all $S'\subseteq S$ and \textbf{(ii)} $\sum_{j\in S}p_j\geq \frac{v(t,S)}{\alpha}$.
\end{definition}

\begin{theorem}\label{thm:multi}
If for any buyer $i$, any type $t_i\in T_i$ and any bundle $S\in [m]$, $v_i(t_i,S)$ has a set of $\alpha$-supporting prices $\{\theta_j^{S}(t_i)\}_{j\in S}$, then for any BIC mechanism $M$ and any constant $b\in (0, 1)$,
\begin{align*}
\textsc{Rev}(M,v,D)\leq 32\alpha \cdot \aperev
+\left(12+\frac{8}{1-b}+\alpha\cdot \left(\frac{16}{b(1-b)}+\frac{96}{1-b}\right)\right)\cdot \prev
\end{align*}

\vspace{0.05in}
If $v_i(t_i,\cdot)$ is an XOS valuation for all $i$ and $t_i\in T_i$, then $\alpha=1$. By setting $b$ to $\frac{1}{4}$, we have $$\rev(M,v,D)\leq 236\cdot\prev+32\cdot\aperev.$$ For general subadditive valuations, $\alpha=O(\log(m))$ by~\cite{BhawalkarR11}, hence $$\textsc{Rev}(M,v,D)\leq O(\log(m))\cdot \max\{\prev,\aperev\}.$$
\end{theorem}
Here is a sketch of the proof for Theorem~\ref{thm:multi}. We show how to upper bound $\single(M,\beta)$ in Lemma~\ref{lem:multi_single}. Then, we decompose $\nf(M,\beta)$ into $\tail(M,\beta)$ and $\core(M,\beta)$ in Lemma~\ref{lem:multi decomposition}. We show how to construct a simple mechanism to approximate $\tail(M,\beta)$ in Section~\ref{subsection:tail} and how to approximate $\core(M,\beta)$ in Section~\ref{subsection:core}.

\vspace{.1in}
 \noindent\textbf{Analysis of $\single(M,\beta)$:} 

\begin{lemma}\label{lem:multi_single}
For any mechanism $M$, $$\textsc{Single}(M, \beta)\leq \copies\leq 6\cdot\prev.$$
\end{lemma}

\begin{proof}
We construct a new mechanism $M'$ in the copies setting based on $M^{(\beta)}$. Whenever $M^{(\beta)}$ allocates item $j$ to buyer $i$ and $t_i\in R_j^{(\beta)}$, $M'$ serves the agent $(i,j)$. Since there is at most one $R_j^{(\beta)}$ that $t_i$ belongs to, $M'$ serves at most one agent $(i,j)$ for each of buyer $i$. Hence, $M'$ is feasible in the copies setting, and $\single(M,\beta)$ is the expected Myerson's ironed virtual welfare of $M'$. Since every agent's value is drawn independently, the optimal revenue in the copies setting is the same as the maximum Myerson's ironed virtual welfare in the same setting. Therefore, $\copies$ is no less than $\single(M,\beta)$.

As showed in~\cite{ChawlaHMS10, KleinbergW12}, a simple posted-price mechanism with the constraint that every buyer can only purchase one item, i.e., an RSPM, achieves revenue at least $\copies/6$ in the original setting. Hence, $\copies\leq 6\cdot\prev$.
\end{proof}

\vspace{.05in}
 \noindent\textbf{Core-Tail Decomposition of $\nf(M,\beta)$:} we decompose $\textsc{Non-Favorite}(M, \beta)$ into two terms $\textsc{Tail}(M, \beta)$ and $\textsc{Core}(M, \beta)$\footnote{In~\cite{CaiDW16}, $\nf$ is decomposed into four different terms $\textsc{Under}$, $\textsc{Over}$, $\core$ and $\tail$. We essentially merge the first three terms into $\core(M, \beta)$ in our decomposition.}. First, we need the following definition.
\begin{definition}\label{def:c_i}
For every buyer $i$, let $c_i :=\inf\big\{x\geq 0:\sum_j \Pr_{t_{ij}}\left[V_i(t_{ij})\geq \beta_{ij}+x\right]\leq \frac{1}{2}\big\}.$ For every $t_i \in T_i$, let $\mathcal{T}_i(t_i)=\{j\ |\ V_i(t_{ij})\geq \beta_{ij}+c_i\}$ and $\mathcal{C}_i(t_i) = [m]\backslash\mathcal{T}_i(t_i)$.
\end{definition}
 Since $v_i(t_i,\cdot)$ is subadditive for all $i$ and $t_i\in T_i$, we have $v_i(t_i,S)\leq v_i\left(t_i,S\cap \mathcal{C}_i(t_i)\right)+\sum_{j\in S\cap \mathcal{T}_i(t_i)}V_i(t_{ij})$. The term $\nf(M,\beta)$ can be decomposed into $\tail(M,\beta)$ and $\core(M,\beta)$ based on the inequality above. The complete proof of Lemma~\ref{lem:multi decomposition} can be found in Appendix~\ref{appx:multi}.

 \begin{lemma}~\label{lem:multi decomposition}
		\begin{align*} &\nf(M,\beta)\\
			\leq&  \sum_i\sum_{t_i}f_i(t_i) \sum_{S\subseteq[m]}\sigma_{iS}^{(\beta)}(t_i)\cdot v_i(t_i,S\cap \mathcal{C}_i(t_i))~~~~~~~~~~(\textsc{Core}(M,\beta))\\
			+&\sum_i\sum_j  \sum_{t_{ij}:V_i(t_{ij})\geq\beta_{ij}+c_i}f_{ij}(t_{ij})\cdot V_i(t_{ij})\cdot\sum_{k\neq j} \Pr_{t_{ik}}\left[ V_i(t_{ik})-\beta_{ik}\geq V_i(t_{ij})-\beta_{ij}\right]~~~~~~~(\textsc{Tail}(M,\beta))
		\end{align*}
		\end{lemma}

\subsection{Analyzing $\tail(M,\beta)$ in the Multi-Bidder Case}\label{subsection:tail}

In this section we show how to bound $\tail(M,\beta)$ with the revenue of an RSPM.
\begin{lemma}\label{lem:multi-tail}
	For any BIC mechanism $M$, $\tail(M, \beta)\leq \frac{2}{1-b}\cdot \prev$.
\end{lemma}

We first fix a few notations. Let $$P_{ij}\in\argmax_{x\geq c_i}(x+\beta_{ij})\cdot \Pr_{t_{ij}}[V_i(t_{ij})-\beta_{ij}\geq x],$$
\begin{align*}
r_{ij}&=(P_{ij}+\beta_{ij})\cdot \Pr[V_i(t_{ij})-\beta_{ij}\geq P_{ij}]=\max_{x\geq c_i}(x+\beta_{ij})\cdot \Pr_{t_{ij}}[V_i(t_{ij})-\beta_{ij}\geq x],
\end{align*}
$r_i=\sum_j r_{ij}$, and $r=\sum_i r_i$. We show in the following Lemma that $r$ is an upper bound of $\tail(M,\beta)$. 
\begin{lemma}\label{lem:tail and r}
For any BIC mechanism $M$, $\tail(M,\beta)\leq r.$
\end{lemma}

\begin{proof}
\begin{equation*}\label{equ:tail1}
\begin{aligned}
\tail(M,\beta)\leq&\sum_i\sum_j\sum_{t_{ij}:V_i(t_{ij})\geq\beta_{ij}+c_i}f_{ij}(t_{ij})\cdot(\beta_{ij}+c_i)\cdot \sum_{k\not=j}\Pr_{t_{ik}}\left[ V_i(t_{ik})-\beta_{ik}\geq V_i(t_{ij})-\beta_{ij}\right]\\
&+\sum_i\sum_j\sum_{t_{ij}:V_i(t_{ij})\geq\beta_{ij}+c_i}f_{ij}(t_{ij})\cdot \left(V_i(t_{ij})-\beta_{ij}\right)\cdot \sum_{k\not=j}\Pr_{t_{ik}}\left[ V_i(t_{ik})-\beta_{ik}\geq V_i(t_{ij})-\beta_{ij}\right]\\
\leq&\frac{1}{2}\cdot\sum_i\sum_j\sum_{t_{ij}:V_i(t_{ij})\geq\beta_{ij}+c_i}f_{ij}(t_{ij})\cdot(\beta_{ij}+c_i)~~\text{(Definition of $c_i$ and $V_i(t_{ij})\geq\beta_{ij}+c_i$)}\\
&+\sum_i\sum_j\sum_{t_{ij}:V_i(t_{ij})\geq\beta_{ij}+c_i}f_{ij}(t_{ij})\cdot \sum_{k\not=j}r_{ik}~~(\text{Definition of $r_{ik}$ and $V_i(t_{ij})\geq\beta_{ij}+c_i$})\\
\leq& \frac{1}{2}\cdot\sum_i\sum_j\Pr_{t_{ij}}[V_i(t_{ij})\geq\beta_{ij}+c_i]\cdot(\beta_{ij}+c_i)+\sum_i r_i \cdot \sum_j\Pr_{t_{ij}}[V_i(t_{ij})\geq\beta_{ij}+c_i]\\
\leq &\frac{1}{2}\cdot\sum_i\sum_j r_{ij}+ \frac{1}{2}\cdot\sum_i r_i~~\text{(Definition of $r_{ij}$ and $c_i$)}\\
 =& r
\end{aligned}
\end{equation*}
In the second inequality, the first term is because $V_{i}(t_{ij})-\beta_{ij}\geq c_i$, so $$\sum_{k\not=j}\Pr_{t_{ik}}\left[ V_i(t_{ik})-\beta_{ik}\geq V_i(t_{ij})-\beta_{ij}\right]\leq \sum_{k} \Pr_{t_{ik}}\left[ V_i(t_{ik})-\beta_{ik}\geq c_i\right]\leq1/2.$$ The second term is because for any $t_{ij}$ such that $V_i(t_{ij})\geq \beta_{ij}+c_i$, $$\left(V_i(t_{ij})-\beta_{ij}\right)\cdot \Pr_{t_{ik}}\left[ V_i(t_{ik})-\beta_{ik}\geq V_i(t_{ij})-\beta_{ij}\right]\leq \left(\beta_{ik}+V_i(t_{ij})-\beta_{ij}\right)\cdot \Pr_{t_{ik}}\left[ V_i(t_{ik})-\beta_{ik}\geq V_i(t_{ij})-\beta_{ij}\right]\leq r_{ik}.$$
\end{proof}

Next, we argue that $r$ can be approximated by an RSPM. Indeed, we prove a stronger lemma, which is also useful for analyzing $\core(M,\beta)$.

\begin{lemma}\label{lem:neprev}
Let $\{x_{ij}\}_{i\in[n], j\in[m]}$ be  a collection of non-negative numbers, such that for any buyer $i$
$$\sum_{j\in [m]} \Pr_{t_{ij}}\left[V_i(t_{ij})\geq x_{ij}+\beta_{ij}\right]\leq 1/2,$$ then
\begin{equation*}
\sum_i\sum_j (x_{ij}+\beta_{ij})\cdot \Pr_{t_{ij}}\left[V_i(t_{ij})\geq x_{ij}+\beta_{ij}\right]\leq \frac{2}{1-b}\cdot \prev.
\end{equation*}
\end{lemma}

\begin{proof}
Consider a RSPM that sells item $j$ to buyer $i$ at price $\xi_{ij}=x_{ij}+\beta_{ij}$. The mechanism
visits the buyers in some arbitrary order. Notice that when it is buyer $i$'s turn, she purchases exactly item $j$ and pays $x_{ij}+\beta_{ij}$ if all of the following three conditions hold: (i) $j$ is still available, (ii) $V_i(t_{ij})\geq x_{ij}+\beta_{ij}$ and (iii) $\forall k\neq j, V_i(t_{ik})< x_{ik}+\beta_{ik}$. The second condition means buyer $i$ can afford item $j$. The third condition means she cannot afford any other item $k\neq j$. Therefore, buyer $i$'s purchases exactly item $j$.

Now let us compute the probability that all three conditions hold. Since every buyer's valuation is subadditive over the items, item $j$ is purchased by someone else only if there exists a buyer $k\neq i$ who has $V_k(t_{kj})\geq \xi_{kj}$. Because $x_{kj}\geq 0$ for all $k$, by the union bound, the event described above happens with probability at most $\sum_{k\neq i} \Pr_{t_{kj}}\left[V_k(t_{kj})\geq \beta_{kj}\right]$, which is less than $b$ by property (i) of Lemma~\ref{lem:requirement for beta}. Therefore, condition (i) holds with probability at least $(1-b)$. Clearly, condition (ii) holds with probability $\Pr_{t_{ij}}\left[V_i(t_{ij})\geq x_{ij}+\beta_{ij}\right]$. Finally, condition (iii) holds with at least probability $1/2$, because according to our assumption of the $x_{ij}$s, the probability that there exists any item $k\neq j$ such that $V_i(t_{ik})\geq x_{ik}+\beta_{ik}$ is no more than $1/2$. Since the three conditions are independent, buyer $i$ purchases exactly item $j$ with probability at least $\frac{(1-b)}{2}\cdot \Pr_{t_{ij}}\left[V_i(t_{ij})\geq x_{ij}+\beta_{ij}\right]$. So the expected revenue of this mechanism is at least $\frac{(1-b)}{2}\cdot \sum_i\sum_j (\beta_{ij}+x_{ij})\cdot \Pr_{t_{ij}}\left[V_i(t_{ij})\geq x_{ij}+\beta_{ij}\right]$.
\end{proof}

\notshow{
\begin{corollary}\label{cor:bound tail }
\begin{equation}
\textsc{Tail}(M,\beta)\leq r\leq \frac{2}{1-b}\cdot \prev.
\end{equation}
\end{corollary}
\begin{proof}
Since $P_{ij}\geq c_i$,  it satisfies the assumption in Lemma~\ref{lem:neprev} due to the choice of $c_i$ 
. Therefore,
$$r= \sum_{i,j}(\beta_{ij}+P_{ij})\cdot \Pr_{t_{ij}}\left[V_i(t_{ij})\geq P_{ij}+\beta_{ij}\right] \leq \frac{2}{1-b}\cdot \prev.$$
Our statement follows from the above inequality and Lemma~\ref{lem:tail and r}.\end{proof}
}

\begin{prevproof}{Lemma}{lem:multi-tail}
Since $P_{ij}\geq c_i$,  it satisfies the assumption in Lemma~\ref{lem:neprev} due to the choice of $c_i$ 
. Therefore,
\begin{equation}\label{r and prev}
r= \sum_{i,j}(\beta_{ij}+P_{ij})\cdot \Pr_{t_{ij}}\left[V_i(t_{ij})\geq P_{ij}+\beta_{ij}\right] \leq \frac{2}{1-b}\cdot \prev.
\end{equation}
Our statement follows from the above inequality and Lemma~\ref{lem:tail and r}.\end{prevproof}


We have done the analysis for $\tail{(M,\beta)}$. Before starting the analysis for $\core{(M,\beta)}$, we show that $r_i$ is within a constant factor of $c_i$. This Lemma is useful for bounding $\core{(M,\beta)}$.

\begin{lemma}\label{lem:c_i}
For all $i\in [n]$, $r_i\geq \frac{1}{2}\cdot c_i$ and $\sum_i c_i/2\leq \frac{2}{1-b}\cdot\prev$.
\end{lemma}
\begin{proof}
By the definition of $P_{ij}$,
\begin{align*}
r_i&= \sum_j (\beta_{ij}+P_{ij})\cdot \Pr[V_i(t_{ij})-\beta_{ij}\geq P_{ij}]
\geq \sum_j (\beta_{ij}+c_i)\cdot \Pr[V_i(t_{ij})-\beta_{ij}\geq c_i]\\
&\geq\sum_j c_i\cdot \Pr[V_i(t_{ij})-\beta_{ij}\geq c_i]\geq\frac{1}{2}\cdot c_i
\end{align*}
The last inequality is because when $c_i>0$,
$\sum_j \Pr_{t_{ij}}\left[V_i(t_{ij})\geq \beta_{ij}+c_i\right]$ is at least $\frac{1}{2}$. As $\sum_i c_i/2 \leq r$, by Inequality~(\ref{r and prev}), 
$\sum_i c_i/2\leq \frac{2}{1-b}\cdot \prev$.
\end{proof}


\subsection{Analyzing $\core(M,\beta)$ in the Multi-Bidder Case}\label{subsection:core}

In this section we upper bound $\core(M,\beta)$. Recall that
$$\textsc{Core}(M,\beta)=\sum_i\sum_{t_i\in T_i}f_i(t_i)\cdot \sum_{S\subseteq[m]}\sigma_{iS}^{(\beta)}(t_i)\cdot v_i(t_i,S\cap \mathcal{C}_i(t_i))$$
We can view it as the welfare of another valuation function $v'$ under allocation $\sigma^{(\beta)}$ where $v'_i(t_i, S) = v_i(t_i,S\cap\mathcal{C}_i(t_i))$. In other words, we ``truncate'' the function at some threshold, i.e., only evaluate the items whose value on its own is less than that threshold. The new function still satisfies monotonicity, subadditivity and no externalities.

We first compare existing methods for analyzing the $\core$ with our approach before jumping into the proofs.

\subsubsection{Comparison between the Existing Methods and Our Approach}\label{sec:core comparison}
As all results in the literature~\cite{ChawlaHMS10, Yao15, CaiDW16,ChawlaM16} only study special cases of constrained additive valuations, we restrict our attention to constrained additive valuations in the comparison, but our approach also applies to XOS and subadditive valuations.

We compare our approach to the state of the art result by Chawla and Miller~\cite{ChawlaM16}. They separate $\core(M,\beta) $ into two parts: (i) the welfare obtained from values below $\beta$, and (ii) the welfare obtained from values between $\beta$ and $\beta+c$\footnote{In particular, if bidder $i$ is awarded a bundle $S$ that is feasible for her, the contribution for the first part is $\sum_{j\in S} \min\left\{\beta_{ij},t_{ij}\right\}\cdot \ind\left[t_{ij}< \beta_{ij}+c_i \right]$ and the contribution to the second part is $\sum_{j\in S} \left(t_{ij}-\beta_{ij}\right)^+\cdot \ind\left[t_{ij}< \beta_{ij}+c_i \right]$ }.
 It is not hard to show that the latter can be upper bounded by the revenue of a sequential posted price with per bidder entry fee mechanism.
  Due to their choice of $\beta$ (similar to the second property of Lemma~\ref{lem:requirement for beta}), the former is upper bounded by $\sum_{i,j} \beta_{ij}\cdot \Pr_{t_{ij}}\left[t_{ij}\geq \beta_{ij}\right]$.
   It turns out when every bidder's feasibility constraint is a matroid, one can use the OCRS from~\cite{FeldmanSZ16} to design a sequential posted price mechanism to approximate this expression.
    However, as we show in Example~\ref{ex:counterexample ocrs}, $\sum_{i,j} \beta_{ij}\cdot \Pr_{t_{ij}}\left[t_{ij}\geq \beta_{ij}\right]$ could be $\Omega\left(\frac{\sqrt{m}}{\log m}\right)$ times larger than the optimal social welfare when the bidders have general downward closed feasibility constraints.
     Hence, such approach cannot yield any constant factor approximation for general constrained additive valuations.

As explained in the intro, we take an entirely different approach. We first construct the posted prices $\{Q_j\}_{j\in[m]}$ for our ASPE (Definition~\ref{def:posted prices}), Feldman et al.~\cite{FeldmanGL15} showed that the anonymous posted price mechanism with these prices achieves welfare $\Omega\left(\core(M,\beta)\right)$. If all bidders have valuations that are subadditive over independent items, for any bidder $i$ and any set of available items $S$, $i$'s surplus for $S$ under valuation $v'_i(t_i, \cdot)$ ($max_{S'\subseteq S}~v'_i(t_i,S') -\sum_{j\in S'} Q_j$) is also subadditive over independent items. According to Talagrand's concentration inequality, the surplus concentrates and its expectation is upper bounded by its median and its Lipschitz constant $a$. One can extract at least half of the median by setting the median of the surplus as the entry fee. How about the Lipschitz constant $a$? Unfortunately, $a$ could be as large as $\frac{1}{2}\max_{j\in[m]}\{\beta_{ij}+c_i\}$, which is too large to be bounded.

Here is how we overcome this difficulty. Instead of considering $v'$, we construct a new valuation $\hat{v}$ that is always dominated by the true valuation $v$. We consider the social welfare induced by $\sigma^{(\beta)}$ under $\hat{v}$ and define it as $\widehat{\core}(M,\beta)$. In Section~\ref{sec:proxy core}, we show that $\widehat{\core}(M,\beta)$ is not too far away from $\core(M,\beta)$, so it suffices to approximate $\widehat{\core}(M,\beta)$ (Lemma~\ref{lem:hat Q}). But why is $\widehat{\core}(M,\beta)$ easier to approximate? The reason is two-fold. \textbf{(i)} For any bidder $i$ and any set of available items $S$,  bidder $i$'s surplus for $S$ under $\hat{v}_i(t_i,\cdot)$ (defined as $\mu_i(t_i,S)$ in Definition~\ref{def:entry fee}, which is $max_{S'\subseteq S}~ \hat{v}_i(t_i,S') -\sum_{j\in S'} Q_j$), is not only subadditive over independent items, but also has a small Lipschitz constant $\tau_i$ (Lemma~\ref{lem:property of mu}). Indeed, these Lipschitz constants are so small that $\sum_i \tau_i$  and can be upper bounded by $\prev$ (Lemma~\ref{lem:tau_i}). \textbf{(ii)} If we set the entry fee of our ASPE to be the median of $\mu_i(t_i,S)$ when $t_i$ is drawn from $D_i$, using a proof inspired by Feldman et al.~\cite{FeldmanGL15}, we can show that our ASPE's revenue collected from the posted prices plus the expected surplus of the bidders (over the randomness of all bidders' types) approximates $\widehat{\core}(M,\beta)$ (implied by Lemma~\ref{lem:lower bounding mu}). Again by Talagrand's concentration inequality, we can bound bidder $i$'s expected surplus by our entry fee and $\tau_i$ (Lemma~\ref{lem:concentration entry fee}). As $\hat{v}$ is always smaller than the true valuation $v$, thus for any type $t_i$ of bidder $i$ and any available items $S$, the surplus for $S$ under $v_i(t_i,\cdot)$ must be larger than $\mu_i(t_i,S)$, and the entry fee is accepted with probability at least $1/2$. Putting everything together, we demonstrate that we can approximate $\core(M,\beta)$ with an ASPE or an RSPM (Lemma~\ref{lem:upper bounding Q}).
\subsubsection{Construction of $\widehat{\core}(M,\beta)$}\label{sec:proxy core}

We first show that if for any $i$ and $t_i\in T_i$ there is a set of $\alpha$-supporting prices for $v_i(t_i,\cdot)$, then there is a set of $\alpha$-supporting prices for $v'_i(t_i,\cdot)$.
\begin{lemma}\label{lem:supporting prices for v'}
	If for any type $t_i$ and any set $S$, there exists a set of $\alpha$-supporting prices $\{\theta_j^S(t_i)\}_{j\in S}$  for $v_i(t_i,\cdot)$, then for any $t_i$ {and $S$} there also exists a set of $\alpha$-supporting prices $\{\gamma_j^S(t_i)\}_{j\in S}$ for $v'_i(t_i,\cdot)$. In particular, $\gamma_j^S(t_i)=\theta^{S\cap \mathcal{C}_i(t_i)}_j(t_i)$ if $j\in S\cap \mathcal{C}_i(t_i)$ and $\gamma_j^S(t_i)=0$ otherwise. Moreover, $\gamma_j^S(t_i)\leq V_i(t_{ij})\cdot\ind[V_i(t_{ij})< \beta_{ij}+c_i]$  for all $i$, $t_i$, $j$ and $S$.
\end{lemma}

\begin{proof}
It suffices to verify that $\{\gamma_j^S(t_i)\}_{j\in S}$ satisfies the two properties of $\alpha$-supporting prices.
For any $S'\subseteq S$, $S'\cap \mathcal{C}_i(t_i)\subseteq S\cap \mathcal{C}_i(t_i)$. Therefore,
\begin{equation*}
v_i'(t_i,S')=v_i(t_i,S'\cap \mathcal{C}_i(t_i))\geq \sum_{j\in S'\cap \mathcal{C}_i(t_i)}\theta^{S\cap \mathcal{C}_i(t_i)}_j(t_i)= \sum_{j\in S'\cap \mathcal{C}_i(t_i)}\gamma_j^S(t_i) = \sum_{j\in S'}\gamma_j^S(t_i)
\end{equation*}

{The last equality is because $\gamma_j^S(t_i)=0$ for $j\in S\backslash\mathcal{C}_i(t_i)$. }Also, we have
\begin{equation*}
\sum_{j\in S}\gamma_j^S(t_i)=\sum_{j\in S\cap \mathcal{C}_i(t_i)}\theta^{S\cap \mathcal{C}_i(t_i)}_j(t_i)\geq\frac{v_i(t_i,S\cap \mathcal{C}_i(t_i))}{\alpha}=\frac{v_i'(t_i,S)}{\alpha}
\end{equation*}

Thus, $\{\gamma_j^S(t_i)\}_{j\in S}$ defined above is a set of $\alpha$-supporting prices for $v_i'(t_i,\cdot)$. Next, we argue that $\gamma_j^S(t_i)\leq V_i(t_{ij})\cdot\ind[V_i(t_{ij})< \beta_{ij}+c_i]$  for all $i$, $t_i$, $j \in S$. If $V_i(t_{ij})\geq\beta_{ij}+c_i$, $j\not\in \mathcal{C}_i(t_i)$, by definition $\gamma_j^S(t_i)=0$. Otherwise if $V_i(t_{ij})<\beta_{ij}+c_i$, then $\{j\}\subseteq S\cap \mathcal{C}_i(t_i)$, by the first property of $\alpha$-supporting prices, $\gamma_j^S(t_i)\leq v'_i(t_i,\{j\})=V_i(t_{ij})$.
\end{proof}

Next, we define the prices of our ASPE.

\begin{definition}\label{def:posted prices}
We define a price $Q_j$ for each item $j$ as follows,
	\begin{equation*}
Q_j=\frac{1}{2}\cdot \sum_i \sum_{t_i\in T_i}f_i(t_i)\cdot\sum_{S:j\in S}\sigma_{iS}^{(\beta)}(t_i)\cdot \gamma_j^{S}(t_i),
\end{equation*}
where $\{\gamma_j^S(t_i)\}_{j\in S}$ are the $\alpha$-supporting prices of $v'_i(t_i,\cdot)$ and set $S$ for any bidder $i$ and type $t_i \in T_i$.
\end{definition}

{ $\core(M,\beta)$ can be upper bounded by $\sum_{j\in [m]}Q_j$. The proof follows from the definition of $\alpha$-supporting prices (Definition~\ref{def:supporting price}) and the definition of $Q_j$ (Definition~\ref{def:posted prices}).}

\begin{lemma}\label{lem:core and q_j}
	$2\alpha\cdot\sum_{j\in [m]}Q_j\geq \core(M,\beta)$.
\end{lemma}
\begin{proof}
	\begin{equation*}\label{equ:core and q_j}
\begin{aligned}
\core(M,\beta)&=\sum_i\sum_{t_i\in T_i}f_i(t_i)\cdot \sum_{S\subseteq[m]}\sigma_{iS}^{(\beta)}(t_i)\cdot v_i'(t_i,S)\\
&\leq \alpha\cdot \sum_i\sum_{t_i\in T_i}f_i(t_i)\cdot \sum_{S\subseteq[m]}\sigma_{iS}^{(\beta)}(t_i)\cdot \sum_{j\in S}\gamma_j^{S}(t_i)\\
&=\alpha\cdot \sum_{j\in[m]}\sum_i\sum_{t_i\in T_i}f_i(t_i)\cdot \sum_{S:j\in S}\sigma_{iS}^{(\beta)}(t_i)\cdot \gamma_j^{S}(t_i)\\
&=2\alpha\cdot \sum_{j\in [m]}Q_j
\end{aligned}
\end{equation*}
\end{proof}

\vspace{0.05in}
In the following definitions, we define $\widehat{\core}(M,\beta)$ which is the welfare of another function $\hat{v}$ under the same allocation $\sigma^{(\beta)}$. 



\begin{definition}\label{def:tau}
Let $$\tau_i := \inf\{x\geq 0: \sum_j \Pr_{t_{ij}}\left[V_i(t_{ij})\geq \max\{\beta_{ij},Q_j+x\}\right]\leq \frac{1}{2}\}.$$
and define $A_i$ to be $\{j\ |\ \beta_{ij}\leq Q_j+\tau_i\}$.
\end{definition}

\begin{definition}\label{def:v hat}
For every buyer $i$ and type $t_i\in T_i$, let $Y_i(t_i)=\{j\ |\ V_i(t_{ij}) < Q_j + \tau_i\}$,  $$ \hat{v}_i(t_i,S) =v_i\left(t_i,S\cap Y_i(t_i)\right)$$
and
$$\hat{\gamma}^S_j(t_i) = \gamma_j^S(t_i)\cdot\ind[V_i(t_{ij})< Q_j+\tau_i]$$
 for any set $S\in [m]$. Moreover, let $$\widehat{\core}(M,\beta)=\sum_i\sum_{t_i\in T_i}f_i(t_i)\cdot \sum_{S\subseteq[m]}\sigma_{iS}^{(\beta)}(t_i)\cdot \hat{v}_i(t_i,S).$$ 
\end{definition}

In the next two Lemmas, we prove some useful properties of $\tau_i$. In particular, we argue that $\sum_{i\in[n]} \tau_i$ can be upper bounded by $\frac{4}{1-b}\cdot \prev$ (Lemma~\ref{lem:tau_i}).
 \begin{lemma}\label{lem:beta_ij}
\begin{align*}
\sum_i\sum_j \max \left\{\beta_{ij},Q_j+\tau_i\right\}\cdot \Pr_{t_{ij}}\left[V_i(t_{ij})\geq \max \left\{\beta_{ij},Q_j+\tau_i\right\}\right]
 \leq \frac{2}{1-b}\cdot \prev
\end{align*}
\end{lemma}
\begin{proof}
According to the definition of $\tau_i$, for every buyer $i$, $\sum_j \Pr_{t_{ij}}\left[V_i(t_{ij})\geq \max \{\beta_{ij},Q_j+\tau_i\}\right]=\frac{1}{2}$, 
 and $\max \{\beta_{ij},Q_j+\tau_i\}\geq \beta_{ij}$. Our statement follows directly from Lemma~\ref{lem:neprev}. 
\end{proof}

\begin{lemma}\label{lem:tau_i}
$$\sum_{i\in[n]} \tau_i\leq \frac{4}{1-b}\cdot \prev.$$
\end{lemma}
\begin{proof}
Since $Q_j$ is nonnegative,  \begin{align*}
  \sum_i\sum_j \max \left\{\beta_{ij},Q_j+\tau_i\right\}\cdot \Pr\left[V_i(t_{ij})\geq \max \{\beta_{ij},Q_j+\tau_i\}\right]
  \geq \sum_i \tau_i\cdot \sum_j \Pr\left[V_i(t_{ij})\geq{\max \{\beta_{ij},Q_j+\tau_i\}}\right].
  \end{align*}
According to the definition of $\tau_i$, when $\tau_i>0$, $$\sum_j \Pr\left[V_i(t_{ij})\geq {\max \{\beta_{ij},Q_j+\tau_i\}}\right]= \frac{1}{2}.$$
Therefore, $\sum_{i\in[n]} \tau_i\leq \frac{4}{1-b}\cdot \prev$ due to Lemma~\ref{lem:beta_ij}.
\end{proof}


In the following two Lemmas, we compare $\widehat{\core}(M,\beta)$ with $\core(M,\beta)$. The proof of Lemma~\ref{lem:hat gamma} is postponed to Appendix~\ref{appx:multi}.
\begin{lemma}\label{lem:hat gamma}
	For every buyer $i$, type $t_i\in T_i$, $\hat{v}_i(t_i,\cdot)$ satisfies monotonicity, subadditivity and no externalities. Furthermore, for every set $S\subseteq[m]$ and every subset $S'$ of $S$, $\hat{v}_i(t_i,S')\geq \sum_{j\in S'}\hat{\gamma}^S_j(t_i).$
\end{lemma}

\begin{lemma}\label{lem:hat Q}
	Let $$\hat{Q}_j = \frac{1}{2}\cdot \sum_i \sum_{t_i\in T_i}f_i(t_i)\cdot \sum_{S:j\in S}\sigma_{iS}^{(\beta)}(t_i)\cdot \hat{\gamma}_j^{S}(t_i).$$ Then, $$Q_j\geq \hat{Q}_j, \text{ for all $j\in[m]$ and}$$
	$$\sum_{j\in[m]}Q_j\leq \sum_{j\in[m]}\hat{Q}_j+\frac{(b+1)}{b\cdot(1-b)}\cdot \prev.$$
\end{lemma}

\begin{proof}
From the definition of $\hat{Q}_j$, it is easy to see that $Q_j\geq \hat{Q}_j$ for every $j$. So we only need to argue that $\sum_{j\in[m]}Q_j\leq \sum_{j\in[m]}\hat{Q}_j+\frac{(b+1)}{b\cdot(1-b)}\cdot \prev$.
\begin{equation}\label{eq:first}
	\begin{aligned}
	&\sum_{j} \left(Q_j- \hat{Q}_j\right) = \frac{1}{2}\cdot \sum_i \sum_j \sum_{t_i\in T_i}f_i(t_i)\cdot \sum_{S: j\in S}\sigma_{iS}^{(\beta)}(t_i)\cdot \left(\gamma_j^S(t_i)- \hat{\gamma}_j^{S}(t_i)\right)\\
	\leq & \frac{1}{2}\cdot \sum_i \sum_j \sum_{t_i\in T_i}f_i(t_i)\cdot \sum_{S: j\in S}\sigma_{iS}^{(\beta)}(t_i)\cdot \left (\beta_{ij}\cdot \ind\left[V_i(t_{ij})\geq Q_j+\tau_i\right]+c_i\cdot \ind\left[V_i(t_{ij})\geq \max\{Q_j+\tau_i,\beta_{ij}\}\right]\right)\\
	= & \frac{1}{2}\cdot \sum_i \sum_j \sum_{t_i\in T_i}f_i(t_i)\cdot \pi_{ij}^{(\beta)}(t_i)\cdot \left (\beta_{ij}\cdot \ind\left[V_i(t_{ij})\geq Q_j+\tau_i\right]+c_i\cdot \ind\left[V_i(t_{ij})\geq \max\{Q_j+\tau_i,\beta_{ij}\}\right]\right)
	\end{aligned}
\end{equation}

	This first inequality is because $\gamma_j^S(t_i)- \hat{\gamma}_j^{S}(t_i)$ is non-zero only when $V_i(t_{ij})\geq Q_j+\tau_i$, and the difference is upper bounded by $\beta_{ij}$ when $V_i(t_{ij})\leq \beta_{ij}$ and upper bounded by $\beta_{ij}+c_i$ when $V_i(t_{ij})> \beta_{ij}$.
	
	We first bound $\sum_i \sum_j \sum_{t_i\in T_i}f_i(t_i)\cdot \pi_{ij}^{(\beta)}(t_i)\cdot \beta_{ij}\cdot \ind[V_i(t_{ij})\geq Q_j+\tau_i]$.
	\begin{equation}\label{eq:second}
	\begin{aligned}
	&\sum_i \sum_j \sum_{t_i\in T_i}f_i(t_i)\cdot \pi_{ij}^{(\beta)}(t_i)\cdot \beta_{ij}\cdot \ind[V_i(t_{ij})\geq Q_j+\tau_i]\\
	\leq &\sum_i \sum_{j\in A_i} \beta_{ij}\cdot \sum_{t_i\in T_i}f_i(t_i)\cdot  \ind[V_i(t_{ij})\geq Q_j+\tau_i]+\sum_i \sum_{j\notin A_i} \beta_{ij}\cdot \sum_{t_i\in T_i}f_i(t_i)\cdot \pi_{ij}^{(\beta)}(t_i)\\
	\leq &\sum_i \sum_{j\in A_i} \beta_{ij}\cdot \Pr_{t_{ij}}[V_i(t_{ij})\geq Q_j+\tau_i]+\sum_i \sum_{j\notin A_i} \beta_{ij}\cdot \Pr_{t_{ij}}[V_i(t_{ij})\geq \beta_{ij}]/b\\
	\leq & (1/b) \cdot \sum_{i}\sum_{j} \max \{\beta_{ij},Q_j+\tau_i\}\cdot \Pr_{t_{ij}}\left[V_i(t_{ij})\geq \max \{\beta_{ij},Q_j+\tau_i\}\right]\\
	\leq & \frac{2}{b\cdot(1-b)}\cdot\prev\end{aligned}
\end{equation}
The set $A_i$ in the first inequality is defined in Definition~\ref{def:tau}. The second inequality is due to property (ii) in Lemma~\ref{lem:requirement for beta}. The third inequality is due to Definition~\ref{def:tau} and the last inequality is due to Lemma~\ref{lem:beta_ij}.

Next, we bound $\sum_i \sum_j \sum_{t_i\in T_i}f_i(t_i)\cdot \pi_{ij}^{(\beta)}(t_i)\cdot c_i\cdot \ind[V_i(t_{ij})\geq \max\{Q_j+\tau_i,\beta_{ij}\}]$.

\begin{equation}\label{eq:third}
		\begin{aligned}
			&\sum_i \sum_j \sum_{t_i\in T_i}f_i(t_i)\cdot \pi_{ij}^{(\beta)}(t_i)\cdot c_i\cdot \ind[V_i(t_{ij})\geq \max\{Q_j+\tau_i,\beta_{ij}\}]\\
			\leq & \sum_i c_i\sum_j\sum_{t_i}f_i(t_i)\cdot \ind[V_i(t_{ij})\geq \max\{Q_j+\tau_i,\beta_{ij}\}]\\
			\leq & \sum_i c_i\sum_j \Pr_{t_{ij}}[V_i(t_{ij})\geq  \max\{Q_j+\tau_i,\beta_{ij}\}]\\
			\leq & \sum_i c_i/2\\
			\leq& \frac{2}{(1-b)}\cdot\prev
		\end{aligned}
	\end{equation}
	
The last inequality is due to Lemma~\ref{lem:c_i}. Combining Inequality~(\ref{eq:first}),~(\ref{eq:second}) and~(\ref{eq:third}), we have proved our claim.
\end{proof}

By Lemma~\ref{lem:core and q_j}, $\sum_{j\in[m]}{Q}_j\leq {\core}(M,\beta)/2\alpha$. Hence, Lemma~\ref{lem:hat Q} shows that to approximate $\core(M,\beta)$, it suffices to approximate $\widehat{\core}(M,\beta)$. Indeed, we will use $\sum_{j\in[m]} \hat{Q}_j$ as an proxy for $\core(M,\beta)$ in our analysis of the ASPE.
\subsubsection{Design and Analysis of Our ASPE}
Consider the sequential post-price mechanism with anonymous posted price $Q_j$ for item $j$. We visit the buyers in the alphabetical order\footnote{We can visit the buyers in an arbitrary order. We use the the alphabetical order here just to ease the notations in the proof.} and charge every bidder an entry fee. We define the entry fee here.

\begin{definition}[Entry Fee]\label{def:entry fee}
For any bidder $i$, any type $t_i\in T_i$ and any set $S$, let $$ \mu_i(t_i,S) = \max_{S'\subseteq S} \big(\hat{v}_i(t_i, S') - \sum_{j\in S'} Q_j\big).$$ For any type profile $t\in T$ and any bidder $i$, let the entry fee for bidder $i$ be $$\delta_i(S_i(t_{<i})) = \Med_{t_i}\left[\mu_i\left(t_i, S_i(t_{<i})\right)\right]\footnote{Here $\Med_x[h(x)]$ denotes the median of a non-negative function $h(x)$ on random variable $x$, i.e.
$\Med_x[h(x)]=\inf\{a\geq 0:\Pr_{x}[h(x)\leq a]\geq\frac{1}{2}\}$.
},$$ where $S_1(t_{<1})=[m]$ and $S_i(t_{<i})$ is the set of items that are not purchased by the first $i-1$ buyers in the ASPE, when buyer $\ell$'s valuation is $v_\ell(t_\ell,\cdot)$ for all $\ell < i$. Notice that even though the seller does not know $t_{<i}$, she can compute the entry fee $\delta_i(S_i(t_{<i}))$, as she observes $S_i(t_{<i})$ after visiting the first $i-1$ bidders.
\end{definition}

In Lemma~\ref{lem:property of mu}, we show that $\tau_i$ is the Lipschitz constant for $\mu_i(\cdot,\cdot)$ and the proof is postponed to Appendix~\ref{appx:multi}. Moreover, $\sum_i \tau_i$ is upper bounded by $\frac{4}{1-b}\cdot \prev$ due to Lemma~\ref{lem:tau_i}.

\begin{lemma}\label{lem:property of mu}
For any $i$, the function $\mu_i(\cdot,\cdot)$ is $\tau_i$-Lipschitz. Moreover, for any type $t_i\in T_i$, $\mu_i(t_i,\cdot)$ satisfies monotonicity, subadditivity and no externalities.
\end{lemma}


The following Lemma is crucial for our proof. We show that in expectation over all type profiles, we can lower bound of the sum of $\mu_i(t_i,S_i(t_{<i}))$ for all bidders. In particular, this lower bound plus our ASPE's revenue from the posted prices already approximates $\widehat{\core}(M,\beta)$. The proof is inspired by Feldman et al.~\cite{FeldmanGL15}. Note that  $\mu_i(t_i,S_i(t_{<i}))$ is a lower bound of the real surplus of buyer $i$ for set $S_i(t_{<i})$. We choose to analyze the sum of $\mu_i(t_i,S_i(t_{<i}))$ because $\mu_i(\cdot,\cdot)$ has a small Lipschitz constant, which allows us to approximate $\mu_i(t_i,S_i(t_{<i}))$ with buyer $i$'s entry fee  $\mu_i(S_i(t_{<i}))$ and $\tau_i$. 
\begin{lemma}\label{lem:lower bounding mu}
	For all $j$, let $Q_j$ (Definition~\ref{def:posted prices}) be the price for item $j$ and every bidder's entry fee be described as in Definition~\ref{def:entry fee}. For every type profile $t\in T$, let $\textsc{SOLD}(t)$ be the set of items sold in the corresponding ASPE when  buyer $i$'s valuation is $v_i(t_i,\cdot)$. Then

\begin{align*}
\E_{t}\left[\sum_{i\in[n]} \mu_i\left(t_i,S_i(t_{< i})\right) \right]&\geq \sum_j \Pr_{t}[j\notin \sold(t)]\cdot(2\hat{Q}_j-Q_j)\\
&\geq \sum_{j\in[m]} \Pr_t\left[j\notin \sold(t)\right]\cdot Q_j - \frac{(2b+2)}{b\cdot(1-b)}\cdot\prev
\end{align*}

\end{lemma}

\begin{proof}
\begin{align*}
&	\E_{t}\left[\sum_i \mu_i\left(t_i,S_i(t_{< i})\right) \right]\\
\geq & \sum_i \E_{t_i, t_{-i}, t'_{-i}}\left[\mu_i\left(t_i,S_i(t_{<i})\cap M^{(\beta)}_i(t_i,t'_{-i})\right)\right]\\
\geq & \sum_i \E_{t_i, t_{-i}, t'_{-i}}\left[\sum_{j\in M^{(\beta)}_i(t_i,t'_{-i})} \ind\left[j\in S_i(t_{<i})\right]\cdot\left(\hat{\gamma}^{M^{(\beta)}_i(t_i,t'_{-i})}_j(t_i)-Q_j\right)^+ \right]\\
=& \sum_i \E_{t_i,t'_{-i}}\left[\sum_{j\in M^{(\beta)}_i(t_i,t'_{-i})} \Pr_{t_{<i}}[j\in S_i(t_{<i})]\cdot\left(\hat{\gamma}^{M^{(\beta)}_i(t_i,t'_{-i})}_j(t_i)-Q_j\right)^+\right]\\
=& \sum_i \E_{t_i}\left[\sum_{S\subseteq[m]} \sigma^{(\beta)}_{iS}(t_i)\cdot \sum_{j\in S} \Pr_{t_{<i}}[j\in S_i(t_{<i})]\cdot\left(\hat{\gamma}^{S}_j(t_i)-Q_j\right)^+\right]\\
=& \sum_i \E_{t_i}\left[\sum_{j\in [m]} \Pr_{t_{<i}}[j\in S_i(t_{<i})]\cdot \sum_{S: j\in S} \sigma^{(\beta)}_{iS}(t_i)\cdot\left(\hat{\gamma}^{S}_j(t_i)-Q_j\right)^+\right]\\
= & \sum_i \sum_j \Pr_{t_{<i}}[j\in S_i(t_{<i})]\cdot \E_{t_i}\left[\sum_{S: j\in S} \sigma^{(\beta)}_{iS}(t_i)\cdot\left(\hat{\gamma}^{S}_j(t_i)-Q_j\right)^+\right]\\
\geq & \sum_i\sum_j \Pr_{t}[j\notin \sold(t)]\cdot \E_{t_i}\left[\sum_{S: j\in S} \sigma^{(\beta)}_{iS}(t_i)\cdot\left(\hat{\gamma}^{S}_j(t_i)-Q_j\right)^+\right]\\
\geq & {\sum_j \Pr_{t}[j\notin \sold(t)] \sum_i \sum_{t_i} f_i(t_i)\cdot \sum_{S: j\in S} \sigma^{(\beta)}_{iS}(t_i)\cdot\left(\hat{\gamma}_j^S(t_i)-Q_j\right)}
\end{align*}

$t_{-i}'$ are fresh samples drawn from $D_{-i}$. The first inequality is because the $\mu_i(t_i,S)$ function is monotone in set $S$ for any $i$ and type $t_i\in T_i$. We use $\left(\hat{\gamma}^{M^{(\beta)}_i(t_i,t'_{-i})}_j(t_i)-Q_j\right)^+$ to denote $\max\left\{\hat{\gamma}^{M^{(\beta)}_i(t_i,t'_{-i})}_j(t_i)-Q_j, 0\right\}$. If we let $S$ be the set of items that are in $S_i(t_{<i})\cap M^{(\beta)}_i(t_i,t'_{-i})$ and satisfy that $\hat{\gamma}^{M^{(\beta)}_i(t_i,t'_{-i})}_j(t_i)-Q_j\geq 0$, then $\mu_i\left(t_i,S_i(t_{<i})\cap M^{(\beta)}_i(t_i,t'_{-i})\right)\geq \hat{v}_i(t_i,S)-\sum_{j\in S}Q_j\geq \sum_{j\in S} \left(\hat{\gamma}^{M^{(\beta)}_i(t_i,t'_{-i})}_j(t_i)-Q_j\right)$ due to the definition of $\mu_i(t_i,\cdot)$ and Lemma~\ref{lem:hat gamma}. This inequality is exactly the second inequality above. The next equality is because $S_i(t_{<i})$ only depends on the types of bidders other than $i$. The second last inequality is because $\Pr_{t_{<i}}[j\in S_i(t_{<i})]\geq \Pr_{t}[j\notin \sold(t)]$ for all $j$ and $i$, as the LHS is the probability that the item is not sold after the seller has visited the first $i-1$ bidders and the RHS is the probability that the item remains unsold till the end of the mechanism. Now, observe that $\sum_i \sum_{t_i} f_i(t_i)\cdot \sum_{S: j\in S} \sigma^{(\beta)}_{iS}(t_i)\cdot\hat{\gamma}_j^S(t_i)=2\hat{Q}_j$ for any $j$ according to the definition in Lemma~\ref{lem:hat Q}. Therefore,

\begin{align*}
&  {\sum_j \Pr_{t}[j\notin \sold(t)] \sum_i \sum_{t_i} f_i(t_i)\cdot \sum_{S: j\in S} \sigma^{(\beta)}_{iS}(t_i)\cdot\left(\hat{\gamma}_j^S(t_i)-Q_j\right)}\\
\geq &\sum_j \Pr_{t}[j\notin \sold(t)]\cdot (2\hat{Q}_j-Q_j)\\
= &  \sum_j \Pr_{t}[j\notin \sold(t)]\cdot Q_j-  \sum_j\Pr_{t}[j\notin \sold(t)]\cdot2(Q_j-\hat{Q}_j)\\
\geq & \sum_j \Pr_{t}[j\notin \sold(t)]\cdot Q_j - \sum_j 2(Q_j-\hat{Q}_j )~~~~~{\text{(Due to Lemma~\ref{lem:hat Q}, $Q_j-\hat{Q}_j\geq 0$ for all $j$)}}\\
\geq & \sum_j \Pr_{t}[j\notin \sold(t)]\cdot Q_j- \frac{(2b+2)}{b\cdot(1-b)}\cdot\prev~~~~~\text{(Lemma~\ref{lem:hat Q})}
\end{align*}
\end{proof}


\vspace{0.05in}
Since entry fee in the ASPE for every bidder as the median of her utility over the available items under $\hat{v}$. Clearly, bidders accept the entry fee with probability at least $1/2$, as their true utilities (under $v$) are always higher than their utilities under $\hat{v}$. Combining the concentration property of the utility under $\hat{v}$ and Lemma~\ref{lem:lower bounding mu}, we can argue that the total revenue from our ASPE is comparable to $\widehat{\core}(M,\beta)$, and therefore is comparable to $\core(M,\beta)$.


\begin{lemma}\label{lem:entry fee acceptance probability}
	For all $i$ and $t_{<i}$, bidder $i$ accepts $\delta_i(t_{<i})$ with probability at least $1/2$ when $t_i$ is drawn from $D_i$. Moreover, $$\aperev\geq \frac{1}{4}\cdot\sum_j Q_j- \left(\frac{5}{2(1-b)}+\frac{(b+1)}{2b\cdot(1-b)}\right)\cdot \prev.$$
\end{lemma}
\begin{proof}
For any bidder $i$, type $t_i\in T_i$ and any set $S$, define bidder $i$'s utility as $u_i(t_i,S) = \max_{S'\subseteq S} \big( v_i(t_i, S') -\sum_{j\in S'} Q_j\big).$ Clearly, $u_i(t_i,S) \geq \mu_i(t_i,S)$ for any type $t_i$ and set $S$. For any $t_{<i}$, as long as  $u_i(t_i, S_i(t_{<i})) \geq \delta_i(S_i(t_{<i}))$, buyer $i$ accepts the entry fee. Since $\delta_i(S_i(t_{<i}))$ is the median of $ \mu_i(t_i,t_{< i})$, $u_i(t_i, S_i(t_{<i})) \geq \delta_i(S_i(t_{<i}))$ with probability at least $1/2$ when $t_i$ is drawn from $D_i$. So the revenue from entry fee is at least $\frac{1}{2}\cdot\sum_i \E_{t_{<i}}\left[\delta_i\left(S_i(t_{<i})\right)\right].$ 

For any $i$ and $t_{<i}$, by Lemma~\ref{lem:property of mu} and Corollary~\ref{corollary:concentrate}, we are able to derive a lower bound for $\delta_i\left(S_i(t_{<i})\right)$, as shown in Lemma~\ref{lem:concentration entry fee}.

\begin{lemma}\label{lem:concentration entry fee}

	For all $i$ and $t_{<i}$, 
	{$$2\delta_i\left(S_i(t_{<i})\right)+\frac{5\tau_i}{2}\geq \E_{t_i}\left[\mu_i\left(t_i,S_i(t_{< i})\right) \right].$$}
\end{lemma}

\begin{proof}
It directly follows from Lemma~\ref{lem:property of mu} and Corollary~\ref{corollary:concentrate}. For any $i$ and $t_{<i}$, let $S_i(t_{<i})$ be the ground set $I$. Therefore, $\mu_i(t_i,\cdot)$  with $t_i\sim D_i$ is a function drawn from a distribution that is subadditive over independent items. Since, $\mu_i(\cdot,\cdot)$ is $\tau_i$-Lipschitz and
$\delta_i(S_i(t_{<i}))=\Med_{t_i}\left[\mu_i\left(t_i, S_i(t_{<i})\right)\right]$,
$$2\delta_i\left(S_i(t_{<i})\right)+\frac{5\tau_i}{2}\geq \E_{t_i}\left[\mu_i\left(t_i,S_i(t_{< i})\right) \right].$$\end{proof}

\vspace{0.05in}

Back to the proof of Lemma~\ref{lem:entry fee acceptance probability}. According to Lemma~\ref{lem:concentration entry fee}, the revenue from the entry fee is at least
$\frac{1}{4}\cdot\sum_i \E_{t_{<i},t_i}\left[\mu_i(t_i,S_i(t_{< i}))\right]- \frac{5}{8}\cdot\sum_i \tau_i$, which is equal to
$\frac{1}{4}\cdot\sum_i \E_{t}\left[\mu_i(t_i,S_i(t_{< i}))\right]- \frac{5}{8}\cdot\sum_i \tau_i$.
Combining Lemma~\ref{lem:tau_i} and Lemma~\ref{lem:lower bounding mu}, we can further show that the revenue from the entry fee is at least $\frac{1}{4}\sum_j \Pr_{t}[j\notin \sold(t)]\cdot Q_j - (\frac{5}{2(1-b)}+\frac{(b+1)}{2b(1-b)})\prev$. Since the revenue from the posted prices is exactly $\sum_j \Pr_t[j\in \sold(t)]\cdot Q_j$, the total revenue of the ASPE is at least $\frac{1}{4}\cdot\sum_j Q_j - \left(\frac{5}{2(1-b)}+\frac{(b+1)}{2b\cdot(1-b)}\right)\cdot \prev.$
\end{proof}

Combining everything together, we have the main result of Section~\ref{subsection:core}.

\begin{lemma}\label{lem:upper bounding Q}
	For any BIC mechanism $M$, {$\core(M,\beta)\leq8\alpha\cdot\aperev+4\alpha\left(\frac{6}{1-b}+\frac{1}{b(1-b)}\right)\prev$.}
\end{lemma}

\begin{proof}
It follows directly from Lemma~\ref{lem:core and q_j} and~\ref{lem:entry fee acceptance probability}.
\end{proof}

Now, we have upper bounded $\single(M,\beta)$, $\tail(M,\beta)$ and $\core(M,\beta)$ using the sum of the revenue of simple mechanisms (RSPM and ASPE). Combining these bounds, we complete the proof of Theorem~\ref{thm:multi}.

\vspace{.1in}
\begin{prevproof}{Theorem}{thm:multi}
	The proof follows from combining {Theorem~\ref{thm:revenue upperbound for subadditive}}, Lemma~\ref{lem:multi_single},~\ref{lem:multi decomposition},~\ref{lem:multi-tail} and~\ref{lem:upper bounding Q}.
\end{prevproof}



\subsubsection{Bad Example for Chawla and Miller's Approach}\label{sec:cs_ocrs}
Let bidders be constrained additive and $\FF_i$ be bidder $i$ feasibility constraint. We use $P_{\mathcal{F}_i}=conv(\{1^S | S\in \mathcal{F}_i\})$ to denote the feasibility polytope of bidder $i$. Let $\{q_{ij}\}_{i\in[n],j\in[m]}$ be a collection of probabilities that satisfy $\sum_i q_{ij}\leq 1/2$ for all item $j$ and $\boldsymbol{q}_i = (q_{i1},\ldots, q_{im})\in b\cdot P_{\mathcal{F}_i}$. Let $\beta_{ij}=F_{ij}^{-1}(q_{ij})$. The analysis by Chawla and Miller~\cite{ChawlaM16} needs to upper bound $\sum_{i, j}\beta_{ij}\cdot q_{ij}$ using the revenue of some BIC mechanism. When $\FF_i$ is a matroid for every bidder $i$, this expression can be upper bounded by the revenue of a sequential posted price mechanism constructed using OCRS from~\cite{FeldmanSZ16}. Here we show that if the bidders have general downward closed feasibility constraints, this expression is gigantic. More specifically, we prove that even when there is only one bidder, the expression could be $\Omega\left(\frac{\sqrt{m}}{\log m}\right)$ times larger than the optimal social welfare.

Consider the following example.
\begin{example}\label{ex:counterexample ocrs}
	The seller is selling $m=k^2$ items to a single bidder. The bidder's value for each item is drawn i.i.d. from distribution $F$, which is the equal revenue distribution truncated at $k$, i.e.,
	\[F(x)=
\begin{cases}
1-\frac{1}{x},&\text{if}~~x<k\\
1,&\text{if}~~x=k
\end{cases}
\]
 Items are divided into $k$ disjoint sets $A_1,...,A_k$, each with size $k$. The bidder is additive subject to feasibility constraint $\mathcal{F}=\left\{S\subseteq [m]|\exists i\in[k], S\subseteq A_i\right\}$.
\end{example}

\begin{lemma}\label{lem:counter example}
Let $P_{\mathcal{F}}=conv(\{1^S | S\in \mathcal{F}\})$ be the feasibility polytope for the bidder in Example~\ref{ex:counterexample ocrs}. Let $SW$ be the optimal social welfare. Then for any constant $b>0$, there exists $q\in b\cdot P_{\mathcal{F}}$ such that for sufficiently large $k$, $$\sum_{j\in[m]}q_j\cdot F^{-1}(1-q_j)=\Theta(\frac{k}{\log k })\cdot SW$$
\end{lemma}

\begin{proof}

For any $b>0$, consider the following feasible allocation rule: w.p. $(1-b)$, don't allocate anything, and w.p. $b$, give the buyer one of the sets $A_i$ uniformly at random. The corresponding ex-ante probability vector $q$ satisfies $q_j=\frac{b}{k}, \forall j\in [m]$. Thus $q\in b\cdot P_{\mathcal{F}}$. Since $q_j<\frac{1}{k}$, $F^{-1}(1-q_j)=k$ for all $j\in [m]$. We have $\sum_{j\in[m]}q_j\cdot F^{-1}(1-q_j)=k^2\cdot \frac{b}{k}\cdot k=b\cdot k^2$. We use $V_i$ to denote the random variable of the bidder's value for set $A_i$. It is not hard to see that $SW=\E[\max_{i\in[k]} V_i]$. 

\begin{lemma}
For any $i\in [k]$,
\[\Pr\left[V_i>3\cdot k\log(k)\right]\leq k^{-3}\]
\end{lemma}
\begin{proof}
Let $X$ be random variable with cdf $F$. Notice $E[X]=\log(k)$, $E[X^2]=2k$, and $|X|\leq k$.
For every $i$, by the Bernstein concentration inequality, for any $t>0$,
\[\Pr\left[V_i-k\log(k)>t\right]\leq \exp\left(-\frac{\frac{1}{2}t^2}{2k^2+\frac{1}{3}kt}\right)\]
Choose $t=2k\log(k)$, we have
\[\Pr\left[V_i>3k\log(k)\right]\leq \exp(-3\log(k))=k^{-3}\]
\end{proof}

By the union bound, $\Pr[\max_{i\in[k]}V_i>3\cdot k\log(k)]\leq k^{-2}$. Therefore, $\E[\max_{i\in[k]} V_i]\leq 3 k\log k +k^2\cdot k^{-2}\leq 4 k\log k$.
\end{proof}

\notshow{In the analysis of the paper by Chawla and Miller~\cite{ChawlaM16}, they rely on the following lemma in single buyer auction.
\begin{lemma}\label{lem:shuchi}
~\cite{ChawlaM16}~\cite{FeldmanSZ16}Suppose the buyer is additive within a matroid constraint $\mathcal{F}$ and let $P_{\mathcal{F}}=conv(\{1^S | S\in \mathcal{F}\})$ be the feasibility polytope. For any constant $b\in (0,1)$ and ex-ante vector $q\in bP_{\mathcal{F}}$, let $\beta$ be the corresponding ex-ante prices. In other words, $\beta_j$ is chosen such that the probability that the value for item $j$
exceeds this price is precisely $q_j$. Then the value $\beta\cdot q$ can be bounded within some constant factor by the revenue of a posted price mechanism with a more strict constraint, which guarantees that the ex-ante probability of the buyer getting item $j$ is at most $q_j$.
\end{lemma}

The result can be generated for $\mathcal{F}$ beyond a matroid~\cite{FeldmanSZ16}. However, Lemma~\ref{lem:shuchi} does not hold for general downward-close $\mathcal{F}$. In this section we provide a counterexample with some general downward-close constraint $\mathcal{F}$, such that the term $\beta\cdot q$ cannot be upper bounded by any single buyer mechanism, within in a constant factor.

\begin{lemma}
Consider the following single buyer auction with $m=k^2$ i.i.d. items. Items are divided into $k$ disjoint sets $A_1,...,A_k$, each with size $k$. The value distribution $F$ for a single item is defined as the equal-revenue distribution truncated at value $k$, i.e.,
\[F(x)=
\begin{cases}
1-\frac{1}{x},&\text{if}~~x<k\\
1,&\text{if}~~x=k
\end{cases}
\]
Define a downward-close feasibility constraint $\mathcal{F}$ for the buyer as $\mathcal{F}=\{S\subseteq [m]|\exists p, S\subseteq A_p\}$. Let $P_{\mathcal{F}}=conv(\{1^S | S\in \mathcal{F}\})$ be the feasibility polytope. Let $SW$ be the optimal social welfare among all BIC mechanisms in this auction. Then for any $b>0$, there exists $q\in bP_{\mathcal{F}}$ such that for sufficiently large $k$,
\begin{equation}\label{equ:q times beta}
\sum_{j\in[m]}q_j\cdot F^{-1}(1-q_j)=\Theta(\frac{k}{log(k)})\cdot SW
\end{equation}
\end{lemma}

\begin{proof}

For any $b>0$, consider the following feasible allocation rule. With probability $(1-b)$, don't allocate anything. With probability $b$, give the buyer one of the sets $A_p$ uniformly at random. The corresponding ex-ante probability vector $q$ satisfies $q_j=\frac{b}{k}, \forall j\in [m]$. Thus $q\in bP_{\mathcal{F}}$.

Since $q_j<\frac{1}{k}$, $F^{-1}(1-q_j)=k$ for all $j\in [m]$. We have
\begin{equation}
\sum_{j\in[m]}q_j\cdot F^{-1}(1-q_j)=k^2\cdot \frac{b}{k}\cdot k=b\cdot k^2
\end{equation}

Consider the optimal social welfare. For every bundle $p$, denote $V_p$ the random variable of the buyer's value for bundle $p$. Notice $A_p\in \mathcal{F}$, $V_p$ is the sum of $k$ independent random variables with cdf $F$. With Bernstein Inequality, $V_p=O(klog(k))$ with high probability.
\notshow{\begin{lemma}\label{lem:bernstein}
(Bernstein Inequality)~\cite{bernstein1924modification}: Suppose $X_1,...,X_n$ are independent random variables with zero mean, and $|X_i|\leq M$ almost surely for all $i$. Then for any $t>0$,
\[\Pr\left[\sum_{i=1}X_i>t\right]\leq exp\left(-\frac{\frac{1}{2}t^2}{\sum_{i=1}^nE[X_i^2]+\frac{1}{3}Mt}\right)\]
\end{lemma}}
\begin{lemma}
For any $p\in [k]$,
\[\Pr\left[V_p>3\cdot k\log(k)\right]\leq k^{-3}\]
\end{lemma}
\begin{proof}
Let $X$ be random variable with cdf $F$. Notice $E[X]=\log(k)$, $E[X^2]=k-1$, and $|X|\leq k$.
For every $p$, by the Bernstein concentration inequality, for any $t>0$,
\[\Pr\left[V_p-k\log(k)>t\right]\leq exp\left(-\frac{\frac{1}{2}t^2}{k^2+\frac{1}{3}kt}\right)\]
Choose $t=2k\log(k)$, we have
\[\Pr\left[V_p>3k\log(k)\right]\leq exp(-3\log(k))=k^{-3}\]
\end{proof}

With union bound,
\[\Pr[\max_{p\in [k]}V_p\geq 3k\log(k)]\leq \sum_{p\in [k]}\Pr\left[V_p>3k\log(k)\right]\leq k^{-2}\]

Notice the social welfare for the mechanism is at most \\
\noindent$\max_{p\in [k]}V_p$ due to the feasible constraint $\mathcal{F}$. Also notice that $\max_{p\in [k]}V_p\leq k^2$, we have
\begin{align*}
SW&\leq E[\max_{p\in [k]}V_p]\leq \bigg(3k\log(k)\cdot \Pr[\max_{p\in [k]}V_p\leq 3k\log(k)]\\
&+k^2\cdot\Pr[\max_{p\in [k]}V_p> 3k\log(k)]\bigg)\\
&\leq 3k\log(k)+k^2\cdot k^{-2}=O(k\log(k))
\end{align*}

\noindent When $k$ is sufficiently large, Equation~\ref{equ:q times beta} holds.

\end{proof}}

\section{Open Questions }
 The main open question following our results is:
 \begin{itemize}
\item \textbf{Open Question:} \em{Can we design simple and approximately revenue-optimal mechanisms for multiple buyers with valuations that are subadditive over independent items?}
\end{itemize}
A large fraction of the proof in this work already applies to subadditive valuations. More specifically, our upper bound for the optimal revenue from Theorem~\ref{thm:revenue upperbound for subadditive} holds for all subadditive valuations, and we have used it to obtain a constant factor approximation for a single subadditive buyer and a $O(\log m)$-approximation for multiple subadditive buyers. Our analysis for the term $\single$ and $\tail$ also applies to subadditive valuations.

 The only component that does not extend to subadditive valuations is the analysis of the $\core$. We heavily used a convenient property of XOS functions, namely, the existence of $1$-supporting prices~\cite{DobzinskiNS05,BhawalkarR11}. Unfortunately, general subadditive functions only permit $\Theta(\log m)$-supporting prices, which is why our approximation ratio degrades to  $O(\log m)$ for subadditive valuations. Our analysis of the $\core$ makes use of the supporting prices in two places: (i) the definition of the proxy core $\widehat{\core}(M,\beta)$ and its comparison to $\core(M,\beta)$ in Section~\ref{sec:proxy core}; (ii) lower bounding the revenue of ASPE. For (ii), our proof is inspired by Feldman et al.~\cite{FeldmanGL15}, who showed that there exists a sequential posted-price mechanism that is an $O(1)$-approximation to the optimal social welfare for bidders with XOS valuations. Their proof also makes heavy use of the supporting prices, and the approximation ratio degrades to $O(\log m)$ for subadditive valuations. To resolve the open question above, it is worthwhile to first consider generalizing the result of~\cite{FeldmanGL15}. In particular,  
\begin{itemize}
\item \textbf{Open Question:} \em{Can sequential posted-price mechanisms obtain a constant fraction of the optimal social welfare when bidders have subadditive valuations?}
\end{itemize}
 

An astute reader may have noticed that the approximation results in this paper are only existential. Luckily, the only nonconstructive part of our argument is finding the right $\beta$, which is essentially the same as finding the ex-ante allocation probabilities of the optimal or an approximately optimal mechanism. In Appendix~\ref{sec:symmetric computation}, we show how to find the right $\beta$ when the bidders are symmetric, but the asymmetric case remains open. \begin{itemize}
\item \textbf{Open Question:} \em{Can we design a polynomial time algorithm to compute these simple and approximately optimal mechanisms for constrained additive and XOS valuations?}
\end{itemize}

\newpage
\appendix
\section*{\huge{Appendix}}
\notshow{
\section{Details from Preliminaries}

We first formally define various valuation classes.
\begin{definition}\label{def:valuation classes}
We define several classes of valuations formally. Let $t$ be the type and $v(t,S)$ be the value for bundle $S\in[m]$.
	
\begin{itemize}[leftmargin=0.5cm]
	\item  \textbf{Constrained Additive} \\
$v(t,S) =\max_{R\subseteq S, R\in \mathcal{I}}\sum_{j\in R} v(t, \{j\})$, where $\mathcal{I}\subseteq 2^{[m]}$ is a downward closed set system over the items specifying the feasible bundles. In particular, when $\mathcal{I}=2^{[m]}$, the valuation is an \textbf{additive function}; when $\mathcal{I}=\{\{j\}\ |\ j\in[m] \}$, the valuation is a \textbf{unit-demand function}; when $\mathcal{I}$ is a matroid, the valuation is a \textbf{matroid-rank function}. An equivalent way to represent any constrained additive valuations is to view the function as additive but the bidder is only allowed to receive bundles that are feasible, i.e., bundles in $\mathcal{I}$. To ease notations, we interpret $t$ as an $m$-dimensional vector $(t_1, t_2,\cdots, t_m)$ such that $t_j =  v(t, \{j\})$.

\item \textbf{XOS/Fractionally Subadditive}\\
 $v(t,S) = \max_{i\in [{K}]} v^{(i)}(t, S)$, where ${K}$ is some finite number and $v^{(i)}(t,\cdot)$ is an additive function for any $i\in[{K}]$.

\item \textbf{Subadditive}\\ 
$v(t,S_1\cup S_2)\leq v(t,S_1)+v(t,S_2)$ for any subset $S_1, S_2\subseteq [m]$.
\end{itemize}

\end{definition}

{The following are a few examples of various valuation distributions which are over independent items (Definition~\ref{def:subadditive independent}):

\begin{example}\label{eg:valuation}\cite{RubinsteinW15}
$t=\{t_j\}_{j\in[m]}$ where $t$ is drawn from $\prod_j D_j$,
\begin{itemize}[leftmargin=0.5cm]
\item Additive: $t_j$ is the value of item $j$. $v(t,S)=\sum_{j\in S}t_j$.
\item Unit-demand: $t_j$ is the value of item $j$. $v(t,S)=\max_{j\in S}t_j$.
\item Constrained Additive: $t_j$ is the value of item $j$.
 $v(t,S)=\max_{R\subseteq S, R\in \mathcal{I}}\sum_{j\in R} t_j$.
\item XOS/Fractionally Subadditive: $t_j=\{t_{j}^{(k)}\}_{k\in[K]}$ encodes all the possible values associated with item $j$, and $v(t,S)=\max_{k\in[K]}\sum_{j\in S}t_{j}^{(k)}$.
\end{itemize}
\end{example}

}
}

\section{Improved Analysis for Constrained Additive Valuation}

In this section, we show that for constrained additive bidders, we do not need to relax the valuations, as applying directly the flow in Section~\ref{sec:flow} already gives an upper bound with the right format. So we can take $M^{(\beta)}$ to simply be $M$. In particular, we can derive the following improved upper bound for $\rev(M,v,D)$ using essentially the same proof as in Section~\ref{sec:flow}.
\begin{theorem}\label{thm:revenue upperbound for constrained additive}
If for any bidder $i$ any type $t_i\in T$, $v_i(t_i, \cdot)$ is a constrained additive valuation, then for any mechanism $M$ and any $\beta=\{\beta_{ij}\}_{i\in[n], j\in[m]}$,
$$\rev{(M,v,D)}\leq \textsc{Non-Favorite}(M,\beta)+\textsc{Single}(M,\beta).$$
\end{theorem}

Combining the same upper bounds we obtained for \\
\noindent$\nf(M,\beta)$ and $\single(M,\beta)$ and the improved upper bound in Theorem~\ref{thm:revenue upperbound for constrained additive}, we can improve the approximation ratio when the bidder(s) have constrained additive valuations.
\begin{theorem}
	For a single buyer whose valuation is constrained additive,
$$\rev(M,v,D)\leq 7\cdot\srev+4\cdot\brev,$$
for any BIC mechanism $M$.
\end{theorem}

\begin{theorem}
	For multiple buyers whose valuations are constrained additive,
\begin{equation}
\begin{aligned}
\rev(M,v,D)&\leq 8\cdot\aperev\\
&+\left(6+\frac{22}{1-b}+\frac{4(b+1)}{(1-b)b}\right)\cdot \prev
\end{aligned}
\end{equation}
for any BIC mechanism $M$. In particular, if we set $b$ to be $\frac{1}{4}$, then $$\rev(M,v,D)\leq 8\cdot\aperev+62\cdot\prev.$$
\end{theorem}

\section{Proof of Theorem~\ref{thm:revenue less than virtual welfare}}\label{sec:proof_duality}
\begin{prevproof}{Theorem}{thm:revenue less than virtual welfare}
When $\lambda$ is useful, we can simplify function $\L(\lambda, \sigma, p)$ by removing the term associated with $p$ and replacing $\sum_{t_{i}'\in T_{i}^{+}}\lambda(t_{i},t_{i}')$ with $f_{i}(t_{i})+\sum_{t_{i}'\in T_{i}} \lambda(t_{i}',t_{i})$. After the simplification, we have
\begin{align*}
&\L(\lambda, \sigma, p) = \sum_{i=1}^{n} \sum_{t_{i}\in T_{i}} f_{i}(t_{i})\cdot \sum_{S\subseteq[m]}\sigma_{iS}(t_{i})\cdot \\
&\left( v_i(t_{i},S)-{1\over f_{i}(t_{i})}\sum_{t_{i}'\in T_{i}} \lambda_{i}(t_{i}',t_{i})\left(v_i(t_{i}',S)-v_i(t_{i},S)\right)\right)\\
&\hspace{1.3cm}=\sum_{i=1}^{n} \sum_{t_{i}\in T_{i}} f_{i}(t_{i})\cdot \sum_{S\subseteq[m]}\sigma_{iS}(t_{i})\cdot\Phi_{i}(t_{i},S),	
\end{align*}
 which is exactly the virtual welfare of $\sigma$ with respect to $\lambda$. Now, we only need to prove that $\L(\lambda, \sigma, p)$ is greater than the revenue of $M$. Let us think of $\L(\lambda, \sigma, p)$ using Expression~(\ref{eq:primal lagrangian}). Since $M$ is a BIC mechanism, $$\sum_{S\subseteq[m]}v_i(t_i,S)\cdot\left(\sigma_{iS}(t_{i})-\sigma_{iS}({t_{i}'})\right)-\left((p_{i}(t_{i})-p_{i}(t_{i}')\right)\geq 0$$ for any $i$, $t_{i}\in T_{i}$ and $t_i'\in T_i^+$. Also, all the dual variables $\lambda$ are nonnegative. Therefore, it is clear that $\L(\lambda, \sigma, p)$ is at least as large as the revenue of $M$.

When $\lambda^{*}$ is the optimal dual variable, by strong duality, we know $\max_{\sigma\in\polytope, p}\L(\lambda^{*}, \sigma, p)$ equals to the revenue of $M^{*}=(\sigma^*,p^*)$. But we also know that $\L(\lambda^{*}, \sigma^{*}, p^{*})$ is at least as large as the revenue of $M^{*}$, therefore $\sigma^{*}$ maximizes the virtual welfare.
\end{prevproof}

\section{Recap: Flow for Additive Valuations}\label{sec:flow_additive}
When the valuations are additive,   we simply view $t_{ij}$ as bidder $i$'s value for receiving item $j$. Although there are many possible ways to define a flow, we focus on a class of simple ones. Every flow in this class $\lambda^{(\beta)}$ is parametrized by a set of parameters $\beta=\{\beta_{ij}\}_{i\in[n], j\in[m]}\in\R^{nm}$. Based on $\beta_i=\{\beta_{ij}\}_{j\in[m]}$, we first partition the type space $T_i$ for each bidder $i$ into $m+1$ regions:
\begin{itemize}[leftmargin=0.7cm]
	\item $R_{0}^{(\beta_i)}$ contains all types $t_i$ such that $t_{ij}<\beta_{ij}$ for all $j\in[m]$.
	\item $R_j^{(\beta_i)}$ contains all types $t_i$ such that $t_{ij}-\beta_{ij}\geq 0$ and $j$ is the smallest index in $\argmax_k\{t_{ik}-\beta_{ik}\}$.
\end{itemize}

We use essentially the same flow as in~\cite{CaiDW16}. Here we provide a partial specification and state some desirable properties of the flow. See Figure~\ref{fig:multiflow} for an example with $2$ items and~\cite{CaiDW16} for a complete description of the flow.
\begin{figure}[ht]
\colorbox{MyGray}{
\begin{minipage}{\textwidth}
{\bf Partial Specification of the flow $\lambda^{(\beta)}$:}
\begin{enumerate}[leftmargin=0.7cm]
 \item For every type $t_{i}$ in region $R^{(\beta_{i})}_{0}$,  the flow goes directly to $\varnothing$ (the super sink).
 \item  For all $j>0$, any flow entering $R^{(\beta_{i})}_{j}$ is from  $s$ (the super source) and any flow leaving $R^{(\beta_{i})}_{j}$ is to $\varnothing$.
 \item For all $t_{i}$ and $t_{i}'$ in $R^{(\beta_{i})}_{j}$ ($j>0$), {$\lambda^{(\beta)}_{i}(t_{i},t_{i}')>0$} only if $t_{i}$ and $t_{i}'$ only differ in the $j$-th coordinate.
\end{enumerate}
\end{minipage}}
\caption{Partial Specification of the flow $\lambda^{(\beta)}$.}
\label{fig:flow specification}
\end{figure}

\begin{figure}
  \centering{\includegraphics[width=0.5\linewidth]{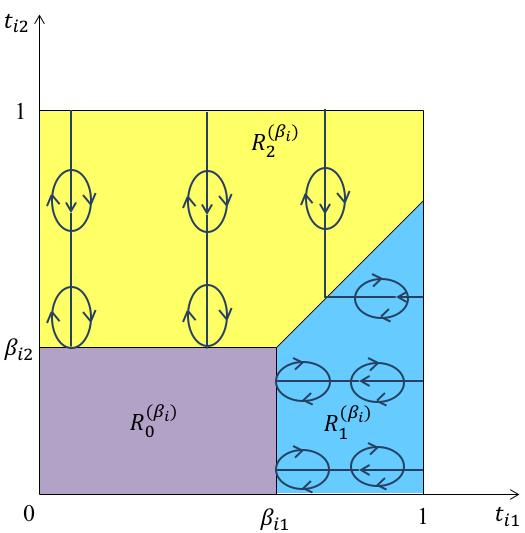}}
  \caption{An example of $\lambda^{(\beta)}_{i}$ for additive bidders with two items.}
  \label{fig:multiflow}
\end{figure}

\begin{lemma}[\cite{CaiDW16}\footnote{Note that this Lemma is a special case of Lemma 3 in~\cite{CaiDW16} when the valuations are additive. }]\label{lem:additive flow properties}
	For any $\beta$, there exists a flow $\lambda^{(\beta)}_{i}$ such that the corresponding virtual value function $\Phi_{i}(t_{i}, \cdot)$ satisfies the following properties:
	\begin{itemize}[leftmargin=0.5cm]
		\item For any $t_{i}\in R^{(\beta_i)}_{0}$, $\Phi_{i}(t_{i},S) = \sum_{k\in S} t_{ik}$.
		\item For any $j>0$, $t_{i}\in R^{(\beta_i)}_{j}$,  $$\Phi_{i}(t_{i},S)\leq \sum_{k\in S \land  k\neq j} t_{ik}+\tp_{ij}(t_{ij})\cdot\ind[j\in S],$$ where $\tp_{ij}(\cdot)$ is Myerson's ironed virtual value function for  $D_{ij}$.
	\end{itemize}
\end{lemma}

The properties above are crucial for showing the approximation results for simple mechanisms in~\cite{CaiDW16}. One of the key challenges in approximating the optimal revenue is how to provide a tight upper bound. A trivial upper bound is the social welfare, which may be arbitrarily bad in the worst case. By plugging the virtual value functions in Lemma~\ref{lem:additive flow properties} into the partial Lagrangian, we obtain a new upper bound that replaces the value of the buyer's favorite item with the corresponding Myerson's ironed virtual value. As demonstrated in~\cite{CaiDW16}, this new upper bound is at most $8$ times larger than the optimal revenue when the buyers are additive, and its appealing structure allows the authors to compare the revenue of simple mechanisms to it. In \notshow{the next section} {Section~\ref{sec:flow}}, we identify some difficulties in directly applying this flow to subadditive valuations. Then we show how to overcome these difficulties by relaxing the subadditive valuations and obtain a similar upper bound.

\section{Proof of Lemma~\ref{lem:subadditive flow properties}}\label{sec:proof_virtual_relaxation}
\begin{lemma}\label{lem:separate the favorite out in virtual value}
	For any flow $\lambda^{(\beta)}_i$ that respects the partial specification in Figure~\ref{fig:flow specification}, the corresponding virtual valuation function $\Phi_i^{(\beta_i)}$ of $v_i^{(\beta_i)}$ for any buyer $i$ is:
\begin{itemize}[leftmargin=0.7cm]
\item $v_i(t_i, S\backslash \{j\})+V_i(t_{ij})-\frac{1}{f_i(t_i)}\sum_{t'_i\in T_i}\lambda(t'_i,t_i)\cdot
    \left(V_i(t'_{ij})-V_i(t_{ij})\right)$, if $t_i\in R_j^{(\beta_i)} \text{ and } j\in S$.
\item $v_i(t_i,S)$, otherwise.
\end{itemize}

\end{lemma}
\begin{prevproof}{Lemma}{lem:separate the favorite out in virtual value}
	The proof follows the definitions of the virtual valuation function (Definition~\ref{def:virtual value}) and relaxed valuation (Definition~\ref{def:relaxed valuation}). We use $t_{i,-j}=\langle t_{i{j'}}\rangle_{j'\not=j}$ to denote bidder $i$'s information for all items except item $j$. If $t_i\in R_j^{(\beta_i)}$ and $j\in S$, $v_i^{(\beta_i)}(t_i,S) = v_i(t_i, S\backslash \{j\})+V_i(t_{ij})$. Since $\lambda(t_i,t_i')>0$ only when $t_{i,-j}=t_{i,-j}'$ and $t_i'\in R_j^{(\beta_i)}$,  $v_i^{(\beta_i)}(t'_i,S) =  v_i(t'_i, S\backslash \{j\})+V_i(t'_{ij})= v_i(t_i, S\backslash \{j\})+V_i(t'_{ij})$. Therefore,
\begin{align*}
    \Phi_i^{(\beta_i)}(t_i, S)=v_i(t_i, S\backslash \{j\})+V_i(t_{ij})
    -\frac{1}{f_i(t_i)}\sum_{t'_i\in T_i}\lambda(t'_i,t_i)\cdot\left(V_i(t'_{ij})-V_i(t_{ij})\right)
\end{align*}
	
	If $t_i\in R_j^{(\beta_i)}$ and $j\notin S$ or $t_i\in R_0^{(\beta_i)}$, then $v_i^{(\beta_i)}(t_i,S) =v_i(t_i, S)$. If $t_i\in R_0^{(\beta_i)}$, there is no flow entering $t_i$ except from the source, so clearly  $\Phi_i^{(\beta_i)}(t_i, S)=v_i(t_i, S)$. If $t_i\in R_j^{(\beta_i)}$, then for any $t'_i$ that only differs from $t_i$ in the $j$-th coordinate, we have $v_i(t'_i, S)=v_i(t_i,S)$, because {$j\not\in S$}. Hence,  $\Phi_i^{(\beta_i)}(t_i, S)=v_i(t_i, S)$.
\end{prevproof}

\begin{prevproof}{Lemma}{lem:subadditive flow properties}

Let $\Psi_{ij}^{(\beta_i)}(t_i)=V_i(t_{ij})-\frac{1}{f_i(t_i)}\sum_{t'_i\in T_i}\lambda(t'_i,t_i)\cdot\left(V_i(t'_{ij})-V_i(t_{ij})\right)$. According to Lemma~\ref{lem:separate the favorite out in virtual value}, it suffices to prove that for any $j>0$, any $t_{i}\in R^{(\beta_i)}_{j}$, $\Psi_{ij}^{(\beta_i)}(t_i)\leq \tp_{ij}(V_i(t_{ij}))$.

\begin{claim}
For any type $t_{i}\in R^{(\beta_i)}_{j}$, if we only allow flow from type $t'_{i}$ to $t_{i}$, where $t_{ik}=t'_{ik}$ for all $k\neq j$ and $t'_{ij}\in \argmin_{s\in T_{ij} \land V_i(s)> V_i(t_{ij})} V_i(s)$, and the flow $\lambda(t_i',t_i)$ equals $\frac{f_{ij}(t_{ij})}{\Pr_{v\sim F_{ij}}[v= V_i(t_{ij})]}$ fraction of the total in flow to $t_i'$, then there exists a flow $\lambda$ such that
\begin{align*}
\Psi_{ij}^{(\beta_i)}(t_i)=\varphi_{ij}(V_i(t_{ij}))
=V_i(t_{ij})-\frac{\left(V_i({t'_{ij}})-V_i(t_{ij})\right)\cdot\Pr_{v\sim F_{ij}}[v>V_i(t_{ij})]}{\Pr_{v\sim F_{ij}}[v= V_i(t_{ij})]},
\end{align*} where $\varphi_{ij}(V_i(t_{ij}))$ is the Myerson virtual value for $V_i(t_{ij})$ with respect to $F_{ij}$. \end{claim}
\begin{proof}
{As the flow only goes from $t_i'$ and $t_i$, where $t_i'$ and $t_i$ only differs in the $j$-th coordinate, and \\
\noindent$t_{ij}\in \argmax_{s\in T_{ij} \land V_i(s)< V_i(t_{ij}')} V_i(s)$. If $t_{ij}$ is a type with the largest $V_i(t_{ij})$ value in $T_{ij}$, then there is no flow coming into it except the one from the source, so $\Psi_{ij}^{(\beta_i)}(t_i)=V_i(t_{ij})$. For every other value of $t_{ij}$, the in flow is exactly
\begin{align*} \frac{f_{ij}(t_{ij})}{\Pr_{v\sim F_{ij}}[v= V_i(t_{ij})]}\prod_{k\neq j} f_{ik}(t_{ik})\cdot  \sum_{x\in T_{ij}:V_i({x})>V_i(t_{ij})} f_{ij}(x) 
=\prod_{k} f_{ik}(t_{ik})\cdot \frac{\Pr_{v\sim F_{ij}}[v>V_i(t_{ij})]}{\Pr_{v\sim F_{ij}}[v=V_i(t_{ij})]}.\end{align*}
 {This is because each type of the form $(x,t_{i,-j})$ with $V_i(x) > V_i(t_{ij})$ is also in $R^{(\beta_i)}_{j}$. So $\frac{f_{ij}(t_{ij})}{\Pr_{v\sim F_{ij}}[v= V_i(t_{ij})]}$ of all flow that enters these types will be passed down to $t_{i}$ (and possibly further, before going to the sink), and the total amount of flow entering all of these types from the source is exactly {$\prod_{k\neq j} f_{ik}(t_{ik})\cdot \sum_{x\in T_{ij}:V_i({x})>V_i(t_{ij})} f_{ij}(x) $}.} Therefore, $\Psi_{ij}^{(\beta_i)}(t_i)=\varphi_{ij}(V_i(t_{ij}))$. Whenever there is no more type $t_i\in R_j^{(\beta_i)}$ with smaller $V_i(t_{ij})$ value, we push all the flow to the sink.}
\end{proof}


If $F_{ij}$ is regular, this completes our proof. When $F_{ij}$ is not regular, we can iron the virtual value function in the same way as in \cite{CaiDW16}. Basically, for two types $t_i,t_i'\in R^{(\beta_i)}_{j}$ that only differ in the $j$-th coordinate, if $\Psi_{ij}^{(\beta_i)}(t_i)>\Psi_{ij}^{(\beta_i)}(t_i')$ but $V_i(t_{ij})<V_i(t_{ij}')$, add a loop between $t_i$ and $t_i'$ with a proper weight to make $\Psi_{ij}^{(\beta_i)}(t_i)=\Psi_{ij}^{(\beta_i)}(t_i')$.
\begin{lemma}\cite{CaiDW16}
For any $\beta$ and $i$, there exists a flow $\lambda_i(\beta)$ such that for any $t_i\in R_j^{(\beta_i)}$, $\Psi_{ij}^{(\beta_i)}(t_i)\leq \tp_{ij}(V_i(t_{ij}))$.
\end{lemma}
\end{prevproof}

\section{Analysis for the Single-Bidder Case}\label{sec:single_appx}

\begin{prevproof}{lemma}{lem:single decomposition}
In $\nf$, since $R^{\beta}_0=\emptyset$, the corresponding term is simply $0$. Notice $v(t,\cdot)$ is a monotone valuation for every $t\in T$,
\begin{align*}
\nf(M)=&\sum_{t\in T}f(t)\cdot \sum_{j\in[m]} \ind\left[t\in R_j^{(\beta)}\right]\cdot \left(\sum_{S:j\in S}\sigma_{S}^{(\beta)}(t)\cdot v(t,S\backslash\{j\})+\sum_{S:j\notin S}\sigma_{S}^{(\beta)}(t)\cdot v(t,S)\right)\\
\leq &\sum_{t\in T}f(t)\cdot \sum_{j\in[m]} \ind\left[t\in R_j^{(\beta)}\right]\sum_S\sigma_{S}^{(\beta)}(t)\cdot v(t,[m]\backslash \{j\})\\
\leq&\sum_{t\in T}f(t)\cdot \sum_{j\in[m]} \ind\left[t\in R_j^{(\beta)}\right]\cdot v(t,[m]\backslash \{j\})~~~~~(\sum_S\sigma_{S}^{(\beta)}(t)\leq 1)
\end{align*}

Recall that for all $t\in T$ and $S\subseteq [m]$, $v(t,S)\leq v\left(t,S\cap \mathcal{C}(t)\right)+\sum_{j\in S\cap \mathcal{T}(t)}V(t_j)$. We will replace $v(t,[m]\backslash \{j\})$ above with $v\left(t,([m]\backslash \{j\})\cap \mathcal{C}(t)\right)+\sum_{k\in ([m]\backslash \{j\})\cap \mathcal{T}(t)}V(t_k)$. First, the contribution from $v\left(t,([m]\backslash \{j\})\cap \mathcal{C}(t)\right)$ is upper bounded by the $\core(M)$.

\begin{align*}
& \sum_{t\in T}f(t)\cdot \sum_{j\in [m]} \ind\left[t\in R_j^{(\beta)}\right]\cdot v\left(t,([m]\backslash \{j\})\cap \mathcal{C}(t)\right)\\
\leq &\sum_{t\in T}f(t)\cdot \sum_{j\in [m]} \ind\left[t\in R_j^{(\beta)}\right]\cdot v\left(t,\mathcal{C}(t)\right)=\sum_{t\in T}f(t)\cdot v\left(t,\mathcal{C}(t)\right)\quad(\textsc{Core}(M))
\end{align*}

The inequality comes from the monotonicity of $v(t,\cdot)$ and the fact that for every $t$ only stays in one region $R_j^{(\beta)}$.

Next, we upper bound the contribution from $\sum_{k\in ([m]\backslash \{j\})\cap \mathcal{T}(t)}V(t_k)$ by the $\tail(M)$.
\begin{align*}
&\sum_{t\in T}f(t)\cdot \sum_{j\in [m]} \ind\left[t\in R_j^{(\beta)}\right]\cdot\sum_{k\in ([m]\backslash \{j\})\cap \mathcal{T}(t)}V(t_k)\\
= &\sum_{t\in T}f(t)\cdot \sum_{j\in \mathcal{T}(t)} V(t_{j})\cdot
{\ind\left[t \not\in R_j^{(\beta)}\right]}\\
\leq &{\sum_{t\in T}f(t)\cdot \sum_{j\in \mathcal{T}(t)} V(t_{j})\cdot \ind\left[\exists k\not=j, V(t_k)\geq V(t_j)\right]~~~\text{(Definition of $R_j^{(\beta)}$)}}\\
=&{\sum_j\sum_{t_j:V(t_j)\geq c}f_j(t_j)\cdot V(t_{j})\cdot \Pr_{t_{-j}}\left[\exists k\not=j, V(t_k)\geq V(t_j)\right]\quad(\tail(M))}
\end{align*}

\end{prevproof}

\begin{prevproof}{Lemma}{lem:single subadditive}
We argue the three properties one by one.

\begin{itemize}[leftmargin=0.7cm]
\item \emph{Monotonicity:} For all $t\in T$ and $U\subseteq V\subseteq [m]$, $U\cap \mathcal{C}(t)\subseteq V\cap \mathcal{C}(t)$. Since $v(t,\cdot)$ is monotone,
$$v'(t,U)=v\left(t,U\cap \mathcal{C}(t)\right)\leq v\left(t,V\cap \mathcal{C}(t)\right)=v'(t,V).$$ Thus, $v'(t,\cdot)$ is monotone.
\item \emph{Subadditivity:} For all $t\in T$ and $U,V\subseteq [m]$, notice $(U\cup V)\cap \mathcal{C}(t)=\left(U\cap\mathcal{C}(t)\right)\cup \left(V\cap\mathcal{C}(t)\right)$, we have
$$v'(t,U\cup V)=v\left(\left(t,(U\cap\mathcal{C}(t)\right)\cup \left(V\cap\mathcal{C}(t)\right)\right)\leq v\left(t,U\cap\mathcal{C}(t)\right)+v\left(t,V\cap\mathcal{C}(t)\right)=v'(t,U)+v'(t,V).$$
\item \emph{No externalities:} For any $t\in T$, $S\subseteq [m]$, and any $t'\in T$ such that $t_{j}=t_{j}'$ for all $j\in S$, to prove $v'(t,S)=v'(t',S)$, it is enough to show $S\cap \mathcal{C}(t)=S\cap \mathcal{C}(t')$. Since $V(t_j)=V(t_j')$ for any $j\in S$, $j\in S\cap \mathcal{C}(t)$ if and only if $j\in S\cap \mathcal{C}(t')$.
\end{itemize}
\end{prevproof}

\begin{prevproof}{Lemma}{lem:single Lipschitz}
For any $t,t'\in T$, and set $X,Y\subseteq [m]$, define set $H=\left\{j\in X\cap Y:t_j=t_j'\right\}$. Since $v'(\cdot,\cdot)$ has no externalities, $v'(t',H)=v'(t,H)$. Therefore,
\begin{align*}
|v'(t,X)-v'(t',Y)|&=\max\left\{v'(t,X)-v'(t',Y),v'(t',Y)-v'(t,X)\right\}\\
&\leq \max\left\{v'(t,X)-v'(t',H),v'(t',Y)-v'(t,H)\right\}\quad\text{(Monotonicity)}\\
&\leq \max\left\{v'(t,X\backslash H),v'(t',Y\backslash H)\right\}\quad\text{(Subadditivity)}\\
& = \max\left\{v\left(t,(X\backslash H)\cap \mathcal{C}(t)\right),v\left(t',(Y\backslash H)\cap\mathcal{C}(t)\right)\right\}\quad\text{(Definition of $v'(\cdot,\cdot)$)}\\
&\leq c\cdot \max\left\{|X\backslash H|,|Y\backslash H|\right\}\\
&\leq c\cdot (|X\Delta Y|+|X\cap Y|-|H|)
\end{align*}
The second last inequality is because both $v(t,\cdot)$ and $v(t',\cdot)$ are subadditive and for any item $j\in \mathcal{C}(t)$ ($\mathcal{C}(t')$) the single-item valuation $V(t_j)$ ($V(t'_j)$) is less than $c$.
\end{prevproof}

\section{Missing Proofs for the Multi-Bidder Case}\label{appx:multi}


\begin{prevproof}{Lemma}{lem:multi decomposition}
We replace every $v_i(t_i,S)$ in $\nf(M,\beta)$ with $v_i\left(t_i,S\cap \mathcal{C}_i(t_i)\right)+$\\
\noindent$\sum_{j\in S\cap \mathcal{T}_i(t_i)}V_i(t_{ij})$. Let the contribution from $v_i\left(t_i,S\cap \mathcal{C}_i(t_i)\right)$ be the first term and the contribution from  $\sum_{j\in S\cap \mathcal{T}_i(t_i)}V_i(t_{ij})$ be the second term.
\begin{align*}
& \sum_i\sum_{t_i\in T_i}f_i(t_i)\cdot \ind\left[t_i\in R_0^{(\beta_i)}\right]\cdot \sum_{S\subseteq[m]}\sigma_{iS}^{(\beta)}(t_i)\cdot v_i(t_i,S\cap \mathcal{C}_i(t_i))+ \\
&\sum_i\sum_{t_i\in T_i}f_i(t_i)\cdot \sum_{j\in [m]} \ind\left[t_i\in R_j^{(\beta_i)}\right]\cdot\\
&~~~~~~~~~~~~~~~~~ \left(\sum_{S:j\in S}\sigma_{iS}^{(\beta)}(t_i)\cdot v_i\left(t_i,(S\backslash\{j\})\cap \mathcal{C}_i(t_i)\right)+\sum_{S:j\not\in S}\sigma_{iS}^{(\beta)}(t_i)\cdot v_i\left(t_i,S\cap \mathcal{C}_i(t_i)\right)\right)\\
\leq& \sum_i\sum_{t_i\in T_i}f_i(t_i)\cdot \sum_{S\subseteq[m]}\sigma_{iS}^{(\beta)}(t_i)\cdot v_i(t_i,S\cap \mathcal{C}_i(t_i))\quad(\core(M,\beta))
\end{align*}

The inequality comes from the Monotonicity of $v_i(t_i,\cdot)$ by replacing $v_i\left(t_i,(S\backslash\{j\})\cap \mathcal{C}_i(t_i)\right)$ with $v_i(t_i,S\cap \mathcal{C}_i(t_i))$.

For the second term, notice that when $t_i\in R_0^{(\beta_i)}$, $\mathcal{T}_i(t_i)=\emptyset$. It can be rewritten as:
\begin{align*}
&\sum_i\sum_{t_i\in T_i}f_i(t_i)\cdot \sum_{j\in [m]} \ind\left[t_i\in R_j^{(\beta_i)}\right]\cdot\left( \sum_{S:j\in S}\sigma_{iS}^{(\beta)}(t_i)\cdot \sum_{k\in (S\backslash\{j\})\cap \mathcal{T}_i(t_i)}V_i(t_{ik})+\sum_{S:j\not\in S}\sigma_{iS}^{(\beta)}(t_i)\cdot \sum_{k\in S\cap \mathcal{T}_i(t_i)}V_i(t_{ik})\right)\\
=&\sum_i\sum_{t_i\in T_i}f_i(t_i)\cdot \sum_{j\in \mathcal{T}_i(t_i)} V_i(t_{ij})\cdot \ind\left[t_i\not\in R_j^{(\beta_i)}\right]\cdot \pi_{ij}^{(\beta)}(t_i)~~~~~~~~\text{(Recall $\pi_{ij}^{(\beta)}(t_i)=\sum_{S:j\in S}\sigma_{iS}^{(\beta)}(t_i)$)}\\
\leq &\sum_i\sum_{t_i\in T_i}f_i(t_i)\cdot \sum_{j\in \mathcal{T}_i(t_i)} V_i(t_{ij})\cdot \ind\left[t_i\not\in R_j^{(\beta_i)}\right]~~~~~~~~\text{($\pi_{ij}^{\beta}(t_i)\leq 1$)}\\
= &\sum_i\sum_{t_i\in T_i}f_i(t_i)\cdot \sum_{j\in \mathcal{T}_i(t_i)} V_i(t_{ij})\cdot \sum_{k\neq j}\ind\left[t_i \in R_k^{(\beta_i)}\right]~~~~~~~~\text{($t_i\notin R_0^{(\beta_i)}$)}\\
\leq&\sum_i\sum_j\sum_{t_{ij}:V_i(t_{ij})\geq\beta_{ij}+c_i}f_{ij}(t_{ij})\cdot V_i(t_{ij})\cdot \sum_{k\neq j} \Pr_{t_{ik}}\left[ V_i(t_{ik})-\beta_{ik}\geq V_i(t_{ij})-\beta_{ij}\right]\quad(\tail(M,\beta))
\end{align*}
\end{prevproof}

\begin{lemma}\label{lem:valuation v_i'}
	Let $\{x_{ij}\}_{i\in[n], j\in[m]}$ be a set of nonnegative numbers. For any buyer $i$, any type $t_i\in T_i$, let $X_i(t_i)=\{j\ |\ V_i(t_{ij})< x_{ij}\}$, and let $$\bar{v}_i(t_i, S) = v_i(t_i,S\cap X_i(t_i)),$$ for any set $S\subseteq[m]$. Then for any bidder $i$, any type $t_i\in T_i$, $\bar{v}_i(t_i,\cdot)$, satisfies monotonicity, subadditivity and no externalities.	
	\end{lemma}
		\begin{prevproof}{Lemma}{lem:valuation v_i'}
		 We will argue these three properties one by one.
	\begin{itemize}
		\item \emph{Monotonicity:} For all $t_i\in T_i$ and $U\subseteq V\subseteq [m]$, since $v_i(t_i,\cdot)$ is monotone, $$\bar{v}_i(t_i,U)=v_i(t_i,U\cap X_i(t_i))\leq v_i(t_i,V\cap X_i(t_i))=\bar{v}(t_i,V)$$ Thus $\bar{v}_i(t_i,\cdot)$ is monotone.
		\item \emph{Subadditivity:} For all $t_i\in T_i$ and $U,V\subseteq [m]$, $(U\cup V)\cap X_i(t_i)=(U\cap X_i(t_i))\cup (V\cap X_i(t_i))$. {Since $v_i(t_i,\cdot)$ is subadditive}, we have
\begin{align*}
&\bar{v}_i(t_i,U\cup V)=v_i(t_i,(U\cap X_i(t_i))\cup (V\cap X_i(t_i)))\\
 &~~~~~~~~~~~~~\leq v_i(t_i,U\cap X_i(t_i))+v_i(t_i,V\cap X_i(t_i))= \bar{v}_i(t_i,U)+\bar{v}_i(t_i,V).
\end{align*}
\item \emph{No externalities:} For any $t_i\in T_i$, $S\subseteq [m]$, and any $t_i'\in T_i$ such that $t_{ij}=t_{ij}'$ for all $j\in S$, to prove $\bar{v}_i(t_i,S)=\bar{v}_i(t_i',S)$, it suffices to show $S\cap X_i(t_i)=S\cap X_i(t_i')$. Since $V_i(t_{ij})=V_i(t_{ij}')$, for any item $j\in S$, $j\in S\cap X_i(t_i)$ if and only if $j\in S\cap X_i(t_i')$.
	\end{itemize}
	\end{prevproof}

\begin{prevproof}{Lemma}{lem:hat gamma}
By Lemma~\ref{lem:valuation v_i'} and Definition~\ref{def:v hat},  $\hat{v}_i(t_i,\cdot)$ satisfies monotonicity, subadditivity and no externalities.
	\notshow{Since for every set $S\subseteq[m]$ and every subset $S'$ of $S$,  $v_i(t_i, S')\geq \sum_{j\in S'}\theta_j^S(t_i)$. Therefore,
	$$\hat{v}_i(t_i,S')= v_i(t_i,\{j\ |\ j\in S' \land V_i(t_{ij})< Q_j+\tau_i\}) \geq v'_i(t_i, \{j\ |\ j\in S' \land V_i(t_{ij})< Q_j+\tau_i\}).$$
	Since $\{j\ |\ j\in S' \land V_i(t_{ij})< Q_j+\tau_i\}$ is a subset of $S'$, it is also a subset of $S$. Therefore,
	$$v'_i(t_i, \{j\ |\ j\in S' \land V_i(t_{ij})< Q_j+\tau_i\}) \geq \sum_{j: j\in S' \land V_i(t_{ij})< Q_j+\tau_i}\gamma_j^S(t_i)= \sum_{j\in S'}\hat{\gamma}^S_j(t_i).$$}

	$$\hat{v}_i(t_i,S')= v_i\left(t_i,S'\cap Y_i(t_i)\right)\geq v_i\left(t_i, \left(S'\cap Y_i(t_i)\right)\cap \mathcal{C}_i(t_i)\right) =v'_i\left(t_i, S'\cap Y_i(t_i)\right).$$
	Since $S'\cap Y_i(t_i)\subseteq S$,
	$$v'_i\left(t_i, S'\cap Y_i(t_i)\right) \geq \sum_{j\in S'\cap Y_i(t_i)}\gamma_j^S(t_i)= \sum_{j\in S'}\hat{\gamma}^S_j(t_i).$$

	for all $j\in S$, we have $\min\{t_{ij}^{(k)}, Q_j+\tau_i\}\geq \hat{\gamma}^S_j(t_i)$. Therefore, $\hat{v}_i(t_i,S)\geq \sum_{j\in S}\hat{\gamma}^S_j(t_i)$.
	\end{prevproof}

\begin{prevproof}{Lemma}{lem:property of mu}
We first prove that $\mu_i(\cdot,\cdot)$ is $\tau_i$-Lipschitz. For any $t_i,t_i'\in T_i$ and set $X,Y\in[m]$, let $X^{*}\in \argmax_{S\subseteq X} \left(\hat{v}_i(t_i,S)-\sum_{j\in S}Q_j \right), Y^{*}\in \argmax_{S\subseteq Y} \left(\hat{v}_i(t'_i,S)-\sum_{j\in S}Q_j \right)$.   Recall that $\hat{v}_i(t_i,X^{*}) =v_i\left(t_i, \left\{j\ |\ (j\in X^{*}) \land (V_i(t_{ij})< Q_j+\tau_i)\right\}\right)$. This means that for every $k\in X^{*}$, $V_i(t_{ik})$ must be less than $Q_k+\tau_i$, because otherwise $\mu_i(t_i,X^{*})<\mu_i(t_i,X^{*}\backslash \{k\})$. Therefore, $\hat{v}_i(t_i,S)=v_i(t_i,S)$ for all $S\subseteq X^{*}$. Since $v_i(t_i,\cdot)$ is subadditive, $v_i(t_i,X^{*})\leq v_i(t_i,X^{*}\backslash \{k\})+V_i(t_{ik})$. So by the optimality of $X^*$, it must be that $V_i(t_{ik})\geq Q_k$ for all $k\in X^*$. Similarly, we can show that for every $k\in Y^{*}$, $V_i(t_{ik}')\in [Q_k,Q_k+\tau_i]$.

Now let set $H=\{j\ |\ j\in X\cap Y \land t_{ij}=t_{ij}'\}$, if $\mu_i(t_i,X)>\mu_i(t_i',Y)$.

\begin{align*}
&\left|\mu_i(t_i,X)-\mu_i(t_i',Y)\right|= \left(\hat{v}_i(t_i,X^{*})-\sum_{j\in X^{*}}Q_j\right)-\left(\hat{v}_i(t_i',Y^{*})-\sum_{j\in Y^{*}}Q_j\right)\\
\leq &\left(\hat{v}_i(t_i,X^{*})-\sum_{j\in X^{*}}Q_j\right)-\left(\hat{v}_i(t_i',X^{*}\cap H)-\sum_{j\in X^{*}\cap H}Q_j\right)\quad\text{(Optimality of $Y^{*}$ and $X^{*}\cap H\subseteq Y$)}\\
\leq &\hat{v}_i(t_i,X^{*})-\hat{v}_i(t_i,X^{*}\cap H)-\sum_{j\in X^{*}\backslash H}Q_j\qquad\qquad\text{(No externalities of $\hat{v}_i(t_i,\cdot)$)}\\
\leq &\hat{v}_i(t_i,X^{*}\backslash H)-\sum_{j\in X^{*}\backslash H}Q_j\qquad\qquad\text{(Subadditivity of $\hat{v}_i(t_i,\cdot)$)}\\
\leq &\tau_i\cdot |X^{*}\backslash H|\qquad\qquad\left(V_i(t_{ij})\in [Q_j,Q_j+\tau_i]\text{ for all } j\in X^{*}\right)\\
\leq &\tau_i\cdot |X\backslash H|
\end{align*}

Similarly, if $\mu_i(t_i,X)\leq \mu_i(t_i',Y)$, $\left|\mu_i(t_i,X)-\mu_i(t_i',Y)\right|\leq \tau_i\cdot |Y\backslash H|$. Thus, $\mu_i(\cdot,\cdot)$ is $\tau_i$-Lipschitz as $$\left|\mu_i(t_i,X)-\mu_i(t_i',Y)\right|\leq \tau_i\cdot \max\left\{|X\backslash H|,|Y\backslash H|\right\}\leq \tau_i\cdot(|X\Delta Y|+|X\cap Y|-|H|).$$

Monotonicity follows directly from the definition of $\mu_i(t_i,\cdot)$. Next, we argue subadditivity. For all {$U, V\subseteq [m]$}, let $S^{*}\in \argmax_{S\subseteq U\cup V} \left(\hat{v}_i(t_i,S)-\sum_{j\in S} Q_j\right)$, $X=S^{*}\cap U\subseteq U$, $Y=S^{*}\backslash X\subseteq V$. Since $\hat{v}_i(t_i,\cdot)$ is a subadditive valuation,
\begin{equation*}
\mu_i(t_i,U\cup V)=\hat{v}_i(t_i, S^{*}) -\sum_{j\in S^{*}} Q_j\leq \left(\hat{v}_i(t_i, X) -\sum_{j\in X} Q_j\right)+\left(\hat{v}_i(t_i, Y) -\sum_{j\in Y} Q_j\right)\leq \mu_i(t_i,U)+\mu_i(t_i,V)
\end{equation*}

Finally, we argue that $\mu_i(t_i,\cdot)$ has no externalities. Consider a set $S$, and types $t_i, t_i'\in T_i$ such that $t_{ij}'=t_{ij}$ for all $j\in S$. For any $S'\subseteq S$, since $\hat{v}_i(t_i,\cdot)$ has no externalities, $\hat{v}_i(t_i,S')-\sum_{j\in S'}Q_j=\hat{v}_i(t_i',S')-\sum_{j\in S'}Q_j$. Thus, $\mu_i(t_i,S)=\mu_i(t_i',S)$.
\end{prevproof}

\section{Efficient Approximation for Symmetric Bidders}\label{sec:symmetric computation}
In this section, we sketch how to compute the RSPM and ASPE to approximate the optimal revenue in polynomial time for symmetric bidders\footnote{Bidders are symmetric if for any two bidders $i$ and $i'$, we have $v_i(\cdot,\cdot) = v_{i'}(\cdot,\cdot)$ and $D_{ij}=D_{i'j}$ for all $j$.}. For any given BIC mechanism $M$, one can follow our proof to construct in polynomial time an RSPM and an ASPE such that the better of the two achieves a constant fraction of $M$'s revenue. We will describe the construction of the RSPM and the ASPE separately in this section. The difficulty of applying the method described above to construct the desired simple mechanisms is that we need to know an (approximately) revenue-maximizing mechanism $M^*$. We will show how to circumvent this difficulty when the bidders are symmetric. 

  Indeed, we can directly construct an RSPM that approximates the $\prev$. As we have restricted the buyers to purchase at most one item in an RSPM, the $\prev$ is upper bounded by the optimal revenue of the unit-demand setting where buyer $i$ has value $V_i(t_{ij})$ for item $j$ when her type is $t_i$. By~\cite{CaiDW16}, we know that the optimal revenue in this unit-demand setting is upper bounded by $4\copies$, so one can simply use the RSPM constructed in~\cite{ChawlaHMS10} to extract revenue at least $\frac{\prev }{24}$. Note that the construction is independent of $M$.

  Unlike the RSPM, our construction for the ASPE heavily relies on $\beta$ which depends on $M$ (Lemma~\ref{lem:requirement for beta}). Given $\beta$, we first compute $c_i$s according to Definition~\ref{def:c_i}. Next, we compute the $Q_j$s (Definition~\ref{def:posted prices}). Finally, we compute the $\tau_i$s (Defintion~\ref{def:tau}) and use them to compute the entry fee (Definition~\ref{def:entry fee}). A few steps of the algorithm above requires sampling from the type distributions, but it is not hard to argue that a polynomial number of samples suffices. The main reason that the information about $M$ is necessary is because our construction crucially relies on the choice of $\beta$. Next, we argue that for symmetric bidders, we can essentially choose a $\beta$ that satisfies all requirements in Lemma~\ref{lem:requirement for beta} for all mechanisms. 

  When bidders are symmetric, the important observation is that the optimal mechanism must also be symmetric, and for any symmetric mechanism we can directly construct a $\beta$ that satisfies all the requirements in Lemma~\ref{lem:requirement for beta}. For every $i\in [n], j\in [m]$, choose $\beta_{ij}$ such that $\Pr_{t_{ij}}\left[V_i(t_{ij})\geq \beta_{ij}\right]=\frac{b}{n}$. Clearly, this choice satisfies property (i) in Lemma~\ref{lem:requirement for beta}. Furthermore, the ex-ante probability for any bidder $i$ to win item $j$ is the same in any symmetric mechanism, and therefore is no more than $1/n$. Hence, property (ii) in Lemma~\ref{lem:requirement for beta} is also satisfied. Given this $\beta$, we can essentially follow the algorithm mentioned above to construct the ASPE. The only difference is that we no longer know the $\sigma$, which is required when computing the $Q_j$s. This can be resolved by considering the welfare maximizing mechanism $M'$ with respect to $v'$. We compute the prices $Q_j$ using the allocation rule of $M'$ and construct our ASPE. As $M'$ is also symmetric, our $\beta$ satisfies all requirements in Lemma~\ref{lem:requirement for beta} with respect to $M'$. Therefore, Lemma~\ref{lem:upper bounding Q} implies that either this ASPE or the RSPM constructed above has at least a constant fraction of $\core(M',\beta)$ as revenue. Since $M'$ is welfare maximizing, $\core(M',\beta)\geq \core(M^*,\beta)$, where $M^*$ is the revenue optimal mechanism. Therefore, we construct in polynomial time a simple mechanism whose revenue is a constant fraction of the optimal BIC revenue.

\section{Proof of Lemma~\ref{lem:relaxed valuation}}\label{sec:proof_relaxed_valuation}
We first prove some properties of $v^{(\beta)}$, which will be useful for proving Lemma~\ref{lem:relaxed valuation}.

\begin{lemma}\label{lem:relaxed larger}
	For any $\beta_i$, $t_i\in T_i$ and $S\in[m]$, $v_i^{(\beta_i)}(t_i,S)\geq v_i(t_i,S)$.
\end{lemma}
\begin{proof}
	This follows from the fact that $v_i(t_i,\cdot)$ is a subadditive function over bundles of items for all $t_i$.
\end{proof}

\begin{lemma}
	For any $\beta_i$ and $t_i\in T_i$, $v_i^{(\beta_i)}(t_i,\cdot)$ is a monotone, subadditive function over the items.
\end{lemma}
\begin{proof}
Monotonicity follows directly from the monotonicity of $v_i(t_i,\cdot)$. We only argue subadditivity here. If $t_i$ belongs to $R_0^{(\beta_i)}$, $v_i^{(\beta_i)}(t_i,\cdot)=v_i(t_i,\cdot)$. So it is clearly a subadditive function. If $t_i$ belongs to $R_j^{(\beta_i)}$ for some $j>0$ and $j$ is not in either $U$ or $V$, then clearly $v_i^{(\beta_i)}(t_i,U\cup V)\leq v_i^{(\beta_i)}(t_i,U)+v_i^{(\beta_i)}(t_i,V)$. If $j$ is in one of the two sets, without loss of generality let's assume it is in $U$. Then $v_i^{(\beta_i)}(t_i,U)+v_i^{(\beta_i)}(t_i,V)=v_i(t_i,U\backslash\{j\})+V_i(t_{ij})+v_i(t_i,V)\geq v_i(t_i,V\cup (U\backslash\{j\}))+V_i(t_{ij})= v_i^{(\beta_i)}(t_i,U\cup V)$.
\end{proof}

Here we prove a stronger version of Lemma~\ref{lem:relaxed valuation}.

\begin{lemma}\label{lem:relaxed valuation stronger}
For any $\beta$, any absolute constant $\eta\in(0,1)$ and any BIC mechanism $M$ for subadditive valuations $\{v_i(t_i,\cdot)\}_{i\in[n]}$ with $t_i\sim D_i$ for all $i$, there exists a BIC mechanism $M^{(\beta)}$ for valuations $\{v_i^{(\beta_i)}(t_i,\cdot)\}_{i\in[n]}$ with $t_i\sim D_i$ for all $i$, such that
	\begin{enumerate}
		\item $\sum_{t_i\in T_i}f_i(t_i)\cdot\sum_{S: j\in S}\sigma^{(\beta)}_{iS}(t_i)\leq \sum_{t_i\in T_i}f_i(t_i)\cdot\sum_{S: j\in S}\sigma_{iS}(t_i)$, for all $i$ and $j$,
		\item $\rev(M, v, D)\leq$\\
$~~~~\frac{1}{1-\eta}\cdot{\rev(M^{(\beta)},v^{(\beta)}, D)}+\frac{1}{\eta}\cdot\sum_i
\sum_{t_i\in T_i}\sum_{S\subseteq[m]} f_i(t_i)\cdot\sigma^{(\beta)}_{iS}(t_i)\cdot \left(v_i^{(\beta_i)}(t_i, S)-v_i(t_i, S)\right)$.
	\end{enumerate}
	$\rev(M, v, D)$ (or $\rev(M^{(\beta)},v^{(\beta)}, D)$) is the revenue of the mechanism $M$ (or $M^{(\beta)}$) while the buyers' types are drawn from $D$ and buyer $i$'s valuation is $v_i(t_i,\cdot)$ (or $v_i^{(\beta_i)}(t_i,\cdot)$). $\sigma_{iS}(t_i)$ (or $\sigma^{(\beta)}_{iS}(t_i)$) is the probability of buyer $i$ receiving exactly bundle $S$ when her reported type is $t_i$ in mechanism $M$ (or $M^{(\beta)}$).
\end{lemma}

\begin{prevproof}{lemma}{lem:relaxed valuation stronger}
Readers who are familiar with the $\epsilon$-BIC to BIC reduction~\cite{HartlineKM11, BeiH11,DaskalakisW12} might have already realized that the problem here is quite similar. Our proof will follow essentially the same approach.

First, we construct mechanism $M^{(\beta)}$, which has two phases:
\vspace{.1in}

\noindent{\bf Phase 1: Surrogate Sale}
\begin{enumerate}
	\item For each buyer $i$, create $\ell-1$ \emph{replicas} and $\ell$ \emph{surrogates} sampled i.i.d. from $D_i$. The value of $\ell$ will be specified later.
	\item Ask each buyer to report her type $t_i$.
	\item For each buyer $i$, create a weighted bipartite graph with the replicas and the buyer $i$ on the left and the surrogates on the right. The edge weight between a replica (or buyer $i$) with type $r_i$ and a surrogate with type $s_i$ is the expected value for a bidder with valuation $v_i^{(\beta_i)}(r_i,\cdot)$ to receive buyer $i$'s interim allocation in $M$ when she reported $s_i$ as her type subtract the interim payment of buyer $i$ multiplied by $(1-\eta)$. Formally, the weight is $\sum_{S} \sigma_{iS}(s_i)\cdot v_i^{(\beta_i)}(r_i,S) - (1-\eta)p_i(s_i)$, where $p_i(s_i)$ is the interim payment for buyer $i$ if she reported $s_i$.
	\item Compute the VCG matching and prices on the bipartite graph created for each buyer $i$. If a replica (or bidder $i$) is unmatched in the VCG matching, match her to a random unmatched surrogate. The surrogate selected for buyer $i$ is whoever she is matched to.
\end{enumerate}

\vspace{.1in}
\noindent{\bf Phase 2: Surrogate Competition}
\begin{enumerate}
	\item Apply mechanism $M$ on the type profiles of the selected surrogates $\vec{s}$. Let $M_i(\vec{s})$ and $P_i(\vec{s})$ be the corresponding allocated bundle and payment of buyer $i$.
	\item If buyer $i$ is matched to her surrogate in the VCG matching, give her bundle $M_i(\vec{s})$ and charge her $(1-\eta)\cdot P_i(\vec{s})$ plus the VCG price. If buyer $i$ is not matched in the VCG matching, award them nothing and charge them nothing.
	\end{enumerate}

\begin{lemma}[\cite{HartlineKM11}]\label{lem:same distribution}
	If all buyers play $M^{(\beta)}$ truthfully, then the distribution of types of the surrogate chosen by buyer $i$ is exactly $D_i$.
\end{lemma}
\begin{proof}
In the mechanism, first the buyer $i$'s type is sampled from the distribution, then we sampled $\ell-1$ replicas and $\ell$ surrogates i.i.d. from the same distribution. Now, imagine a different order of sampling. We first sample the $\ell$ replicas and $\ell$ surrogates, then we pick one replica to be buyer $i$ uniformly at random. The two different orders above provide exactly the same joint distribution over the replicas, surrogates and buyer $i$. So we only need to argue that in the second order of sampling, the distribution of types of the surrogate chosen by buyer $i$ is exactly $D_i$. Note that the perfect matching (VCG matching plus the uniform random matching with the leftover replicas/surrogates) only depends on the types but not the identity of the node (replica or buyer $i$). So we can decide who is buyer $i$ after we have decided the perfect matching. Since buyer $i$ is chosen uniformly at random among the replicas, the chosen surrogate is also uniformly at random. Clearly, the distribution of the types of a surrogate chosen uniformly at random is also $D_i$. The assumption that buyer $i$ is reporting truthfully is crucial, because otherwise the distribution of buyer $i$'s reported type will be different from the type of a replica, and in that case, we cannot use the second sampling order.
\end{proof}

\begin{lemma}
	$M^{(\beta)}$ is a BIC mechanism with respect to valuation $v^{(\beta)}$.
\end{lemma}
\begin{proof}
	We need to argue that for every buyer $i$ reporting truthfully is a best response, if every other buyer is truthful. In the VCG mechanism, buyer $i$ faces a competition with the replicas to win a surrogate.  If buyer $i$ has type $t_i$, then her value for winning a surrogate with type $s_i$ in the VCG mechanism is  $\sum_{S} \sigma_{iS}(s_i)\cdot v_i^{(\beta_i)}(t_i,S) - (1-\eta)p_i(s_i)$ due to Lemma~\ref{lem:same distribution}. Clearly, if buyer $i$ reports truthfully, the weights on the edges between her and all the surrogates will be exactly her value for winning those surrogates. Since buyer $i$ is in a VCG mechanism, reporting the true edge weights is a dominant strategy for her, therefore reporting truthfully is also a best response for her assuming the other buyers are truthful. It is critical that the other buyers are reporting truthfully, otherwise we cannot invoke Lemma~\ref{lem:same distribution} and buyer $i$'s value for winning a surrogate with type $s_i$ may be different from the weight on the corresponding edge.
	\end{proof}
	
\begin{lemma}
	For any $i$ and $j$, $\sum_{t_i\in T_i}f_i(t_i)\cdot\sum_{S: j\in S} \sigma^{(\beta)}_{iS}(t_i)\leq \sum_{t_i\in T_i}f_i(t_i)\cdot\sum_{S: j\in S}\sigma_{iS}(t_i)$.
\end{lemma}

\begin{proof}
	The LHS is the ex-ante probability for buyer $i$ to win item $j$ in $M^{(\beta)}$, and the RHS is the corresponding probability in $M$. By Lemma~\ref{lem:same distribution}, we know the surrogate selected by buyer $i$ is participating in $M$ against all other surrogates whose types are drawn from $D_{-i}$. Therefore, the ex-ante probability for the surrogate chosen by buyer $i$ to win item $j$ is the same as RHS. Clearly, the chosen surrogate's ex-ante probability for winning any item should be at least as large as the ex-ante probability for  buyer $i$ to win the item in $M^{(\beta)}$.
	\end{proof}
	
Next, we want to compare $\rev(M^{(\beta)},v^{(\beta)}, D)$ with $\rev(M,v, D)$. The following simple Lemma relates both quantities to the expected prices charged to the surrogates by mechanism $M$. As in the proof of Lemma~\ref{lem:same distribution}, we change the order of the sampling. We first sample $\ell$ replicas and $\ell$ surrogates then select a replica uniformly at random to be buyer $i$.
Let $s_i^{k}$ and $r_i^{k}$ be the type of the $k$-th surrogate and replica, $\Bs= (s_i^{1},\ldots, s_i^{\ell})$, $\Br=(r_i^{1},\ldots, r_i^{\ell})$ and $V(\Bs,\Br)$ be the VCG matching between surrogates and replicas with types $\Bs$ and $\Br$. We will slightly abuse notation by using $s_i^k$ (or $r_i^j$) $\in V(\Bs,\Br)$ to denote that $s_i^k$  (or $r_i^j$) is matched in the VCG matching $V(\Bs,\Br)$.
\begin{lemma}\label{lem:revenue by surrogates}
For every buyer $i$, her expected payments in $M^{(\beta)}$ is at least $$(1-\eta)\cdot\E_{\Bs,\Br}\left[\sum_{s_i^k\in V(\Bs,\Br)} \frac{ p_i(s_i^k)}{\ell}\right],$$ and her expected payments in $M$ is $$\E_{\Bs}\left[\sum_{k\in[\ell]} \frac{ p_i(s_i^k)}{\ell}\right].$$
\end{lemma}
\begin{proof}
	The revenue of $M^{(\beta)}$ contains two parts -- the prices paid by the chosen surrogates and the revenue of the VCG mechanism. Let's compute the first part. For buyer $i$ and each realization of $\Br$ and $\Bs$ only when the buyer $i$'s chosen surrogate is in $ V(\Bs,\Br)$, she pays the surrogate price. Since each surrogate is selected with probability $1/\ell$, the expected surrogate price paid by buyer $i$ is exactly $(1-\eta)\cdot\E_{\Bs,\Br}\left[\sum_{s_i^k\in V(\Bs,\Br)} \frac{ p_i(s_i^k)}{\ell}\right]$. Since the VCG payments are nonnegative, we have proved our first statement.
	
	The expected payment from buyer $i$ in $M$ is $\E_{t_i\sim D_i}\left[p_i(t_i)\right]$. Since all $s_i^k$ is drawn from $D_i$, this is exactly the same as $\E_{\Bs}\left[\sum_{k\in[\ell]} \frac{ p_i(s_i^k)}{\ell}\right]$.
\end{proof}

If the VCG matching is always perfect, then Lemma~\ref{lem:revenue by surrogates} already shows that the revenue of $M^{(\beta)}$ is at least $(1-\eta)$ fraction of the revenue of $M$. But since the VCG matching may not be perfect, we need to show that the total expected price from surrogates who are not in the VCG matching is small. We prove this in two steps. First, we consider another matching $X(\Bs,\Br)$ -- a maximal matching that only matches replicas and surrogates that have the same type, and show that the expected cardinality of $X(\Bs,\Br)$ is close to $\ell$. Then we argue that for any realization $\Br$ and $\Bs$ the total payments from surrogates that are in $X(\Bs, \Br)$ but not in $V(\Bs,\Br)$ is small.

\begin{lemma}[\cite{HartlineKM11}]\label{lem:equal type matching}
For every buyer $i$, the expected cardinality of a maximal matching that only matches replicas and surrogates with the same type is at least $\ell-\sqrt{|T_i|\cdot \ell}$.
\end{lemma}

The proof can be found in Hartline et al.~\cite{HartlineKM11}.
\begin{corollary}\label{cor:bound revenue by X}
Let $\mathcal{R} = \max_{i,t_i\in T_i}\max_{S\in[m]} v_i(t_i,S)$, then
$$\E_{\Bs,\Br}\left[\sum_{s_i^k\in X(\Bs,\Br)} \frac{ p_i(s_i^k)}{\ell}\right]\geq \E_{\Bs}\left[\sum_{k\in[\ell]} \frac{ p_i(s_i^k)}{\ell}\right]- \sqrt{\frac{|T_i|}{\ell}}\cdot\mathcal{R}.$$
\end{corollary}
\begin{proof}
	Since $M$ is a IR mechanism when the buyers' valuations are $v$, $\mathcal{R}\geq p_i(t_i)$ for any buyer $i$ and any type $t_i$ of $i$. Our claim follows from Lemma~\ref{lem:equal type matching}.
\end{proof}

Now we implement the second step of our argument. The plan is to show the total prices from surrogates that are unmatched by going from  $X(\Bs,\Br)$ to $V(\Bs,\Br)$. For any $\Bs,\Br$, $V(\Bs,\Br)\cup X(\Bs,\Br)$ can be decompose into a disjoint collection augmenting paths and cycles.  If a surrogate is matched in $X(\Bs,\Br)$ but not in $V(\Bs,\Br)$, then it must be the starting point of an augmenting path. The following Lemma upper bounds the price of this surrogate.
\begin{lemma}[Adapted from~\cite{DaskalakisW12}]\label{lem:bounding the price for each augmenting path}
	For any buyer $i$ and any realization of $\Bs$ and $\Br$, let $P$ be an augmenting path that starts with a surrogate that is matched in $X(\Bs, \Br)$ but not in $V(\Bs,\Br)$. It has the form of either (a) $\left(s_i^{\rho(1)},r_i^{\theta(1)},s_i^{\rho(2)},r_i^{\theta(2)},\ldots, s_i^{\rho{(k)}}\right)$ when the path ends with a surrogate, or\\ (b) $\left(s_i^{\rho(1)},r_i^{\theta(1)},s_i^{\rho(2)},r_i^{\theta(2)},\ldots, s_i^{\rho{(k)}},r_i^{\theta(k)}\right)$ when the path ends with a replica, where $r_i^{\theta(j)}$ is matched to $s_i^{\rho(j)}$ in $X(\Bs, \Br)$ and matched to $s_i^{\rho(j+1)}$ in $V(\Bs,\Br)$ (whenever $s_i^{\rho(j+1)}$ exists) for any $j$.
	\begin{align*}&\sum_{s_i^{\rho(j)}\in P\cap X(\Bs,\Br)} p_i \left(s_i^{\rho(j)}\right)-\sum_{s_i^{\rho(j)}\in P\cap V(\Bs,\Br)} p_i \left(s_i^{\rho(j)}\right)\leq\\
	 &~~~~~~~~~~~~~~~~~~~~~~~~~~~~~\frac{1}{\eta}\cdot\sum_{j=1}^{k-1} \sum_S \sigma_{iS}\left(s_i^{\rho(j+1)}\right)\cdot \left(v_i^{(\beta_i)}(r_i^{\theta(j)},S)-v_i(r_i^{\theta(j)},S)\right).
	\end{align*}
\end{lemma}
\begin{proof}
	Since $r_i^{\theta(j)}$ is matched to $s_i^{\rho(j)}$ in $X(\Bs, \Br)$, $r_i^{\theta(j)}$ must be equal to $s_i^{\rho(j)}$. $M$ is a BIC mechanism when buyers valuations are $v$, therefore the expected utility for reporting the true type is better than lying. Hence, the following holds for all $j$:
	\begin{equation}\label{eq:BIC for M}\sum_S\sigma_{iS}\left(s_i^{\rho(j)}\right)\cdot v_i\left(r_i^{\theta(j)},S\right)-p_i\left(s_i^{\rho(j)}\right)\geq \sum_S\sigma_{iS}\left(s_i^{\rho(j+1)}\right)\cdot v_i\left(r_i^{\theta(j)},S\right)-p_i\left(s_i^{\rho(j+1)}\right)	
	\end{equation}

The VCG matching finds the maximum weight matching, so the total edge weights in path $P \cap V(\Bs,\Br)$ is at least as large as the total edge weights in path $P\cap X(\Bs,\Br)$. Mathematically, it is the following inequalities.
\begin{itemize}
\item If $P$ has format (a): \begin{align}\label{eq:VCG great a}
&\sum_{j=1}^{k-1} \left(\sum_S\sigma_{iS}\left(s_i^{\rho(j+1)}\right)\cdot v_i^{(\beta_i)}\left(r_i^{\theta(j)},S\right)-(1-\eta)\cdot p_i\left(s_i^{\rho(j+1)}\right)\right) \geq	\\
&~~~~~~~~~~~~~~~~~~~~~~~~~~~~~~~~~~~~~~\sum_{j=1}^{k-1}\left( \sum_S\sigma_{iS}\left(s_i^{\rho(j)}\right)\cdot v_i^{(\beta_i)}\left(r_i^{\theta(j)},S\right)-(1-\eta)\cdot p_i\left(s_i^{\rho(j)}\right)\right) \nonumber
\end{align}
\item If $P$ has format (b): \begin{align}\label{eq:VCG great b}
&\sum_{j=1}^{k-1} \left(\sum_S\sigma_{iS}\left(s_i^{\rho(j+1)}\right)\cdot v_i^{(\beta_i)}\left(r_i^{\theta(j)},S\right)-(1-\eta)\cdot p_i\left(s_i^{\rho(j+1)}\right)\right) \geq	\\
&~~~~~~~~~~~~~~~~~~~~~~~~~~~~~~~~~~~~~~\sum_{j=1}^{k}\left( \sum_S\sigma_{iS}\left(s_i^{\rho(j)}\right)\cdot v_i^{(\beta_i)}\left(r_i^{\theta(j)},S\right)-(1-\eta)\cdot p_i\left(s_i^{\rho(j)}\right)\right) \nonumber
\end{align}

\end{itemize}

Next, we further relax the RHS of inequality~(\ref{eq:VCG great a}) using inequality~(\ref{eq:BIC for M}).
\begin{align*}
	&\text{RHS of inequality~(\ref{eq:VCG great a})}\\
	\geq& \sum_{j=1}^{k-1}\left( \sum_S\sigma_{iS}\left(s_i^{\rho(j)}\right)\cdot v_i\left(r_i^{\theta(j)},S\right)-p_i\left(s_i^{\rho(j)}\right)\right)+ \eta\cdot\sum_{j=1}^{k-1}p_i\left(s_i^{\rho(j)}\right)~~\text{(Lemma~\ref{lem:relaxed larger})}\\
	\geq & \sum_{j=1}^{k-1}\left( \sum_S\sigma_{iS}\left(s_i^{\rho(j+1)}\right)\cdot v_i\left(r_i^{\theta(j)},S\right)-p_i\left(s_i^{\rho(j+1)}\right)\right)+ \eta\cdot\sum_{j=1}^{k-1}p_i\left(s_i^{\rho(j)}\right)~~\text{(Inequality~\ref{eq:BIC for M})}\\
\end{align*}
We can obtain the following inequality by combining the relaxation above with the LHS of inequality~(\ref{eq:VCG great a}) and rearrange the terms.
$$\frac{1}{\eta}\cdot\sum_{j=1}^{k-1} \sum_S\sigma_{iS}\left(s_i^{\rho(j+1)}\right)\cdot \left(v_i^{(\beta_i)}\left(r_i^{\theta(j)},S\right)-v_i\left(r_i^{\theta(j)},S\right)\right)\geq  p_i\left(s_i^{\rho(1)}\right)-p_i\left(s_i^{\rho(k)}\right).$$
The inequality above is exactly the inequality in the statement of this Lemma when $P$ has format (a).

Similarly, we have the following relaxation when $P$ has format (b):
\begin{align*}
	&\text{RHS of inequality~(\ref{eq:VCG great b})}\\
	\geq& \sum_{j=1}^{k}\left( \sum_S\sigma_{iS}\left(s_i^{\rho(j)}\right)\cdot v_i\left(r_i^{\theta(j)},S\right)-p_i\left(s_i^{\rho(j)}\right)\right)+ \eta\cdot\sum_{j=1}^{k}p_i\left(s_i^{\rho(j)}\right)~~\text{(Lemma~\ref{lem:relaxed larger})}\\
	\geq & \sum_{j=1}^{k-1}\left( \sum_S\sigma_{iS}\left(s_i^{\rho(j+1)}\right)\cdot v_i\left(r_i^{\theta(j)},S\right)-p_i\left(s_i^{\rho(j+1)}\right)\right)+ \eta\cdot\sum_{j=1}^{k}p_i\left(s_i^{\rho(j)}\right)~~\text{(Inequality~\ref{eq:BIC for M} and $M$ is IR)}\\
\end{align*}
Again, by combining the relaxation with the LHS of inequality~(\ref{eq:VCG great b}), we can prove our claim when $P$ has format (b).
$$\frac{1}{\eta}\cdot\sum_{j=1}^{k-1} \sum_S\sigma_{iS}\left(s_i^{\rho(j+1)}\right)\cdot \left(v_i^{(\beta_i)}\left(r_i^{\theta(j)},S\right)-v_i\left(r_i^{\theta(j)},S\right)\right)\geq  p_i\left(s_i^{\rho(1)}\right).$$
\end{proof}

\begin{lemma}\label{lem: gap between X and V}
For any $\beta$,
	\begin{align*}
&\E_{\Bs,\Br}\left[\sum_{s_i^k\in X(\Bs,\Br)} \frac{ p_i(s_i^k)}{\ell}\right]\leq\\
&~~~~~~~~~~~~~~~\E_{\Bs,\Br}\left[\sum_{s_i^k\in V(\Bs,\Br)} \frac{ p_i(s_i^k)}{\ell}\right]+\frac{1}{\eta}\cdot\sum_{t_i\in T_i}\sum_{S\subseteq[m]} f_i(t_i)\cdot\sigma^{(\beta)}_{iS}(t_i)\cdot \left(v_i^{(\beta_i)}(t_i, S)-v_i(t_i, S)\right).
	\end{align*}
\end{lemma}
\begin{proof}
	Due to Lemma~\ref{lem:bounding the price for each augmenting path}, for any buyer $i$ and any realization of $\Br$ and $\Bs$,  we have
	$$\sum_{s_i^k\in X(\Bs,\Br)} \frac{ p_i(s_i^k)}{\ell}-\sum_{s_i^k\in V(\Bs,\Br)} \frac{ p_i(s_i^k)}{\ell}\leq \frac{1}{\eta\cdot\ell}\cdot\sum_{s_i^k \in V(\Bs,\Br)} \sum_S \sigma_{iS}\left(s_i^{k}\right)\cdot \left(v_i^{(\beta_i)}(r_i^{\omega(k)},S)-v_i(r_i^{\omega(k)},S)\right),$$ where $r_i^{\omega(k)}$ is the replica that is matched to $s_i^k$ in $ V(\Bs,\Br)$. If we take expectation over $\Br$ and $\Bs$ on the RHS, the expectation means whenever mechanism $M^{(\beta)}$ awards buyer $i$ (with type $t_i$)  bundle $S$, $\frac{1}{\eta}\cdot\left(v_i^{(\beta_i)}(t_i, S)-v_i(t_i, S)\right)$ is contributed to the expectation. Therefore, the expectation of the RHS is the same as $$\frac{1}{\eta}\cdot\left(
\sum_{t_i\in T_i}\sum_{S\subseteq[m]} f_i(t_i)\cdot\sigma^{(\beta)}_{iS}(t_i)\cdot \left(v_i^{(\beta_i)}(t_i, S)-v_i(t_i, S)\right)\right).$$ This completes the proof of the Lemma.
\end{proof}

Now, we are ready to prove Lemma~\ref{lem:relaxed valuation stronger}.
\begin{align*}
	&\rev(M, v, D)\\
	=& \sum_i \E_{\Bs}\left[\sum_{k\in[\ell]} \frac{ p_i(s_i^k)}{\ell}\right]~~\text{(Lemma~\ref{lem:revenue by surrogates})}\\
	\leq & \sum_i\left(\E_{\Bs,\Br}\left[\sum_{s_i^k\in X(\Bs,\Br)} \frac{ p_i(s_i^k)}{\ell}\right] +\sqrt{\frac{|T_i|}{\ell}}\cdot\mathcal{R}\right)~~\text{(Corollary~\ref{cor:bound revenue by X})}\\
	\leq  &\sum_i \E_{\Bs,\Br}\left[\sum_{s_i^k\in V(\Bs,\Br)} \frac{ p_i(s_i^k)}{\ell}\right]\\
	&~~~~~~~~~~~~+\frac{1}{\eta}\cdot\sum_i \sum_{t_i\in T_i}\sum_{S\subseteq[m]} f_i(t_i)\cdot\sigma^{(\beta)}_{iS}(t_i)\cdot \left(v_i^{(\beta_i)}(t_i, S)-v_i(t_i, S)\right)+\sum_i\sqrt{\frac{|T_i|}{\ell}}\cdot\mathcal{R} ~~~\text{(Lemma~\ref{lem: gap between X and V})}\\
	\leq & \frac{1}{1-\eta}\cdot \rev(M^{(\beta)},v^{(\beta)},D)\\
	&~~~~~~~~~~~~+\frac{1}{\eta}\cdot\sum_i \sum_{t_i\in T_i}\sum_{S\subseteq[m]} f_i(t_i)\cdot\sigma^{(\beta)}_{iS}(t_i)\cdot \left(v_i^{(\beta_i)}(t_i, S)-v_i(t_i, S)\right)+\sum_i\sqrt{\frac{|T_i|}{\ell}}\cdot\mathcal{R} ~~~\text{(Lemma~\ref{lem:revenue by surrogates})}
\end{align*}

Since $|T_i|$ and $\mathcal{R}$ are finite numbers, we can take $\ell$ to be sufficiently large, so that $\sum_i\sqrt{\frac{|T_i|}{\ell}}\cdot\mathcal{R} < \epsilon$ for any $\epsilon$. Let $P^{(\beta)}$ be the set of all BIC mechanisms that satisfy the first condition in Lemma~\ref{lem:relaxed valuation stronger}. Clearly, $P^{(\beta)}$ is a compact set and contains all $M^{(\beta)}$ we constructed (by choosing different values for $\ell$). Notice that both $\rev(M^{(\beta)},v^{(\beta)},D)$ and $\sum_i \sum_{t_i\in T_i}\sum_{S\subseteq[m]} f_i(t_i)\cdot\sigma^{(\beta)}_{iS}(t_i)\cdot \left(v_i^{(\beta_i)}(t_i, S)-v_i(t_i, S)\right)$ are linear functions over the allocation/price rules of mechanism $M^{(\beta)}$. Therefore, \begin{align*}
 	&\rev(M, v, D)\\
 	\leq &\max_{M^{(\beta)}\in P^{(\beta)}} \left(\frac{1}{1-\eta}\cdot \rev(M^{(\beta)},v^{(\beta)},D)+\frac{1}{\eta}\cdot\sum_i \sum_{t_i\in T_i}\sum_{S\subseteq[m]} f_i(t_i)\cdot\sigma^{(\beta)}_{iS}(t_i)\cdot \left(v_i^{(\beta_i)}(t_i, S)-v_i(t_i, S)\right)\right).
 \end{align*}
This completes the proof of Lemma~\ref{lem:relaxed valuation stronger}.
\end{prevproof}

\newpage
\bibliographystyle{plain}
\bibliography{Yang}


\end{document}